\documentclass[11pt]{article}
\usepackage[utf8]{inputenc}
\usepackage{amsmath,amssymb}

\usepackage{amsthm}
\usepackage{comment}

\usepackage[version=4]{mhchem}  
\usepackage{graphicx}
\usepackage{xcolor}

\newcommand{\bsub}{\begin{subequations}}
  \newcommand{\esub}{\end{subequations}$\!$}

\renewcommand\theequation{\thesection.\arabic{equation}}

\usepackage[bbgreekl]{mathbbol}
\usepackage{bbm}

\renewcommand{\theequation}{\arabic{section}.\arabic{equation}}

\usepackage{epsfig}
\usepackage{subcaption}		   
\graphicspath{{pics/}}		   

\usepackage{placeins}
\usepackage{float}
\usepackage{authblk}
\usepackage[shortlabels]{enumitem}

\graphicspath{{pics/}}					

\def\bR{\mathbb{R}}
\def\bC{\mathbb{C}}

\def\1{\mathbbm{1}}
\def\G{\mathcal{G}}

\newtheorem{definition}{Definition}[section]

\def\K{{\rm K}}

\def\Re{{\rm Re}}

\def\diag{{\rm diag}}

\def\O{\mathcal{O}}

\newcommand{\R}{{\mathbb{R}}}

\def\m{\rm m}
\def\s{\rm s}
\def\mol{\rm moles}

\usepackage{geometry} \title{Symmetry-Breaking Bifurcations for Compartmental Reaction Kinetics
  Coupled by Two Bulk Diffusing Species with Comparable Diffusivities
  in 2-D}
\addtocounter{footnote}{1}
\author[]{Merlin Pelz\footnote{merlinpelz@math.ubc.ca} }
\addtocounter{footnote}{1}
\author[]{Michael J. Ward\footnote{ward@math.ubc.ca}}

\affil[]{Dept. of Mathematics, University of British Columbia, Vancouver, BC, Canada}

\begin{document}

\maketitle

\let\thefootnote\relax\footnote{Submitted to Frontiers in Applied Mathematics and Statistics.}

\begin{abstract}
  For a 2-D coupled PDE-ODE bulk-cell model, we investigate
  symmetry-breaking bifurcations that can emerge when two bulk
  diffusing species are coupled to two-component nonlinear
  intracellular reactions that are restricted to occur only within a
  disjoint collection of small circular compartments, or ``cells'', of
  a common small radius that are confined in a bounded 2-D
  domain. Outside of the union of these cells, the two bulk species
  with comparable diffusivities and bulk degradation rates diffuse and
  globally couple the spatially segregated intracellular reactions
  through Robin boundary conditions across the cell boundaries, which
  depend on certain membrane reaction rates. In the singular limit of
  a small common cell radius, we construct steady-state solutions for
  the bulk-cell model and formulate a nonlinear matrix eigenvalue
  problem that determines the linear stability properties of the
  steady-states. For a certain spatial arrangement of cells for which
  the steady-state and linear stability analysis become highly
  tractable, we construct a symmetric steady-state solution where the
  steady-states of the intracellular species are the same for each
  cell.  As regulated by the ratio of the membrane reaction rates on
  the cell boundaries, we show for various specific prototypical
  intracellular reactions, and for a specific two-cell arrangement,
  that our 2-D coupled PDE-ODE model admits symmetry-breaking
  bifurcations from this symmetric steady-state, leading to linearly
  stable asymmetric patterns, even when the bulk diffusing species
  have comparable or possibly equal diffusivities. Overall, our
  analysis shows that symmetry-breaking bifurcations can occur without
  the large diffusivity ratio requirement for the bulk diffusing
  species as is well-known from a Turing stability analysis applied to a
  spatially uniform steady-state for typical two-component
  activator-inhibitor systems.  Instead, for our theoretical
  compartmental-reaction diffusion bulk-cell model, our analysis shows
  that the emergence of stable asymmetric steady-states can be
  controlled by the ratio of the membrane reaction rates for the two
  species. Bifurcation theoretic results for symmetric and asymmetric
  steady-state patterns obtained from our asymptotic theory are
  confirmed with full numerical PDE simulations.
\end{abstract}

\setcounter{equation}{0}
\setcounter{section}{0}
\section{Introduction}

A central issue in many chemical and biological systems that involve
the coupling of diffusive processes and nonlinear reactions is to
determine conditions for which spatio-temporal patterns can form from
either a patternless or a pre-patterned state. In a pioneering
theoretical study, Alan Turing \cite{turing} established that
diffusing morphogens with different diffusivities can destablilize a
spatially uniform and stable steady-state of the nonlinear reaction
kinetics.  As applied to two-component activator-inhibitor
reaction-diffusion (RD) systems, this Turing stability analysis shows
that a sufficiently large diffusivity ratio is typically needed to
obtain spatial pattern formation from the destabilization of a
spatially uniform state, unless the nonlinear reaction kinetics are
finely tuned (cf.~\cite{pearson1}, \cite{baker}, \cite{diambra}).  For
certain chemical systems, this large diffusivity ratio requirement
needed for pattern formation may be feasible to achieve in situations
where one of the chemical species can bind to a substrate, which has
the consequence of reducing the {\em effective} diffusivity of this
species (cf.~\cite{epstein}, \cite{french}).  However, in many
cellular processes related to developmental biology and morphogenesis,
the theoretical large diffusivity ratio threshold needed for freely
diffusing morphogens to create symmetry-breaking patterns is often
unrealistic as different small molecules typically have very
comparable diffusivities (cf.~\cite{morphogen}, \cite{rauch}). In
\cite{morphogen}, various modifications of the simple ``freely
diffusing'' morphogen paradigm such as, {\em facilitated diffusion},
{\em transient binding}, {\em immobilization} and {\em transcytosis},
among others, have been postulated to play a central role in specific
applications of diffusive transport at the cellular
level. Qualitatively, the postulated overall effect of these
mechanisms is to modify an {\em effective} diffusivity ratio of the
morphogens, which can, therefore, lead to the emergence of spatial
patterns and symmetry-breaking behavior in cellular processes related
to developmental biology and early morphogenesis
(cf.~\cite{symmbreak}).

As a result, one key long-standing theoretical question in RD theory
is how to modify the two-component RD paradigm so as to robustly
generate stable spatial patterns from a spatially homogeneous state
when the time scales for diffusion of the interacting species are
comparable. By including an additional non-diffusible component, which
roughly models either membrane-bound proteins or an immobile
chemically active substrate, it has been shown (cf.~\cite{pearson2},
\cite{klika}, \cite{korv}) that this ``$2+1$'' extension of the
two-component RD framework can yield stable spatial patterns even when
the two diffusible species have a common diffusivity. In another
direction, which is based on graph-theoretic properties associated
with nonlinear reactions between multiple species that are either
immobile or freely diffusing, it has been shown that with certain
activating and inhibiting feedback relations in the chemical kinetics,
spatial patterns can form without the large diffusivity ratio
requirement (cf. \cite{macron}, \cite{macron1}, \cite{landge}). More
recently, the authors in \cite{goldstein} have revealed that in random,
multi-component, RD systems the required diffusivity threshold for
pattern formation typically decreases as the number of interacting and
diffusing species increases. 

From a theoretical viewpoint, in specific applications where a large
diffusivity ratio is a realistic assumption, it has been shown both
analytically and from numerical simulations (cf.~\cite{vanag},
\cite{ward}, \cite{frey1}, \cite{frey2}) that two-component RD systems
admit a wide range of spatially localized patterns and instabilities
that occur in the ``far-from-equilibrium'' regime, far from where a
Turing linear stability analysis will provide any insight into
pattern-forming properties.

The goal of this paper is to formulate and quantitatively analyze a
new theoretical model in a 2-D setting that robustly leads to pattern
formation even when the two diffusing species have a comparable or
equal diffusivity. More specifically, we analyze symmetry-breaking
pattern formation for a 2-D PDE-ODE bulk-cell RD model in which
spatially segregated localized reaction compartments, referred to as
``cells'', are coupled to a two-component linear bulk diffusion field
with constant bulk degradation rates. In the cells, which are assumed
to have a common radius that is small compared to the domain
length-scale and the inter-cell distances, two-component intracellular
activator-inhibitor reaction kinetics are specified. The intracelluar
species undergo an exchange with the two bulk species across the cell
boundaries, as mediated by membrane reaction rates in a Robin boundary
condition that is specified on each cell boundary. The two
extracellular diffusing bulk species, with comparable diffusivities
and degradation rates, provide the mechanism that couples the
nonlinear intracellular reactions that occur in the union of the
spatially segregated cells. We refer to this modeling framework as a
{\bf compartmental-reaction diffusion system.}

The numerical implementation of our theoretical analysis for this
model for various specific intracellular reaction kinetics reveals
that it is the ratio of the reaction rate of the inhibitor component
to that of the activator component on the compartment boundaries that
plays a central role in the initiation of symmetry-breaking
bifurcations of a symmetric steady-state. The magnitude of this ratio
ultimately controls whether linearly stable asymmetric steady-states
for the bulk-cell model can occur even when the bulk diffusivities are
comparable or equal. The bifurcation threshold condition for this key
membrane reaction rate ratio parameter is distinct from the usual
large diffusivity ratio threshold that is required for pattern
formation from a spatially uniform state for typical two-component
activator-inhibitor RD systems (cf.~\cite{nguyen}, \cite{krause}).  We
emphasize that our linear stability analysis predicting
symmetry-breaking bifurcations for the bulk-cell model, as regulated
by the membrane reaction rate ratio, is significantly more challenging
than performing a simple Turing stability analysis \cite{turing} since
it is based on the linearization of the bulk-cell model around a {\em
  spatially non-uniform} symmetric steady-state. In our previous 1-D
study \cite{1dturing}, where nonlinear reactions were restricted
either to domain boundaries or at lattice site on a 1-D periodic
chain, it has been shown for some specific nonlinear kinetics that
symmetry-breaking bifurcations can occur from a symmetric steady-state
when the ratio of membrane reaction rates exceeds a threshold.

We remark that our 2-D study, and related 1-D analysis in
\cite{1dturing}, is largely inspired by the agent-based numerical
computations in \cite{rauch} where it was shown that nonlinear kinetic
reactions restricted to lattice sites on a 2-D lattice can generate
stable Turing-type spatial patterns when coupled through a spatially
discretized two-component bulk diffusion field in which the two
diffusible species have a comparable diffusivity.

In a broader context, the study of novel pattern-forming properties
associated with compartmentalized reactions interacting through a
passive bulk diffusion field originates from the 1-D analysis in
\cite{gomez2007} for the FitzHugh-Nagumo model and the bulk-membrane
analysis of \cite{levine2005} in disk-shaped domains. In a 1-D
context, and with one bulk diffusing species, this
compartmental-reaction diffusion system modeling paradigm has been
shown to lead to triggered oscillatory instabilities for various
reaction kinetics involving conditional oscillators
(cf.~\cite{gou2015}, \cite{gou2016}, \cite{gou2017}). Amplitude
equations characterizing the local branching behavior for these
triggered oscillations have been derived in \cite{paquin_1d} using a
weakly nonlinear analysis. Applications of this framework have been
used to model intracellular polarization and oscillations in fission
yeast (cf.~\cite{xu}, \cite{xu2018}). In a 2-D domain, similar
bulk-cell models, but with only one diffusing bulk species, have been
formulated and used to model quorum-sensing behavior
(cf.~\cite{gou2d}, \cite{smjw_diff}, \cite{ridgway},
\cite{q_survey}). With regards to bulk-membrane RD models in a
multi-spatial dimensional context, where nonlinear kinetics are
restricted to the membrane, the associated pattern-forming properties
have been studied both theoretically (cf.~(\cite{ratz2015},
\cite{elliott}, \cite{madzvamuse2015}, \cite{madzvamuse2016},
\cite{paquin_memb}), and for some specific biological applications
(cf.~\cite{keshet}, \cite{ratz2012}, \cite{ratz2014}, \cite{holst},
\cite{paquin_model}).

The outline of this paper is as follows. In \S \ref{sec:2d} we
formulate our bulk-cell model and use a singular perturbation approach
in the limit of a small common cell radius to derive a nonlinear
algebraic system characterizing all steady-state solutions of the
model. In \S \ref{sec:stab} we show that the discrete eigenvalues of
the linearization of the bulk-cell model around a steady-state
solution are determined by a root-finding condition on a nonlinear
matrix eigenvalue problem. For a certain type of spatial configuration
of the cells, the bulk-cell model is shown to admit a symmetric
steady-state solution in which the steady-states of the intracellular
reactions are identical.  The possibility of symmetry-breaking
bifurcations along this symmetric steady-state solution branch,
leading to the existence of linearly stable asymmetric patterns, are
analyzed by applying solution path continuation software to our
bifurcation-theoretic analytical results. For a certain two-cell
configuration in the unit disk, and for either Gierer-Meinhardt
\cite{gm}, Rauch-Millonas \cite{rauch}, or FitzHugh-Nagumo
\cite{gomez2007} intracellular reactions, we show in \S
\ref{sec:examples} that it is the magnitude of the ratio of the
reaction rates for the two bulk species on the cell membranes that
controls whether linearly stable asymmetric patterns can bifurcate from
the symmetric steady-state. Our theoretical predictions of
symmetry-breaking behavior, leading to stable asymmetric steady-states
even when the two bulk species have comparable or equal diffusivities,
are confirmed from full PDE numerical simulations. For a
closely-spaced arrangement of cells as is typical in biological
tissues, and where our asymptotic theory no longer applies, the PDE
numerical simulations shown in \S \ref{sec:gm_close} illustrate that
symmetry-breaking bifurcations can still be controlled by the reaction
rate ratio on the cell boundaries. In particular, our numerical
results suggest that such bifurcations occur with a smaller membrane
reaction-rate ratio than for the situation where the cells are more
spatially segregated. In \S \ref{sec:discussion} we discuss our
theoretical results in a wider context, and suggest a few open
directions.

\setcounter{equation}{0}
\setcounter{section}{1}
\section{Compartmental-reaction diffusion system in
  2-D}\label{sec:2d}

\subsection{Model formulation}\label{sec:model}
We consider a bounded 2-D domain with length scale $L$, denoted by
$\Omega^L \subset \bR^2$, that contains $m$ disconnected circular
compartments $\Omega_j^L$, for $j\in\{1,...,m\}$, referred to as
``cells''. We will assume that these cells have a common radius that
is small in comparison with the length scale $L$ of the domain.
The bulk or extracellular medium is the region
$\Omega^L\backslash \bigcup_{j=1}^m \Omega_j^L$.

In the bulk we assume that there are two extracellularly diffusing and
degrading chemical species with concentrations $U$ and $V$. These
messenger molecules are synthesized on the ``cell'' membranes through
the interaction with two corresponding intracellular species $M_{j}$
and $H_{j}$. With the molecule counts
$\mathfrak{U}, \mathfrak{V}, \mathfrak{M}_{j}$ and $\mathfrak{H}_{j}$
corresponding to respectively $U, V, M_{j}$ and $H_{j}$, the chemical
equations are
\begin{equation}
  \ce{\mathfrak{U} <=>[$\beta_U$][$\beta_U$] \mathfrak{M}_{j}}\,,
\qquad
  \ce{\mathfrak{V} <=>[$\beta_V$][$\beta_V$] \mathfrak{H}_{j}}\,.
\end{equation}
Here we made the assumption that the exponential forward reaction
rates equal the backward reaction rates and that all compartments are
identical in that they have common membrane reaction rates. The
intra-compartmental species, in turn, are produced by certain reaction
kinetics, denoted by $f(M, H)$ and $g(M,H)$, that are assumed to be
identical in each compartment.

More precisely, in dimensional variables, our bulk-cell coupled model is
\begin{subequations} \label{dim:pde}
\begin{eqnarray}
        \text{bulk} &&
        \begin{cases}
          \partial_T U = D_U \,\Delta_X U - \kappa_U \,U\,, & {\bf X}\in
          \Omega^L\backslash\bigcup_{j=1}^m \Omega_j^L \,, \\
          \partial_T V = D_V \, \Delta_X V - \kappa_V \,V\,, & {\bf X}\in
          \Omega^L\backslash\bigcup_{j=1}^m \Omega_j^L \,, \\
            \partial_{\tilde{n}_X} U = \partial_{\tilde{n}_X} V=0\,, & {\bf
X}\in
            \partial\Omega^L \,,\qquad \qquad \text{(Neumann condition)}
        \end{cases} \\
        \text{reaction fluxes} &&
        \begin{cases}
          D_U \,\partial_{n_{j,X}} U = \beta_{U,1}\, U - \beta_{U,2}\, M_{j}\,,
& {\bf X}\in
    \partial\Omega_j^L  \,, \qquad \quad \text{(Robin condition)}\\
    D_V \,\partial_{n_{j,X}} V = \beta_{V,1}\, V - \beta_{V,2}\, H_{j}\,, &
{\bf X}\in
    \partial\Omega_j^L\,,
    \\
        \end{cases}\\
        \text{compartments} &&
        \begin{cases}
          \frac{d}{dT} M_j = \kappa_R\, \mu_c \;
          f\left(\frac{1}{\mu_c}M_j,\frac{1}{\mu_c}H_j\right) +
          \int_{\partial \Omega_j^L} (\beta_{U,1}\, U - \beta_{U,2}\,
M_{j})\;dS_X\,, \\
          \frac{d}{dT} H_j = \kappa_R\, \mu_c \;
          g\left(\frac{1}{\mu_c}M_j, \frac{1}{\mu_c} H_j\right) +
          \int_{\partial \Omega_j^L} (\beta_{V,1}\, V - \beta_{V,2}\,
H_{j})\;dS_X \quad
          \text{(reaction kinetics)}\,,
        \end{cases}
\end{eqnarray}
\end{subequations}
with $j\in\{1,...,m\}$ and where $n_{j,X}$ is the outward unit normal
vector to $\Omega_j^L$ while $\tilde{n}_X$ is the outward unit normal
vector to $\Omega^L$. The diffusivities (diffusion coefficients) for
$U$ and $V$ are $D_U$ and $D_V$, and $U$ and $V$ are degrading in the
bulk with exponential rate constants $\kappa_U$ and $\kappa_V$,
respectively. The exponential reaction rates on the compartment
boundaries are $\beta_U$ and $\beta_V$ with corresponding rates $\beta_{U,1}$
and $\beta_{V,1}$ per area times length and $\beta_{U,2}$ and $\beta_{V,2}$ per
length and time units, and $\mu_c$ is a normalizing constant for the
intracellular species. Lastly, $\kappa_R$ is a dimensional reaction rate for
the intracellular reactions.

\begin{figure}
    \centering
    \def\svgwidth{0.65\textwidth}
    \def\svgheight{4.2cm}
	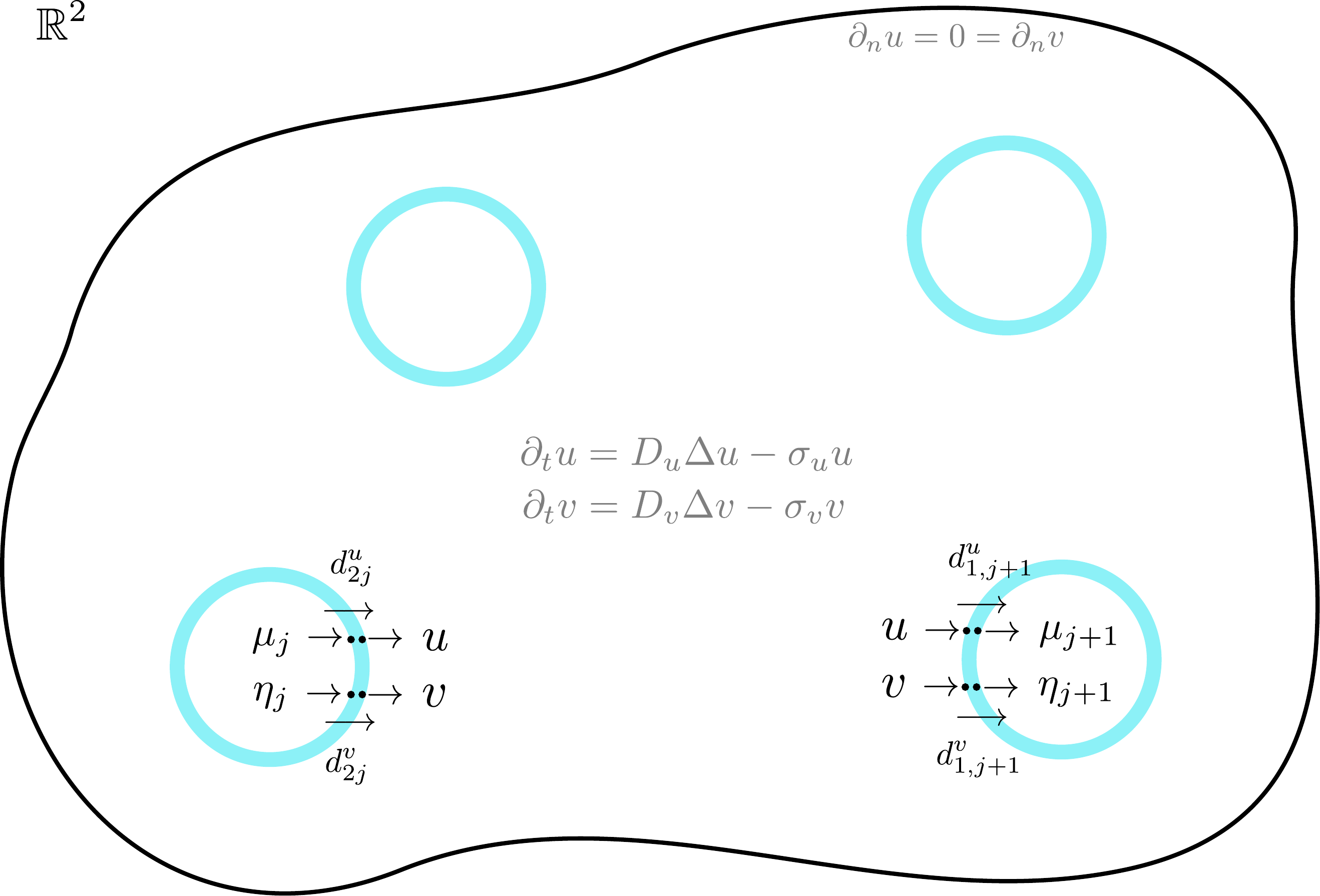
        \caption{A 2-D bounded domain with four diffusion-coupled
          circular cells of a common radius. In the $j^{\mbox{th}}$
          cell, activator-inhibitor reaction kinetics occur for the
          activator $\mu_j$ and inhibitor $\eta_j$.  Across the cell
          membrane, there is an exchange between the intracellular and
          bulk species. In the bulk, $u$ and $v$ diffuse and undergo
          degradation. The cells are assumed to be small but are drawn
          larger here for illustration only.}
    \label{fig:generaldomain}
\end{figure}

In Appendix \ref{app:nondim} we non-dimensionalize (\ref{dim:pde})
to obtain the dimensionless PDE-ODE model
\begin{subequations} \label{eqsys:full}
    \begin{eqnarray}
	\text{bulk} &&
	\begin{cases}
          \partial_t u = D_u \Delta u - \sigma_u u\,, & {\bf x} \in
          \Omega\backslash\bigcup_{j=1}^m \Omega_j \,, \\
          \partial_t v = D_v \Delta v - \sigma_v v\,, & {\bf x} \in
          \Omega\backslash\bigcup_{j=1}^m \Omega_j \,, \\ 
          \partial_{\tilde{n}} u  = \partial_{\tilde{n}} v=0 \,, &
          {\bf x} \in \partial\Omega\,,
	\end{cases} \label{eqsys:bulk}\\
	\text{reaction fluxes} &&
	\begin{cases}
          \varepsilon D_u \partial_{n_j} u = d_{1}^u u - d_{2}^u \mu_{j}\,, &
          {\bf x} \in \partial\Omega_j  \,,\\
    \varepsilon D_v \partial_{n_j} v = d_{1}^v v - d_{2}^v \eta_{j}\,,
    & {\bf x} \in \partial\Omega_j \,, \\
	\end{cases} \label{eqsys:fluxes}\\
	\text{compartments} &&
	\begin{cases}
          \frac{d\mu_j}{dt} = f(\mu_j, \eta_j) + \frac{1}{\varepsilon}
          \int_{\partial\Omega_j} (d_{1}^u u - d_{2}^u \mu_{j})\;dS \,, \\
          \frac{d\eta_j}{dt} = g(\mu_j, \eta_j) + \frac{1}{\varepsilon}
          \int_{\partial\Omega_j} (d_{1}^v v - d_{2}^v \eta_{j})\;dS \,,
	\end{cases} \label{eqsys:comp}
\end{eqnarray}
\end{subequations}
for $j\in\lbrace{1,\ldots,m\rbrace}$. Here $n$ and $\tilde{n}$ are the
outward unit normal vectors to $\Omega_j$ and $\Omega$, respectively,
and we have dropped the label ``$x$" for $\Delta$ and the outward unit
normal vectors. In (\ref{eqsys:full}), the compartments are disks of a
common radius $\varepsilon\ll 1$ centered at ${\bf x}_j\in \Omega$,
i.e.
$\Omega_j\equiv \lbrace{ {\bf x}\, \vert \, |{\bf x}-{\bf x}_j|\le
  \varepsilon \rbrace}$. We will refer to $d_1^{u}$, $d_1^{v}$,
$d_2^{u}$, and $d_2^{v}$ as dimensionless membrane reaction rates. An
illustration of the bulk-cell model is shown in Figure
\ref{fig:generaldomain}.

We will use strong localized perturbation theory \cite{ward} to
construct the steady-state solutions of (\ref{eqsys:full}) and to
analyze their linear stability properties in the asymptotic limit
$\varepsilon\ll 1$ and under the assumption that $m$ circular cells
are well-separated in the sense that the cell centers satisfy
$|{\bf x}_i-{\bf x}_j|=\O(1)$, for $i,j \in \{1, ..., m\}$ and
$i\neq j$.

\subsection{Asymptotic construction of the steady-states}\label{sec:steady-state}

Our main goal is to construct a symmetric steady-state solution for
(\ref{eqsys:full}) in which the concentration of each species is the
same inside and in the local vicinity of each compartment. We will
show below that even when the bulk diffusing species have comparable
diffusivities this symmetric steady-state is unstable to 
symmetry-breaking perturbations that occur beyond a pitchfork bifurcation
point associated with the membrane reaction rate ratio
$\rho\equiv {d_1^v/d_1^u} = {d_2^v/d_2^u}$. This leads to the
existence of linearly stable asymmetric steady-state solutions to
(\ref{eqsys:full}).

In the absence of diffusion, the ODE system for the
intra-compartmental species is decoupled from the bulk medium and
reduces to
\begin{equation} \label{eq:uncoupledeq}
    \dot{\mu}(t) = f(\mu,\eta)\,, \qquad \dot{\eta}(t) = g(\mu,\eta)\,.
\end{equation}
Let $(\mu_e, \eta_e)$ be an equilibrium point for
\eqref{eq:uncoupledeq} and label
$F(\mu, \eta) \equiv (f (\mu, \eta), g(\mu, \eta))$. For a specific
parameter set, the linear stability property of the equilibrium state
is characterized by whether the eigenvalues $\lambda$ of the Jacobian
matrix $DF(\mu_e, \eta_e)$ have positive (unstable, exponentially
growing perturbations) or negative (stable, exponentially decaying
perturbations) real parts $\Re(\lambda)$. However, when there is bulk
diffusion and the compartments are coupled through the bulk, the 
steady-state solution in the compartments depends on the bulk
diffusivities, the membrane reaction rates, and the spatial
configuration of the cells.

We now use the method of matched asymptotic expansions to construct
steady-state solutions for (\ref{eqsys:full}).  In the $j^{\text{th}}$
inner region, defined within an $\mathcal{O} (\varepsilon)$
neighborhood of the boundary of the $j^{\text{th}}$ cell, we introduce
the local variables ${\bf y}_j =\varepsilon^{-1} ({\bf x}-{\bf x}_j)$,
$u_j({\bf x}) = u(\varepsilon {\bf y}_j + {\bf x}_j)$, and
$v_j({\bf x}) = v(\varepsilon {\bf y}_j + {\bf x}_j)$, where
$p_j\equiv |{\bf y}_j|$.  Upon writing the steady-state of
(\ref{eqsys:bulk}) in terms of the inner variables, for
$\varepsilon \to 0$ the steady-state problem in the $j^{\text{th} }$
inner region is $\Delta u_j = 0$ and $\Delta v_j=0$, for
$p_j\geq 1$, subject to
$D_u\, \partial_{p_j} u_j = d_{1}^{u} u_j - d_{2}^{u} \mu_j$ and
$D_v\, \partial_{p_j} v_j = d_{1}^{v} v_j - d_{2}^{v} \eta_j$ on
$p_j=1$. The radially symmetric solutions to these problems are
\begin{equation} \label{Inner_Solution}
  u_j(p_j) = A_j^{u} \log p_j + \frac{1}{d_{1}^{u}} \left( D_u\, A_j^{u} +
    d_{2}^{u}\mu_j \right)\,, \qquad
  v_j(p_j) = A_j^{v} \log p_j + \frac{1}{d_{1}^{v}} \left( D_v\, A_j^{v} +
    d_{2}^{v}\eta_j \right)\,, 
\end{equation}
for $j\in\lbrace{1, \ldots ,m\rbrace}$, where $A_j^{u}$ and $A_j^{v}$
for $j=1,\ldots,m $ are constants to be determined.  Upon substituting
(\ref{Inner_Solution}) into the steady-state problem of
(\ref{eqsys:comp}), we obtain for the $j^{\mbox{th}}$ cell that
\begin{equation}\label{Intra}
  f(\mu_j,\eta_j) + 2 \pi D_u \, A_j^{u} =0 \,, \qquad
  g(\mu_j,\eta_j) + 2 \pi D_v\, A_j^{v} =0 \,, \qquad j \in \lbrace{1,
    \ldots, m\rbrace}\,.
\end{equation}
Next, we must determine $A_j^{u}$ and $A_j^{v}$ by matching the far-field
behavior of the inner solutions \eqref{Inner_Solution} to the outer
solutions defined in the bulk region.

In the limit $\varepsilon\to 0$, in the bulk region the compartments
formally shrink to points and from the far-field behavior of
(\ref{Inner_Solution}), when written in outer variables, we obtain
that the steady-state bulk species $U$ satisfies
\begin{subequations}\label{Outer_U} 
\begin{eqnarray}
  && \Delta U - \omega_u^2 \,U =0 \,, \quad {\bf x} \in \Omega \setminus
     \{{\bf x}_1,\ldots,{\bf x}_m \} \,; \qquad \partial_n U =0 \,,\quad
     {\bf x} \in \partial \Omega \,; \label{Outer_Ua}   \\
  && U \sim  A_j^u \log |{\bf x}-{\bf x}_j|  + \frac{A_j^{u}}{\nu} +
     \frac{1}{d_{1}^{u}}(D_uA_j^{u} + d_{2}^{u} \mu_j)\,, \quad \text{as} \quad
     {\bf x} \to {\bf x}_j\,, \qquad j \in \lbrace{1,\ldots,m\rbrace}\,,
     \label{Outer_Ub} 
\end{eqnarray}
\end{subequations}
where $\nu \equiv {-1/\log\varepsilon}\ll 1$ and
$\omega_u\equiv \sqrt{\sigma_u/D_u}$.  Similarly,
with $\omega_v\equiv \sqrt{\sigma_v/D_v}$, for the bulk species
$V$ we have that
\begin{subequations}\label{Outer_V} 
\begin{eqnarray}
&&  \Delta V - \omega_v^2 \,V =0\,, \quad {\bf x} \in \Omega \setminus
  \{{\bf x}_1,\ldots,{\bf x}_m \} \,; \qquad \partial_n V =0 \,,\quad
  {\bf x} \in \partial \Omega \,; \label{Outer_Va} \\
&&  V \sim  A_j^v \log |{\bf x}-{\bf x}_j| +\frac{A_j^{v}}{\nu} +
  \frac{1}{d_{1}^{v}}(D_vA_j^{v} + d_{2}^{v} \eta_j)\,, \quad \text{as} \quad
   {\bf x} \to {\bf x}_j\,, \qquad j\in\lbrace{1,\ldots,m\rbrace}\,.
   \label{Outer_Vb} 
\end{eqnarray}
\end{subequations}

To represent solutions to (\ref{Outer_U}) and (\ref{Outer_V}), we
introduce the \emph{reduced-wave} Green's function
$G_\omega({\bf x},{\bf x}_j)$ that satisfies
\begin{subequations}\label{R_Green}
\begin{eqnarray}
&&  \Delta G_\omega\, - \,\omega^2  G_\omega = -\delta({\bf x} - {\bf x}_j)\,,
  \quad
  {\bf x} \in \Omega\,; \qquad  \partial_n G_\omega =0 \,,\quad {\bf x} \in
  \partial \Omega\,; \label{R_Green_a}\\
&&  G_\omega \sim - \frac{1}{2\pi} \log|{\bf x}-{\bf x}_j|
  + R_\omega({\bf x}_j) + o(1)\,,\quad \text{as} \quad {\bf x}\to{\bf x}_j\,.
                                \label{R_Green_b}
\end{eqnarray}
\end{subequations}
Here $R_{\omega}({\bf x}_j)$ is the regular, or non-singular, part of
the singularity at ${\bf x}={\bf x}_j$.  The solutions to
(\ref{Outer_U}) and (\ref{Outer_V}) are represented as
\begin{equation}\label{U_Green}
  U({\bf x}) =  -2 \pi \sum_{i=1}^{m} A_i^{u} G_{\omega_u}({\bf x}; {\bf x}_i)\,,
  \qquad
  V({\bf x}) =  -2 \pi \sum_{i=1}^{m} A_i^{v} G_{\omega_v}({\bf x}; {\bf x}_i)\,.
\end{equation}

The pre-specification of the regular part of each singularity
condition in (\ref{Outer_U}) and (\ref{Outer_V}) yields a
constraint. These constraints provide algebraic systems for $A_j^u$
and $A_j^v$ for $j\in\lbrace{1,\ldots,m\rbrace}$.  By expanding
(\ref{U_Green}) as ${\bf x} \to {\bf x}_j$, we enforce that that the
non-singular terms in the resulting expression agree with the
conditions that are required in (\ref{Outer_Ub}) and (\ref{Outer_Vb})
for each $j\in\lbrace{1,\ldots,m\rbrace}$. This leads to linear
algebraic systems for
${\bf {\mathcal{A}}}^{u} \equiv (A_1^u, \ldots, A_m^u)^T$ and
${\bf {\mathcal{A}}}^{v} \equiv (A_1^v, \ldots, A_m^v)^T$, given in
matrix form by
\begin{equation}\label{Linear_System} 
  \left( \left(1+\frac{\nu D_u}{d_1^u}\right) I + 2\pi \nu
  \mathcal{G}_{\omega_u}
  \right){\bf {\mathcal{A}}}^u =  - \frac{\nu d_2^u}{d_1^u}\, {\bf \mu}\,,
  \qquad
  \left( \left(1+\frac{\nu D_v}{d_1^v}\right) I + 2\pi \nu
  \mathcal{G}_{\omega_v} \right){\bf {\mathcal{A}}}^v = - \frac{\nu
    d_2^v}{d_1^v}\, {\bf \eta}\,,
\end{equation}
where ${\bf \mu}\equiv (\mu_1, \ldots, \mu_m)^T$ and
${\bf \eta}\equiv (\eta_1, \ldots, \eta_m)^T$. In
\eqref{Linear_System}, $\mathcal{G}_{\omega}$ with either
$\omega=\omega_u$ or $\omega=\omega_v$ is the symmetric reduced-wave
Greens' interaction matrix defined by
\begin{equation}\label{GreensMatrix}
\mathcal{G}_{\omega} \equiv
\begin{pmatrix}
R_{\omega 1}    & G_{\omega 12} & \dots & G_{\omega 1m}\\
G_{\omega 21} &  R_{\omega 2}   & \dots & G_{\omega 2m}\\
\vdots & \vdots & \ddots &  \vdots \\
G_{\omega m1} & G_{\omega m2} & \dots & R_{\omega m}  
\end{pmatrix}\,.
\end{equation}
Here
$G_{\omega ji} = G_{\omega ij} \equiv G_\omega({\bf x}_j;{\bf x}_i)$
for $i \neq j$, and $R_{\omega j} \equiv R_\omega({\bf x}_j)$ for
$j \in \lbrace{1, \ldots,m\rbrace}$, are obtained from the solution to
(\ref{R_Green}).

To determine a nonlinear algebraic system that characterizes our
steady-state solution, we solve (\ref{Linear_System}) for
${\bf {\mathcal{A}}}^v$ and ${\bf {\mathcal{A}}}^u$, and substitute
the resulting expressions into (\ref{Intra}). In this way, we obtain
a $2m$ dimensional nonlinear algebraic system for $\mu_j$ and $\eta_j$,
for $j=1,\ldots,m$, given by
\begin{subequations}\label{non:full}
\begin{equation}
  f(\mu_j,\eta_j) -{\bf e}_j^T \Theta_u {\bf \mu}= 0 \,, \qquad
  g(\mu_j,\eta_j) -{\bf e}_j^T \Theta_v {\bf \eta}= 0 \,, \quad
  \mbox{for} \quad j\in\lbrace{1,\ldots,m\rbrace} \,, \label{non:full_1}
\end{equation}
where ${\bf e}_j\equiv (0,\ldots,0,1,0,\ldots,0)^T$ is the unit vector
in the $j^{\mbox{th}}$ direction. In (\ref{non:full_1}), $\Theta_u$
and $\Theta_v$ are defined by
\begin{equation}\label{non:full_mat}
  \Theta_u \equiv 2\pi \nu D_u \frac{d_2^{u}}{d_1^{u}} \left[
      \left(1+\frac{\nu D_u}{d_1^u}\right) I + 2\pi \nu
      \mathcal{G}_{\omega_u} \right]^{-1} \,, \qquad
    \Theta_v \equiv 2\pi \nu D_v \frac{d_2^{v}}{d_1^{v}} \left[
      \left(1+\frac{\nu D_v}{d_1^v}\right) I + 2\pi \nu
      \mathcal{G}_{\omega_v} \right]^{-1} \,.
\end{equation}
\end{subequations}  

We can simplify our steady-state analysis for the special case where
$g(\mu, \eta)$ is linear and inhibiting in $\eta$, with the form
\begin{equation}\label{g:linear}
  g(\mu, \eta) = g_1(\mu) - g_2\eta\,,
\end{equation}
where $g_2\geq 0$ is a constant. This specific form applies to
Gierer-Meinhardt \cite{gm}, Rauch-Millonas \cite{rauch}, and
FitzHugh-Nagumo \cite{gomez2007} reaction kinetics, and is relevant
for the illustrations of the theory given in \S \ref{sec:examples}. In
this case, we obtain from the second equation in (\ref{non:full_1})
that
\begin{equation}\label{g:eta_solve}
  {\bf \eta} = \left[g_2 I + \Theta_v\right]^{-1} {\bf g}_1 \qquad
  \mbox{where} \qquad {\bf g}_1\equiv (g_1(\mu_1),\ldots,
  g_1(\mu_m))^T \,.
\end{equation}
Then, from the first equation in (\ref{non:full_1}) we obtain
an $m$ dimensional nonlinear algebraic system for
${\bf \mu}=(\mu_1,\ldots,\mu_m)^T$ given by
\begin{equation}\label{g:mu_solve}
  f\left(\mu_j,{\bf e}_j^T(g_2I+\Theta_v)^{-1}{\bf g}_1\right) -
  {\bf e}_j^T \Theta_u {\bf \mu}= 0 \,, \qquad j\in\lbrace{1,\ldots,m\rbrace}
  \,.
\end{equation}
Next, we define a symmetric cell arrangement for which the
steady-state analysis can be further simplified.

\begin{definition}\label{def:symm} A {\bf symmetric cell arrangement} is
  defined by the condition that the symmetric Green's matrix
  ${\mathcal G}_{\omega}$ satisfies the following two properties:
\begin{itemize}
\item {\bf Property 1:} ${\bf e}\equiv (1,\ldots,1)^T$ is an eigenvector
  of ${\mathcal G}_{\omega}$ for all $\omega>0$:
\item {\bf Property 2:} The eigenspace of ${\mathcal G}_{\omega}$ orthogonal
  to ${\bf e}$ is independent of $\omega$.
\end{itemize}
\end{definition}

These two properties certainly hold when ${\mathcal G}_\omega$ is a
circulant matrix. In particular, ${\mathcal G}_{\omega}$ is a circulant
matrix when $m$ small cells are equidistantly spaced on a ring that is
concentric within a circular domain $\Omega$. Such an arrangement of cells
is called a {\bf ring pattern}.

For a symmetric cell arrangement, ${\mathcal G}_{\omega_u}$ and
${\mathcal G}_{\omega_v}$ have a common eigenspace, and so we can seek
a symmetric solution to (\ref{non:full}) of the form
\begin{equation}\label{symm:scal}
  {\bf \mu} =\mu_c {\bf e} \,, \quad   {\bf \eta} =\eta_c {\bf e} \,, \quad
  {\bf {\mathcal{A}}}^{u} = A_c^u{\bf e} \,, \quad
  {\bf {\mathcal{A}}}^{v} = A_c^v{\bf e} \,,
\end{equation}
where the scalars $\mu_c$, $\eta_c$, $A_c^u$, and $A_c^v$ are to be found.
Upon substituting (\ref{symm:scal}) into (\ref{non:full}), we obtain that
$\mu_c$ and $\eta_c$ satisfy the nonlinear algebraic system
\begin{equation}\label{symm:scal_sol1}
  f(\mu_c,\eta_c)-\alpha_u \mu_c =0 \,, \qquad
  g(\mu_c,\eta_c)-\alpha_v \eta_c =0 \,,
\end{equation}
where $\alpha_u$ and $\alpha_v$, denoting the eigenvalues of $\Theta_u$ and
$\Theta_v$ for the eigenvector ${\bf e}$, respectively, are defined by
\begin{subequations}\label{symm:alpkap}
\begin{equation}\label{symm:alpha}
  \alpha_u \equiv \frac{2\pi \nu D_u {d_2^u/d_1^u}}{1+{\nu D_u/d_1^u}+
    2\pi \nu \kappa_u } \,, \qquad
  \alpha_v \equiv \frac{2\pi \nu D_v {d_2^v/d_1^v}}{1+{\nu D_v/d_1^v}+
    2\pi \nu \kappa_v } \,.
\end{equation}
Here $\kappa_u$ and $\kappa_v$ are the eigenvalues of the Green's
matrices for the eigenvector ${\bf e}$, given by
\begin{equation}\label{symm:kappa}
  {\mathcal G}_{\omega_u} {\bf e} = \kappa_u {\bf e} \,, \qquad
      {\mathcal G}_{\omega_v} {\bf e} = \kappa_v {\bf e} \,.
\end{equation}
\end{subequations}
Moreover, if $g(\mu,\eta)$ has the specific form in (\ref{g:linear}),
we obtain from (\ref{g:mu_solve}) that for a symmetric pattern of cells,
there is a symmetric steady-state solution whenever there is a root $\mu_c$
to the scalar nonlinear algebraic equation
\begin{equation} \label{symm:scalar}
    f\left(\mu_c, \frac{g_1(\mu_c)}{g_2+\alpha_v}\right) - \alpha_u \mu_c = 0\,.
\end{equation}

In summary, for a symmetric pattern of cells, the asymptotic
construction of a symmetric steady-state solution for
(\ref{eqsys:full}) is reduced to the much simpler problem of
determining a solution to the two-dimensional nonlinear algebraic
problem (\ref{symm:scal_sol1}) for general reaction kinetics, or to
(\ref{symm:scalar}) when $g$ has the specific form in
(\ref{g:linear}). In these algebraic problems, the eigenvalues
$\kappa_u$ and $\kappa_v$, as needed in (\ref{symm:alpha}), are the
constant row sums of the Green's matrices for the two bulk
species. The bulk diffusivities, the membrane reaction rates, and the
spatial configuration of the cells all influence $\alpha_u$ and
$\alpha_v$.

\subsection{Symmetry-breaking bifurcations}\label{sec:symm-break}

To detect any symmetry-breaking pitchfork bifurcation points along the
symmetric steady-state solution branch we can perform a linear stability
analysis of (\ref{eqsys:full}) around the steady-state solution and
seek $\lambda=0$ eigenvalue crossings. An equivalent, but simpler,
approach to detect zero-eigenvalue crossings for the linearized
problem is to determine bifurcation points associated with the
linearization of the nonlinear algebraic system (\ref{non:full})
around a symmetric steady-state.

To do so, we introduce the perturbations
\begin{equation}\label{break:pert}
  {\bf \mu} =\mu_c {\bf e} + {\bf \tilde{\mu}} \,, \qquad
  {\bf \eta} =\eta_c {\bf e} + {\bf \tilde{\eta}} \,, \qquad
  {\bf {\mathcal{A}}}^{u} = A_c^u{\bf e} + {\bf \tilde{{\mathcal{A}}}}^{u} 
  \,, \qquad {\bf {\mathcal{A}}}^{v} = A_c^v{\bf e} +
  {\bf \tilde{{\mathcal{A}}}}^{v} \,,
\end{equation}
into (\ref{non:full}) and linearize the resulting system. In this way,
we obtain that a symmetry-breaking bifurcation occurs if and only if
there is a non-trivial solution to the $2m\times 2m$ homogeneous
linear system
\begin{equation}\label{break:full}
  \begin{pmatrix}
     f_{\mu}^{c} I - \Theta_u & f_{\eta}^c I \\
       g_{\mu}^{c} I  & g_{\eta}^c I - \Theta_v 
    \end{pmatrix}
    \begin{pmatrix}
      {\bf \tilde{\mu}} \\
      {\bf \tilde{\eta}}
    \end{pmatrix} = \begin{pmatrix}
      {\bf 0} \\
      {\bf 0}
    \end{pmatrix}\,,
\end{equation}
at some point along the symmetric solution branch given by
(\ref{symm:scal_sol1}).  In (\ref{break:full}) we have labeled
$f_{\mu}^c$ by $f_{\mu}^{c}\equiv \partial_\mu f(\mu,\eta)$ when
evaluated at $\mu=\mu_c$ and $\eta=\eta_c$, while $I$ is the
$m\times m$ identity matrix. For the special case where $g$ has the
specific form in (\ref{g:linear}), we can solve (\ref{break:full}) for
${\bf \tilde{\eta}}$ and reduce (\ref{break:full}) to the
$m$-dimensional homogeneous linear system
\begin{equation}\label{break:simp}
  \left( f_{\mu}^c I + f_{\eta}^c g_{1}^{\prime}(\mu_c)
    \left(g_2 I + \Theta_v\right)^{-1}
    -\Theta_u\right) {\bf {\tilde \mu}}= {\bf 0} \,.
\end{equation}

Next, by Property 2 for a symmetric cell arrangement, it follows that
${\mathcal G}_{\omega_u}$ and ${\mathcal G}_{\omega_v}$ have a common
orthogonal eigenspace
${\mathcal Q}^{\perp}\equiv \mbox{span}\lbrace{ {\bf q}_2,\ldots,{\bf
    q}_m\rbrace}$, where ${\bf q}_j^T{\bf e}=0$ for
$j\in\lbrace{2,\ldots,m\rbrace}$ and ${\bf q}_i^T {\bf q}_j=0$ for
$i\neq j$. The eigenvalues of ${\mathcal G}_{\omega_u}$ and
${\mathcal G}_{\omega_v}$ in this common eigenspace are labeled by
\begin{equation}\label{break:kappa_perp}
  {\mathcal G}_{\omega_u} {\bf q}_j = \kappa_{u,j}^{\perp} {\bf q}_j \,,
  \qquad
  {\mathcal G}_{\omega_v} {\bf q}_j = \kappa_{v,j}^{\perp} {\bf q}_j \,,
  \qquad j\in\lbrace{2,\ldots,m\rbrace} \,,
\end{equation}
so that
\begin{equation}\label{break:alpha_perp}
  \Theta_{u} {\bf q}_j = \alpha_{u,j}^{\perp} {\bf q}_j \,,
  \qquad
  \Theta_{v} {\bf q}_j = \alpha_{v,j}^{\perp} {\bf q}_j \,,
    \qquad j\in\lbrace{2,\ldots,m\rbrace} \,,
\end{equation}
with
\begin{equation}\label{break:alpha_def}
  \alpha_{u,j}^{\perp} \equiv \frac{2\pi \nu D_u {d_2^u/d_1^u}}{1+{\nu D_u/d_1^u}+
    2\pi \nu \kappa_{u,j}^{\perp} } \,, \qquad
  \alpha_{v,j}^{\perp} \equiv \frac{2\pi \nu D_v {d_2^v/d_1^v}}{1+{\nu D_v/d_1^v}+
    2\pi \nu \kappa_{v,j}^{\perp}} \,. 
\end{equation}

By setting ${\bf \tilde \mu}={\tilde \mu}_c {\bf q}_j$ and
${\bf \tilde \eta}={\tilde \eta}_c {\bf q}_j$ in
(\ref{break:full}), we conclude that a symmetry-breaking bifurcation
occurs for the $j^{\mbox{th}}$ mode with $j\in\lbrace{2,\ldots,m\rbrace}$
whenever
\begin{equation}\label{break:red}
  \begin{pmatrix}
     f_{\mu}^{c}  - \alpha_{u,j}^{\perp} & f_{\eta}^c  \\
       g_{\mu}^{c}  & g_{\eta}^c  - \alpha_{v,j}^{\perp}
    \end{pmatrix}
    \begin{pmatrix}
      \tilde{\mu}_c \\
      \tilde{\eta}_c
    \end{pmatrix} = \begin{pmatrix}
      0 \\
      0
    \end{pmatrix}\,,
\end{equation}
has a nontrivial solution. This is equivalent to the condition that
\begin{equation}\label{break:red_new}
  \left(f_{\mu}^c - \alpha_{u,j}^{\perp}\right)   \left(g_{\eta}^c -
    \alpha_{v,j}^{\perp}\right) - f_{\eta}^cg_{\mu}^c = 0 \,, \qquad
  j\in\lbrace{2,\ldots, m \rbrace}\,,
\end{equation}
is satisfied at some point along the symmetric solution branch
defined by the solution to (\ref{symm:scal_sol1}).

Finally, for the special case where $g$ has the specific form in
(\ref{g:linear}), we obtain that there is a symmetry-breaking
bifurcation for the $j^{\mbox{th}}$ mode, with
$j\in\lbrace{2,\ldots,m\rbrace}$, when there is a root to the scalar
problem
\begin{equation}\label{break:red_simp}
  f_{\mu}^{c} + \frac{f_{\eta}^c g_{1}^{\prime}(\mu_c)}{g_2 + \alpha_{v,j}^{\perp}}
  - \alpha_{u,j}^{\perp} = 0 \,,
\end{equation}
whenever $\eta_c={g_1(\mu_c)/(g_2+\alpha_v)}$ where $\mu_c$ satisfies
(\ref{symm:scalar}).

In the examples shown in \S \ref{sec:examples} we will use
$\rho\equiv {d_1^v/d_1^u} = {d_2^v/d_2^u}$ as the bifurcation
parameter to detect whether symmetry-breaking bifurcations can occur
along the symmetric solution branch.

\subsection{A symmetric cell arrangement with two cells} \label{sec:twocell}

Consider a symmetric cell arrangement with two cells, i.e. $m=2$, for the
special case where $g$ has the form in (\ref{g:linear}). Then, to
determine all steady-state solutions we need only solve the nonlinear
algebraic system (\ref{g:mu_solve}) for $\mu_1$ and $\mu_2$. The
symmetric steady-state solution, for which $\mu_c\equiv\mu_1=\mu_2$,
is obtained by solving the scalar problem (\ref{symm:scalar}). To
detect whether symmetry-breaking bifurcations can occur, we note that
${\bf q}_2=(1,-1)^T$ spans the common eigenspace of
${\mathcal G}_{\omega_u}$ and ${\mathcal G}_{\omega_v}$ orthogonal to
${\bf e}$, and that $\kappa_u=R_{\omega_u 1}-G_{\omega_u 12}$ and
$\kappa_v=R_{\omega_v 1}-G_{\omega_v 12}$ are the associated
eigenvalues for ${\bf q}_2$.  This yields that the root-finding condition
(\ref{break:red_simp}) becomes
\begin{equation}\label{two:red_simp}
  f_{\mu}^{c} + \frac{f_{\eta}^c g_{1}^{\prime}(\mu_c)}{g_2 + \alpha_{v,2}^{\perp}}
  - \alpha_{u,2}^{\perp} = 0 \,,
\end{equation}
where in terms of the entries of the Green's matrices we have
\begin{equation}\label{two:alpha_def}
  \alpha_{u,2}^{\perp} \equiv \frac{2\pi \nu D_u {d_2^u/d_1^u}}{1+{\nu D_u/d_1^u}+
    2\pi \nu \left[R_{\omega_u 1}-G_{\omega_u 12}\right] } \,, \qquad
  \alpha_{v,2}^{\perp} \equiv \frac{2\pi \nu D_v {d_2^v/d_1^v}}{1+{\nu D_v/d_1^v}+
    2\pi \nu \left[ R_{\omega_v 1}-G_{\omega_v 12}\right]} \,. 
\end{equation}

To detect any pitchfork bifurcation points on the symmetric
steady-state branch parameterized by
$\rho={d_1^v/d_1^u}={d_2^v/d_2^u}$ we numerically solve
(\ref{symm:scalar}) together with (\ref{two:red_simp}). In \S
  \ref{sec:examples} we illustrate this approach for certain reaction
  kinetics when $\Omega$ is the unit disk. The advantage of
  considering a disk-shaped confining domain is that the reduced-wave
  Green's function is known analytically by using separation of
  variables (see Appendix \ref{app:green}). We remark that it would
  also be readily feasible to illustrate our asymptotic theory for a
  rectangular-shaped confining domain, since the reduced-wave Green's
  function is also available analytically for such a domain.

\setcounter{equation}{0}
\setcounter{section}{2}
\section{The Linear Stability Analysis}\label{sec:stab}

In this section, we formulate the linear stability problem for the
steady-state solutions constructed in \S \ref{sec:steady-state}. We
denote the bulk steady-state solutions of \S \ref{sec:steady-state} by
$u_e({\bf x})$ and $v_{e}({\bf x})$, and the steady-state vector of
intracellular steady-states by
${\bf \mu}_e=(\mu_{e1},\ldots,\mu_{em})^T$ and
${\bf \eta}_e=(\eta_{e1},\ldots,\eta_{em})^T$.

To formulate the linear stability problem, we first introduce the
perturbations
\begin{eqnarray*}
 &&  u(t,{\bf x})=u_e({\bf x})+ e^{\lambda t} \phi({\bf x}) \,, \qquad
  v(t,{\bf x})=v_e({\bf x})+ e^{\lambda t} \psi({\bf x}) \,, \\
  && \mu_j(t)=\mu_{ej}+ e^{\lambda t} \xi_j \,, \qquad
     \eta_j(t)=\eta_{ej}+ e^{\lambda t} \zeta_j \,, \qquad \mbox{for}
     \quad j\in \lbrace{1,\ldots, m\rbrace}\,,
\end{eqnarray*}
into (\ref{eqsys:full}) and linearize the resulting system. This yields
the eigenvalue problem
\begin{subequations} \label{eqsys:pertfull}
    \begin{eqnarray}
	\text{bulk} &&
	\begin{cases}
          \Delta \phi - \Omega_u^2 \phi = 0\,, & {\bf x} \in
          \Omega\backslash\bigcup_{j=1}^m \Omega_j \,, \\
          \Delta \psi - \Omega_v^2 \psi = 0\,, & {\bf x} \in
          \Omega\backslash\bigcup_{j=1}^m \Omega_j\,, \\ 
          \partial_{\tilde{n}} \phi = \partial_{\tilde{n}} \psi =0 \,, &
  {\bf x} \in \partial\Omega \,,
	\end{cases} \label{eqsys:pertbulk}\\
	\text{reaction fluxes} &&
	\begin{cases}
 \varepsilon D_u \partial_{n_j} \phi = d_{1}^u \phi - d_{2}^u \xi_{j}\,, & {\bf x}
   \in \partial\Omega_j  \,, \; \; \\
 \varepsilon D_v \partial_{n_j} \psi = d_{1}^v \psi - d_{2}^v \zeta_{j}, & {\bf x}
   \in \partial\Omega_j \,, \\
	\end{cases} \label{eqsys:pertfluxes}\\
	\text{compartments} &&
	\begin{cases}
          \left(\lambda I - J_j \right)
          \begin{pmatrix}
            \xi_j \\
            \zeta_j
          \end{pmatrix} = \varepsilon^{-1}
          \begin{pmatrix}
             \int_{\partial\Omega_j}
             (d_{1}^u \phi - d_{2}^u \xi_{j})\;dS \\
             \int_{\partial\Omega_j} (d_{1}^v \psi -d_{2}^v \zeta_{j})\;dS
           \end{pmatrix}\,, \qquad j\in \lbrace{1,\ldots,m\rbrace} \,.
	\end{cases} \label{eqsys:pertcomp}
\end{eqnarray}
\end{subequations}
Here the Jacobian matrix $J_j$ of the intracellular kinetics, as well as
$\Omega_u$ and $\Omega_v$ are defined by
\begin{equation}\label{eig:def}
  J_j \equiv \begin{pmatrix}
     \partial_\mu f(\mu,\eta) & \partial_\eta f(\mu,\eta) \\
     \partial_\mu g(\mu,\eta) & \partial_\eta g(\mu,\eta)
    \end{pmatrix}\Big{\vert}_{\mu=\mu_{ej},\eta=\eta_{ej}}\,, \qquad
  \Omega_u \equiv \sqrt{\frac{\lambda+\sigma_u}{D_u}} \,, \qquad
  \Omega_v \equiv \sqrt{\frac{\lambda+\sigma_v}{D_v}}\,.
\end{equation}

We now use strong localized perturbation theory \cite{ward} to analyze
(\ref{eqsys:pertfull}) in the limit $\varepsilon \to 0$. In this
way we will derive a nonlinear matrix eigenvalue problem, referred
to as the globally coupled eigenvalue problem (GCEP), for the discrete
eigenvalues $\lambda$ of the linearization. This GCEP will be used to
investigate various instabilities of the steady-state solutions constructed
in \S \ref{sec:steady-state}. 

In the $\mathcal{O}(\varepsilon)$ inner region near the
$j^{\text{th}}$ cell we introduce the local variables
${\bf y}_j= \varepsilon^{-1}({\bf x} -{\bf x}_j)$,
$\phi_j({\bf x}) \equiv \phi({\bf x}_j + \varepsilon {\bf y}_j)$ and
$\psi_j({\bf x}) \equiv \psi({\bf x}_j + \varepsilon {\bf y}_j)$, with
$p_j=|{\bf y}_j|$.  Upon writing (\ref{eqsys:pertbulk}) in terms of the
inner variables, for $\varepsilon \to 0$ we obtain in the
$j^{\text{th} }$ inner region that $\Delta \phi_j = 0$ and
$\Delta \psi_j=0$, for $p_j\geq 1$, subject to
$D_u\, \partial_{p_j} \phi_j = d_{1}^{u} \phi_j - d_{2}^{u} \xi_j$ and
$D_v\, \partial_{p_j} \psi_j = d_{1}^{v} \psi_j - d_{2}^{v} \zeta_j$ on
$p_j=1$. The radially symmetric solutions to these problems are
\begin{equation} \label{lin:Inner}
  \phi_j(p_j) = c_j^{u} \log p_j + \frac{1}{d_{1}^{u}} \left( D_u\, c_j^{u} +
    d_{2}^{u}\xi_j \right)\,, \qquad
  \psi_j(p_j) = c_j^{v} \log p_j + \frac{1}{d_{1}^{v}} \left( D_v\, c_j^{v} +
    d_{2}^{v}\zeta_j \right)\,, 
\end{equation}
for $j\in\lbrace{1, \ldots ,m\rbrace}$, where $c_j^{u}$ and $c_j^{v}$
for $j\in\lbrace{1,\ldots,m\rbrace}$ are constants to be determined.
Upon substituting (\ref{lin:Inner}) into (\ref{eqsys:pertcomp}) we
obtain, in terms of the Jacobian $J_j$ of (\ref{eig:def}), that
\begin{equation} \label{eqsys:linintrasys}
    \left(\lambda I - J_j\right) 
    \begin{pmatrix}
        \xi_j \\ 
        \zeta_j
    \end{pmatrix}
    = 
    \begin{pmatrix}
        2\pi D_u c_{j}^u \\
        2\pi D_v c_{j}^v
      \end{pmatrix}\,, \qquad \mbox{for} \quad
      j\in\lbrace{1,\ldots,m\rbrace}\,.
\end{equation}

To determine $c_j^{u}$ and $c_j^{v}$ we must match the far-field
behavior of the inner solutions \eqref{lin:Inner} to the outer
solutions defined in the bulk region. Similar to the analysis of the
steady-state solution, we obtain that
\begin{subequations}\label{lin:Outer_U} 
\begin{eqnarray}
 && \Delta \phi - \Omega_u^2 \,\phi =0 \,, \quad {\bf x} \in \Omega \setminus
  \{{\bf x}_1,\ldots,{\bf x}_m \} \,; \qquad \partial_n \phi =0 \,,\quad
  {\bf x} \in \partial \Omega \,; \label{lin:Outer_Ua}   \\
 && U \sim  c_j^u \log |{\bf x}-{\bf x}_j|  + \frac{c_j^{u}}{\nu} +
  \frac{1}{d_{1}^{u}}(D_uc_j^{u} + d_{2}^{u} \xi_j)\,, \quad \text{as} \quad
    {\bf x} \to {\bf x}_j\,, \qquad j \in\lbrace{1,\ldots,m\rbrace}\,,
    \label{lin:Outer_Ub} 
\end{eqnarray}
\end{subequations}
where $\nu \equiv {-1/\log\varepsilon}\ll 1$. Similarly, for the
perturbation of the other bulk species we obtain
\begin{subequations}\label{lin:Outer_V} 
\begin{eqnarray}
  &&  \Delta \psi - \Omega_v^2 \,\psi =0\,, \quad {\bf x} \in \Omega \setminus
     \{{\bf x}_1,\ldots,{\bf x}_m \} \,; \qquad \partial_n \psi =0 \,,\quad
     {\bf x} \in \partial \Omega \,; \label{lin:Outer_Va} \\
  &&  \psi \sim  c_j^v \log |{\bf x}-{\bf x}_j| +\frac{c_j^{v}}{\nu} +
     \frac{1}{d_{1}^{v}}(D_vc_j^{v} + d_{2}^{v} \zeta_j)\,, \quad \text{as} \quad
     {\bf x} \to {\bf x}_j\,, \qquad j \in\lbrace{1,\ldots,m\rbrace}\,. \label{lin:Outer_Vb} 
\end{eqnarray}
\end{subequations}

The solutions to (\ref{lin:Outer_U}) and (\ref{lin:Outer_V}) are represented
as
\begin{equation}\label{lin:U_Green}
  \phi(x) =  -2 \pi \sum_{i=1}^{m} c_i^{u} G_{u,\lambda}({\bf x}; {\bf x}_i)\,,
  \qquad
  \psi(x) =  -2 \pi \sum_{i=1}^{m} c_i^{v} G_{v,\lambda}({\bf x}; {\bf x}_i)\,,
\end{equation}
where, to simplify the notation and emphasize the dependence on the
eigenvalue parameter $\lambda$, we have defined
\begin{equation}\label{lin:green_lam}
  G_{u,\lambda}({\bf x};{\bf x}_j) \equiv G_{\Omega_u}({\bf x};{\bf x}_j) \,,
  \qquad
  G_{v,\lambda}({\bf x};{\bf x}_j) \equiv G_{\Omega_v}({\bf x};{\bf x}_j) \,,
\end{equation}
where $G_{\omega}({\bf x};{\bf x}_j)$ is defined by the solution to
(\ref{R_Green}). Upon letting ${\bf x}\to {\bf x}_j$ in (\ref{lin:U_Green})
and ensuring that the singularity conditions in (\ref{lin:Outer_Ub}) and
(\ref{lin:Outer_Vb}) are satisfied, we obtain a linear algebraic system
for the vectors ${\bf c}^{u} \equiv (c_1^u, \ldots, c_m^u)^T$ and
${\bf c}^{v} \equiv (c_1^v, \ldots, c_m^v)^T$, given in matrix form by
\begin{equation}\label{lin:Linear_System} 
  \left( \left(1+\frac{\nu D_u}{d_1^u}\right) I + 2\pi \nu
  \mathcal{G}_{u,\lambda}
  \right){\bf c}^u =  - \frac{\nu d_2^u}{d_1^u}\, {\bf \xi}\,,
  \qquad
  \left( \left(1+\frac{\nu D_v}{d_1^v}\right) I + 2\pi \nu
  \mathcal{G}_{v,\lambda} \right){\bf c}^v = - \frac{\nu
    d_2^v}{d_1^v}\, {\bf \zeta}\,,
\end{equation}
where ${\bf \xi}\equiv (\xi_1, \ldots, \xi_m)^T$ and
${\bf \zeta}\equiv (\zeta_1, \ldots, \zeta_m)^T$. In
\eqref{lin:Linear_System}, $\mathcal{G}_{u,\lambda}$ and
$\mathcal{G}_{v,\lambda}$ denote the reduced-wave Green's matrix
given in (\ref{GreensMatrix}) with either $\omega=\Omega_u$ or
$\omega=\Omega_v$, respectively. Here $\Omega_u$ and $\Omega_v$ are defined
in terms of $\lambda$ by (\ref{eig:def}).

Assuming that $\lambda$ is not an eigenvalue of $J_j$ for any $j\in
\lbrace{1,\ldots,m\rbrace}$, we obtain upon inverting
(\ref{eqsys:linintrasys}) and writing the system in matrix form that
\begin{equation}\label{lin:vectors}
  {\bf \xi} = 2\pi D_u \K_{11} {\bf c}^u + 2\pi D_v \K_{12}
                {\bf c}^v\,, \qquad
  {\bf \zeta} = 2\pi D_u \K_{21} {\bf c}^u + 2\pi D_v \K_{22}
                  {\bf c}^v\,,
\end{equation}
where ${\bf \xi}=(\xi_1,\ldots,\xi_m)^T$ and
${\bf \zeta}=(\zeta_1,\ldots, \zeta_m)^T$. Here
$\K_{11}$, $\K_{12}$, $\K_{21}$, and $\K_{22}$ are the diagonal matrices
defined by
\begin{subequations}\label{lin:kdef}
\begin{equation}
 \K_{11} \equiv \diag(K_{11j})\,, \qquad \K_{12} \equiv \diag(K_{12j})\,, \qquad
 \K_{21} \equiv \diag(K_{21j})\,, \qquad \K_{22} \equiv \diag(K_{22j})\,,
\end{equation}
with diagonal entries given by
\begin{equation}
  K_{11j} \equiv {\bf e}_1^T(\lambda I - J_j)^{-1} {\bf e}_1\,, \quad
  K_{12j} \equiv {\bf e}_1^T(\lambda I - J_j)^{-1} {\bf e}_2\,, \quad
  K_{21j} \equiv {\bf e}_2^T(\lambda I - J_j)^{-1} {\bf e}_1\,,
    \quad K_{22j} \equiv {\bf e}_2^T(\lambda I - J_j)^{-1} {\bf e}_2\,,
\end{equation}
\end{subequations}
where ${\bf e}_1=(1,0)^T$ and ${\bf e}_2=(0,1)^T$.

Then, upon substituting (\ref{lin:Linear_System}) into (\ref{lin:vectors}),
we obtain the $2m\times 2m$ homogeneous algebraic system, which we write
in block matrix form as
\begin{subequations}\label{lin:mat_eig}
\begin{equation}
  {\mathcal M}(\lambda) \begin{pmatrix}
    {\bf c}^u \\
    {\bf c}^v
  \end{pmatrix} = \begin{pmatrix}
    {\bf 0} \\
    {\bf 0}
  \end{pmatrix} \,, \qquad \mbox{where} \qquad
  {\mathcal M}(\lambda) \equiv \begin{pmatrix}
    {\mathcal M}_u(\lambda) &  {\mathcal H}_u(\lambda) \\
    {\mathcal M}_v(\lambda) &  {\mathcal H}_v(\lambda)
  \end{pmatrix}\,,
\end{equation}
with
\begin{eqnarray}
  && {\mathcal M}_u(\lambda) \equiv
     \left(1+\frac{\nu D_u}{d_1^u}\right) I +2\pi\nu D_u \frac{d_2^u}{d_1^u}
     \K_{11} +  2\pi \nu \G_{u, \lambda} \,, \qquad
     {\mathcal H}_u(\lambda) \equiv 2\pi \nu D_v \frac{d_2^u}{d_1^u}\K_{12}\,,\\
  && {\mathcal H}_v(\lambda) \equiv 2\pi \nu D_u \frac{d_2^v}{d_1^v}\K_{21}\,,
     \qquad
     {\mathcal M}_v(\lambda) \equiv  \left(1+ \frac{\nu D_v}{d_1^v}\right)I +
     2\pi\nu D_v\frac{d_2^v}{d_1^v}\K_{22} +2\pi \nu \G_{v,\lambda}\,.
\end{eqnarray}
\end{subequations}

The nonlinear matrix eigenvalue problem (\ref{lin:mat_eig}) is
referred to as the globally coupled eigenvalue problem (GCEP). The
GCEP has a nontrivial solution
$({\bf c}^u,{\bf c}^v)^T\neq ({\bf 0},{\bf 0})^T$, if and only
$\lambda$ satisfies $\det \mathcal{M}(\lambda) = 0$. The set
$\Lambda({\mathcal M})$, defined by
\begin{equation}\label{TransDent}
 \Lambda({\mathcal M}) \equiv \lbrace{ \lambda \, \vert \,
    \det \mathcal{M}(\lambda)=0 \rbrace} \,,
\end{equation}
is the union of all such roots. Any element
$\lambda\in \Lambda({\mathcal M})$ satisfying $\mbox{Re}(\lambda)>0$
provides an approximation, valid as $\varepsilon\to 0$, for an
unstable discrete eigenvalue of the linearized problem
(\ref{eqsys:pertfull}).  However, if for all
$\lambda\in \Lambda({\mathcal M})$ we have $\mbox{Re}(\lambda)<0$, then
the steady-state solution is linearly stable.

When there is a large number of cells $m$, the determination of the
discrete eigenvalues comprising $\Lambda({\mathcal M})$ is in general
a very challenging numerical problem. A survey of nonlinear matrix
eigenvalue problems and available solution strategies that apply to
only certain classes of matrices is given in \cite{guttel} and
\cite{betcke}.  Specific applications of nonlinear matrix problems in
simpler contexts where ${\mathcal M}(\lambda)$ is either a polynomial
or a rational function of $\lambda$ are discussed in
\cite{betcke2}. Since for our problem, ${\mathcal M}(\lambda)$ is
not symmetric and has a complicated dependence on the eigenvalue
parameter through the Green's matrices and from the diagonal $\K$
matrices of (\ref{lin:kdef}), these previously developed numerical
strategies are not applicable for computing the set
$\Lambda({\mathcal M})$ in (\ref{TransDent}) for a steady-state
solution with an arbitrary collection of cells.

For a symmetric steady-state solution corresponding to a symmetric cell
arrangement, we now verify that the condition $\det({\mathcal M}(0))=0$ in
(\ref{lin:mat_eig}) is equivalent to the zero-eigenvalue crossing
condition derived in (\ref{break:full}), which was based on a linearization
of the nonlinear algebraic system around the symmetric steady-state. When
$\lambda=0$, we have
\begin{equation*}
  2\pi D_u {\bf c}^u = -\Theta_{u} {\bf \xi} \,, \qquad
  2\pi D_v {\bf c}^v = -\Theta_{v} {\bf \zeta} \,,
\end{equation*}
where $\Theta_{u}$ and $\Theta_v$ are defined in (\ref{non:full_mat}).
Since $\K_{11}$, $\K_{12}$, $\K_{22}$, and $\K_{21}$ are all multiples
of the identity for a symmetric steady-state, we obtain from
(\ref{lin:mat_eig}) together with (\ref{lin:Linear_System})
that when $\lambda=0$ we have in block matrix form
\begin{equation}\label{lin:zero_check}
  \begin{pmatrix}
    {\bf \xi} \\
    {\bf \zeta}
   \end{pmatrix} + \begin{pmatrix}
               K_{11}I & K_{12} I\\
               K_{12}I & K_{22}I
             \end{pmatrix}
             \begin{pmatrix}
                  \Theta_u &  0 \\
                     0  & \Theta_v
             \end{pmatrix}
               \begin{pmatrix}
                {\bf \xi} \\
                {\bf \zeta}
              \end{pmatrix}
              = \begin{pmatrix}
                {\bf 0} \\
                {\bf 0}
              \end{pmatrix}\,.
\end{equation}
Here we have re-defined the scalars $K_{11}$, $K_{12}$, $K_{21}$, and
$K_{22}$ by $\K_{11}=K_{11}I$, $\K_{12}=K_{12}I$, $\K_{21}=K_{21}I$,
and $\K_{22}=K_{22}I$.  When $\lambda=0$, we calculate that
\begin{equation*}
  {\mathcal K} = \begin{pmatrix}
    K_{11}I & K_{12}I \\
    K_{21}I & K_{22}I
  \end{pmatrix} = -J_c^{-1} \,, \quad \mbox{where}\quad
  J_c \equiv \begin{pmatrix}
    f_{\mu}^cI & f_{\eta}^cI \\
    g_{\mu}^cI & g_{\eta}^cI
    \end{pmatrix}\,.
\end{equation*}
Finally, upon multiplying (\ref{lin:zero_check}) by $J_c$, and using
$J_c{\mathcal K}=-I$, we readily obtain that
\begin{equation}\label{lin:zero_final}
  \begin{pmatrix}
          f_{\mu}^c I - \Theta_u  & f_{\eta}^c I\\
              g_{\mu}^c I & g_{\eta}^c I - \Theta_v
             \end{pmatrix} \begin{pmatrix}
                {\bf \xi} \\
                {\bf \zeta}
              \end{pmatrix} = \begin{pmatrix}
                {\bf 0} \\
                {\bf 0}
              \end{pmatrix}\,,
\end{equation}
which is precisely the same as in  (\ref{break:full}).

\subsection{Re-formulation of the linear stability problem}
A simpler formulation of the linear stability problem that applies to
both symmetric and asymmetric steady-state solutions can be done when
$g$ has the specific form in (\ref{g:linear}). In this situation, we
can write (\ref{eqsys:linintrasys}) in the form
\begin{equation*}
  \zeta_j = \frac{g_1^{\prime}(\mu_j)}{\lambda + g_2} \xi_j +
  \frac{2\pi D_v}{\lambda + g_2} c_{j}^v \,, \qquad
  \left(\lambda - f_{\mu}(\mu_j,\eta_j)\right) \xi_j -
  f_{\eta}(\mu_j,\eta_j)\zeta_j = 2\pi D_u c_j^{u}\,.
\end{equation*}
Then, upon relating $c_j^v$ and $c_j^u$ to $\zeta_j$ and $\xi_j$ by using
(\ref{lin:Linear_System}), we obtain in matrix form that
\begin{equation}\label{lin:mat_new}
  {\bf \zeta} =\frac{1}{\lambda + g_2}\Lambda_2{\bf \xi} - \frac{1}{\lambda+g_2}
  \Theta_{v,\lambda}{\bf \zeta} \,, \qquad
  \Lambda_3 {\bf \xi} - {\Lambda}_4{\bf \zeta} = -\Theta_{u,\lambda}{\bf \xi}\,,
\end{equation}
where $\Theta_{u,\lambda}$, $\Theta_{v,\lambda}$, and the diagonal matrices
$\Lambda_2$, $\Lambda_3$, and $\Lambda_4$ are defined by
\begin{subequations}\label{lin:theta_mat}
  \begin{eqnarray}
 && \Theta_{u,\lambda} \equiv 2\pi \nu D_u \frac{d_2^{u}}{d_1^{u}} \left[
      \left(1+\frac{\nu D_u}{d_1^u}\right) I + 2\pi \nu
      \mathcal{G}_{u,\lambda} \right]^{-1} \,, \quad
    \Theta_{v,\lambda} \equiv 2\pi \nu D_v \frac{d_2^{v}}{d_1^{v}} \left[
      \left(1+\frac{\nu D_v}{d_1^v}\right) I + 2\pi \nu
      \mathcal{G}_{v,\lambda} \right]^{-1} \,, \\
    &&\Lambda_2 \equiv \diag( g_1^{\prime}(\mu_j))\,, \qquad \Lambda_3
       \equiv \diag( \lambda - f_{\mu}(\mu_j,\eta_j)) \,, \qquad
   \Lambda_4 \equiv  \diag( f_{\eta}(\mu_j,\eta_j))\,.
  \end{eqnarray}
\end{subequations}
Upon eliminating ${\bf \zeta}$ in (\ref{lin:mat_new}), we obtain the
following nonlinear eigenvalue problem for the case where $g$ has the
form in (\ref{g:linear}):
\begin{equation}\label{lin:glin_non}
  {\mathcal N}(\lambda) {\bf \xi}={\bf 0} \,,
  \qquad \mbox{where} \qquad
  {\mathcal N}(\lambda) \equiv \Lambda_3 - \Lambda_4 \left[(\lambda + g_2)I
    +\Theta_{v,\lambda}\right]^{-1}\Lambda_2 + \Theta_{u,\lambda} \,.
\end{equation}
Observe that setting $\det({\mathcal N}(\lambda))=0$ involves
root-finding on the determinant of a matrix of size $m\times m$ rather
than that for the larger $2m\times 2m$ matrix, as needed for
(\ref{TransDent}).

The characterization (\ref{lin:glin_non}) is particularly useful for
determining the linear stability properties of a symmetric steady-state
for a symmetric cell arrangement when $g$ has the form in (\ref{g:linear}).
For a symmetric steady-state with $\lambda=0$, we obtain from
(\ref{lin:theta_mat}) that $\Lambda_2=g_1^{\prime}(\mu_c)I$,
$\Lambda_3=-f_{\mu}^{c}I$, and $\Lambda_4=f_{\eta}^{c}I$. From
(\ref{lin:glin_non}), this yields that
\begin{equation*}
  {\mathcal N}(0) =-f_{\mu}^cI - f_{\eta}^c g_1^{\prime}(\mu_c)
  \left[g_2 I + \Theta_{v,0}\right]^{-1} + \Theta_{u,0}\,.
\end{equation*}
Since $\Theta_{u,\lambda}=\Theta_u$ and
$\Theta_{v,\lambda}=\Theta_{v}$ when $\lambda=0$, where $\Theta_u$ and
$\Theta_v$ were defined in (\ref{non:full_mat}), we obtain that the
condition $\det({\mathcal N}(0))=0$ is equivalent to the formulation
(\ref{break:simp}) derived in \S \ref{sec:symm-break}, which was based
on linearizing the nonlinear algebraic system around the symmetric
steady-state solution.

For a symmetric steady-state solution of a symmetric cell arrangement, one
key advantage of the re-formulation (\ref{lin:glin_non}) is that the
discrete eigenvalues of the linearization (\ref{eqsys:pertfull})
can be determined by finding
the union of the roots of $m$ scalar problems. This is done by using
$\det({\mathcal N}(\lambda))=\prod_{j=1}^{m}\sigma_j(\lambda)$, where
$\sigma_j(\lambda)$ for $j=\lbrace{1,\ldots,m\rbrace}$ are the
eigenvalues of ${\mathcal N}(\lambda)$.  More specifically, since
${\mathcal G}_{u,\lambda}$ and ${\mathcal G}_{v,\lambda}$ have the
common eigenspace
\begin{subequations}\label{lin:common}
\begin{equation}
  {\mathcal G}_{u,\lambda} {\bf e} = \kappa_{u,\lambda} {\bf e} \,, \quad
  {\mathcal G}_{v,\lambda} {\bf e} = \kappa_{v,\lambda} {\bf e} \,; \qquad
  {\mathcal G}_{u,\lambda} {\bf q}_j = \kappa_{u,\lambda j}^{\perp}
  {\bf q}_j \,,  \quad
  {\mathcal G}_{v,\lambda} {\bf q}_j = \kappa_{v,\lambda j}^{\perp}
  {\bf q}_j \,, \qquad j\in \lbrace{2,\ldots,m\rbrace} \,,
\end{equation}
\end{subequations}
the eigenvalue $\sigma_1(\lambda)$ corresponding to ${\bf e}$ and the
eigenvalues $\sigma_j(\lambda)$ corresponding to ${\bf q}_j$, for
$j\in\lbrace{2,\ldots,m\rbrace}$ are readily calculated. A simple
calculation yields that
\begin{subequations}\label{lin:root}
  \begin{eqnarray}\
  &&  \sigma_1(\lambda)= \lambda - f_{\mu}^c -
    \frac{f_{\eta}^c g_1^{\prime}(\mu_c)}{\lambda + g_2 + \alpha_{v,\lambda}}
    + \alpha_{u,\lambda} \,, \\
  &&  \sigma_j(\lambda)= \lambda - f_{\mu}^c -
    \frac{f_{\eta}^c g_1^{\prime}(\mu_c)}{\lambda + g_2 +
    \alpha_{v,\lambda j}^{\perp}}
    + \alpha_{u,\lambda j}^{\perp} \,, \qquad j\in \lbrace{2,\ldots,m\rbrace}\,,
\end{eqnarray}
where we have defined
\begin{eqnarray}\label{lin:root_2}
 && \alpha_{u,\lambda} \equiv \frac{2\pi \nu D_u {d_2^u/d_1^u}}{1+{\nu D_u/d_1^u}+
  2\pi \nu \kappa_{u,\lambda}} \,, \qquad
  \alpha_{v,\lambda} \equiv \frac{2\pi \nu D_v {d_2^v/d_1^v}}{1+{\nu D_v/d_1^v}+
  2\pi \nu \kappa_{v,\lambda}} \,, \\
  &&  \alpha_{u,\lambda j}^{\perp} \equiv \frac{2\pi \nu D_u {d_2^u/d_1^u}}
     {1+{\nu D_u/d_1^u}+
    2\pi \nu \kappa_{u,\lambda j}^{\perp}} \,, \qquad
     \alpha_{v,\lambda j}^{\perp} \equiv
     \frac{2\pi \nu D_v {d_2^v/d_1^v}}{1+{\nu D_v/d_1^v}+
 2\pi \nu \kappa_{v,\lambda j}^{\perp}} \,, \qquad j\in \lbrace{2,\ldots,m\rbrace}
     \,.
\end{eqnarray}
\end{subequations}

With this re-formulation, for a symmetric steady-state of a symmetric
cell arrangement, and with $g$ of the form in (\ref{g:linear}), the
set of discrete eigenvalues of the linearization, $\Lambda({\mathcal M})$, in
(\ref{TransDent}) can be written conveniently as
\begin{equation}\label{new:TransDent}
  \Lambda({\mathcal M}) \equiv \lbrace{ \lambda \,\, \vert \,
    \sigma_1(\lambda)=0\rbrace} \,\,\cup\,\,
  \bigcup_{j=2}^{m} \,\,\lbrace{ \lambda \,\, \vert \sigma_j(\lambda)=0
    \rbrace}\,.
\end{equation}
In summary, to determine the linear stability properties of this symmetric
steady-state solution we need only solve $m$ scalar root-finding problems
and determine whether there are any roots in $\mbox{Re}(\lambda)>0$.
This is considerably more tractable numerically than performing a
root-finding based on the determinant of the GCEP in (\ref{lin:glin_non}).

\setcounter{equation}{0}
\setcounter{section}{3}
\section{Illustrations of the theory: A ring pattern of cells}\label{sec:examples}

In this section we illustrate the steady-state and linear stability
theory developed in \S \ref{sec:steady-state} and \S \ref{sec:stab}
for a ring pattern of cells inside unit disk $\Omega$, for which the
Green's function is given analytically in Appendix \ref{app:green}. We
will show that symmetry-breaking bifurcations can occur for the
Gierer-Meinhardt \cite{gm}, Rauch-Millonas \cite{rauch} and
FitzHugh-Nagumo \cite{gomez2007} reaction kinetics. The theoretical
prediction of stable asymmetric patterns will be confirmed through
full time-dependent numerical simulations of (\ref{eqsys:full})
computed using FlexPDE \cite{flexpde}.

For a ring pattern of $m$ cells in the unit disk with ring radius $r$,
with $0<r<1$, the cell centers are located at
\begin{equation}\label{ex:ring_centers}
  {\bf x}_{k} = r\left(\cos\left(\frac{2\pi (k-1)}{m}\right),\,
    \sin\left(\frac{2\pi (k-1)}{m}\right)\right) \,, \qquad k\in
  \lbrace{1,\ldots,m\rbrace}\,.
\end{equation}
For a ring pattern of cells, all Green matrices are symmetric and
circulant and have the common eigenspace
\begin{equation*}
  {\bf v}_k = (1, Z_k, Z_k^2, ..., Z_k^{m-1})^{T} \quad \text{with} \quad
  Z_k\equiv  \mbox{exp}\left(\frac{2\pi i (k-1)}{m}\right) \quad \text{and}
  \quad k\in\{1,...,m\}\,,
\end{equation*}
which are a basis of $\bC^m$. In (\ref{app:cyclic_mat}) of Appendix
\ref{app:green} we summarize how to obtain the matrix spectrum of a
symmetric and circulant matrix that has a real-valued basis for the
eigenspace.

\begin{figure}
    \centering
    \def\svgwidth{0.5\textwidth}
        \def\svgheight{4.5cm}
	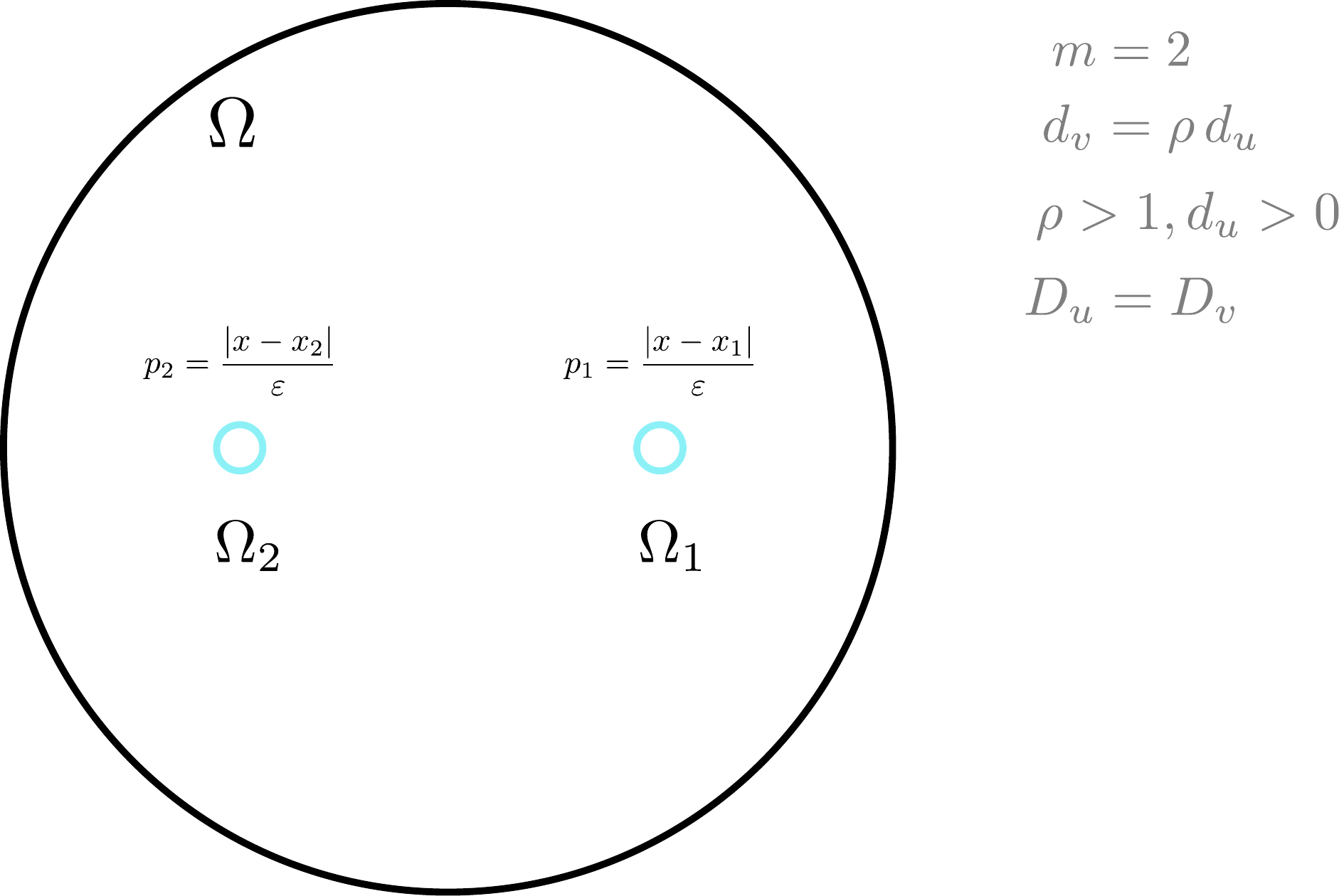
        \caption{A schematic plot of a ring pattern in the unit disk
          with two cells.  The bifurcation parameter for
          symmetry-breaking is $\rho$, while the diffusivities satisfy
          $D_u=D_v$.}
    \label{fig:2cellsring}
\end{figure}

In our illustrations of the theory below, we will assume for simplicity that
the membrane reaction rates satisfy
\begin{equation}\label{ex:dall}
  d_1^u=d_2^u\equiv d_u \qquad  d_1^v=d_2^v\equiv d_v \,, \qquad
  \mbox{with} \quad \rho \equiv \frac{d_v}{d_u} \,,
\end{equation}
and that $g(\mu,\eta)$ has the specific form in (\ref{g:linear}). We
will focus on a two-cell ring pattern in the unit disk, as shown
schematically in Figure \ref{fig:2cellsring}, for three specific
reaction kinetics.

To numerically implement our asymptotic theory, the steady-state
solution branches are computed from (\ref{g:mu_solve}) with $m=2$
using the parameter continuation software MatCont \cite{matcont}, while
the symmetric solution branch is obtained from
(\ref{symm:scalar}). Symmetry-breaking bifurcation points in $\rho$
along the symmetric branch are identified by numerically solving
(\ref{two:red_simp}) together with (\ref{symm:scalar}). Finally, to
determine the linear stability properties of the symmetric branch we
need only determine if there exists a $\lambda$ with
${\mbox{Re}}(\lambda)>0$ in the set $\Lambda({\mathcal M})$ given in
(\ref{new:TransDent}). For $m=2$, this is done by calculating all the
roots of $\sigma_1(\lambda)=0$ and $\sigma_2(\lambda)=0$ by using
(\ref{lin:root}) and the explicit expressions for the eigenvalues of
the Green's matrices as can be obtained from Appendix \ref{app:green}.

\subsection{Gierer-Meinhardt reaction kinetics}

We consider a prototypical Gierer-Meinhardt model (GM), where the
nonlinear reaction kinetics are confined within the compartments. The
original GM model, introduced in \cite{gm} and
\cite{gierer} to model pattern formation in biological morphogenesis,
has the form
\begin{equation*}
  \partial_t u =  D_u \Delta u  - \sigma_u u + \varrho_0(x) +
  c_u\varrho_u(x) \frac{u^2}{v} \,, \qquad
      \partial_t v = D_v \Delta v - \sigma_v v + c_v\varrho_v(x) u^2\,.
\end{equation*}
This model disregards that in biological tissues morphogen-producing
reactions mostly occur intracellularly and on the membranes of
cells. For simplicity, we illustrate our compartmental-reaction
diffusion theory for the specific case where
$\varrho_0\equiv 0, c_u\varrho_u(x)\equiv 1$, and
$c_v\varrho_v\equiv 1$. When there is no bulk diffusion, the
compartments are uncoupled from the bulk and we impose the
intracellular reaction kinetics
\begin{equation}\label{cell:GM}
        \dot{\mu}(t) = f(\mu,\eta) \equiv \frac{\mu^2}{\eta}\,, \qquad
        \dot{\eta}(t) = g(\mu,\eta) \equiv \mu^2\,.
\end{equation}
The uncoupled equilibrium for (\ref{cell:GM}) given by $\mu_e=0$, and where
$\eta_e$ is an arbitrary constant, is non-hyberbolic in all
directions.

To apply the bulk-cell steady-state analysis of \S \ref{sec:2d} for a
two-cell ring pattern, we first identify that
$g(\mu,\eta)=g_1(\mu)-g_2\eta$, where $g_1=\mu^2$ and $g_2=0$. For
$m=2$, we conclude from (\ref{g:mu_solve}) that all steady-states of
the bulk-cell system are associated with the nonlinear algebraic
problem
\begin{equation} \label{gm:2cells_all}
    \begin{array}{rcl}
      f(\mu_{e1}, {\bf e}_1^T\Theta_v^{-1}((\mu_{e1})^2, (\mu_{e2})^2)^T) -
      {\bf e}_1^T \Theta_u(\mu_{e1},\mu_{e2})^T &=& 0 \\ 
      f(\mu_{e2}, {\bf e}_2^T\Theta_v^{-1}((\mu_{e1})^2, (\mu_{e2})^2)^T) -
      {\bf e}_2^T \Theta_u(\mu_{e1},\mu_{e2})^T &=& 0\,.
    \end{array}
\end{equation}
  
The symmetric equilibrium $(\mu_e,\eta_e)$, which satisfies (\ref{symm:scalar}),
is readily calculated as
\begin{equation}\label{gm:2cells_symm}
  \mu_e = \frac{\alpha_v}{\alpha_u}\,, \qquad \eta_e =
  \frac{\alpha_v}{\alpha_u^2} \,,
\end{equation}
where $\alpha_u$ and $\alpha_v$ are defined in (\ref{symm:alpkap}).  By
combining (\ref{gm:2cells_symm}) with (\ref{two:red_simp}), we
conclude that a symmetry-breaking bifurcation from the symmetric
steady-state occurs whenever the condition
\begin{equation}\label{gm:2cells_symmb}
  \frac{\alpha_v}{\alpha_{v,2}^{\perp}} +
  \frac{\alpha_{u,2}^{\perp}}{2 \alpha_u} -1=0\,,
\end{equation}
is satisfied at some point along the symmetric solution branch. Here
$\alpha_{u,2}^{\perp}$ and $\alpha_{v,2}^{\perp}$ were defined in
(\ref{two:alpha_def}).

\begin{figure}[H]
  \centering
  \includegraphics[width=0.48\textwidth,height=4.9cm]{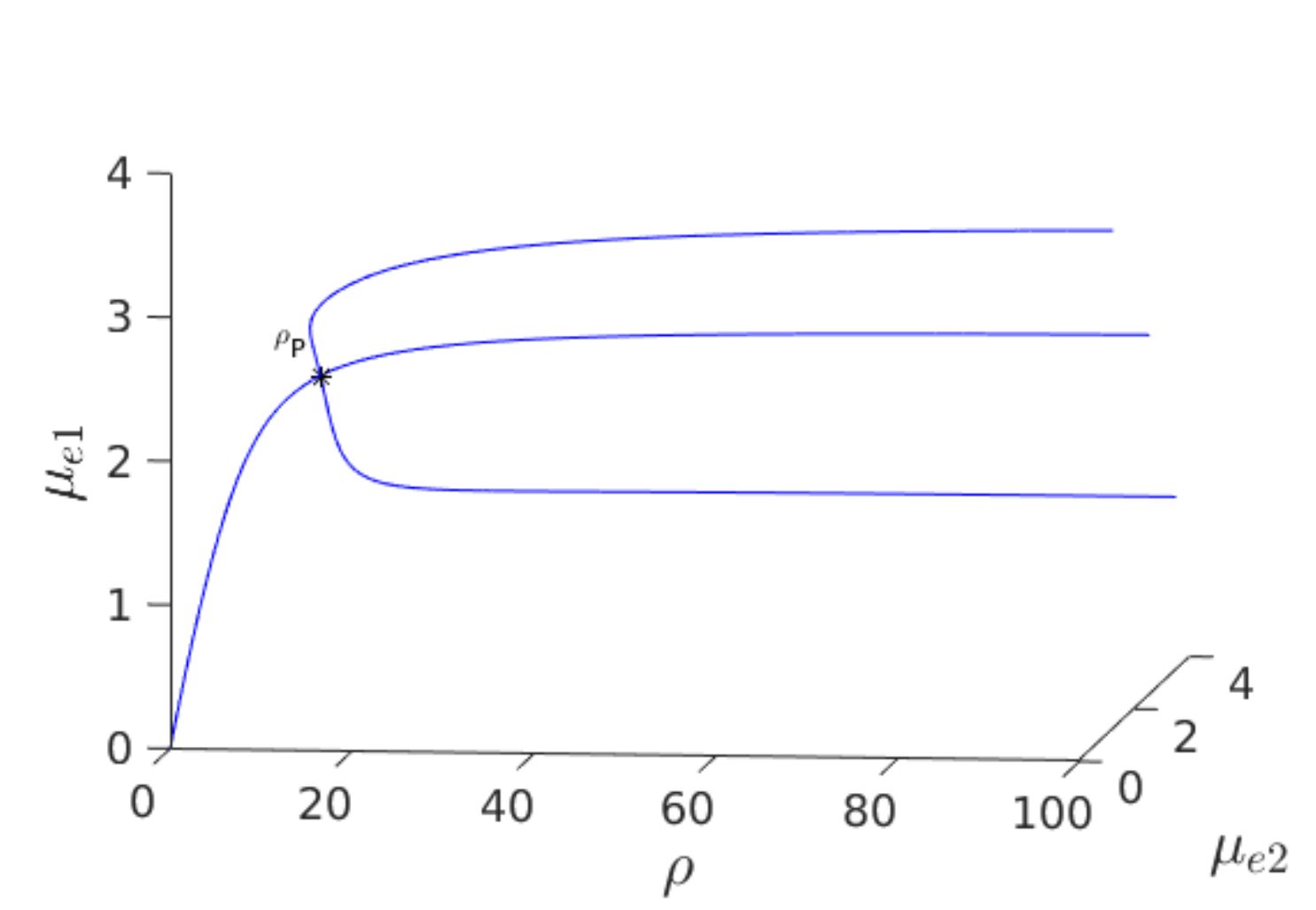}
  \includegraphics[width=0.48\textwidth,height=4.7cm]{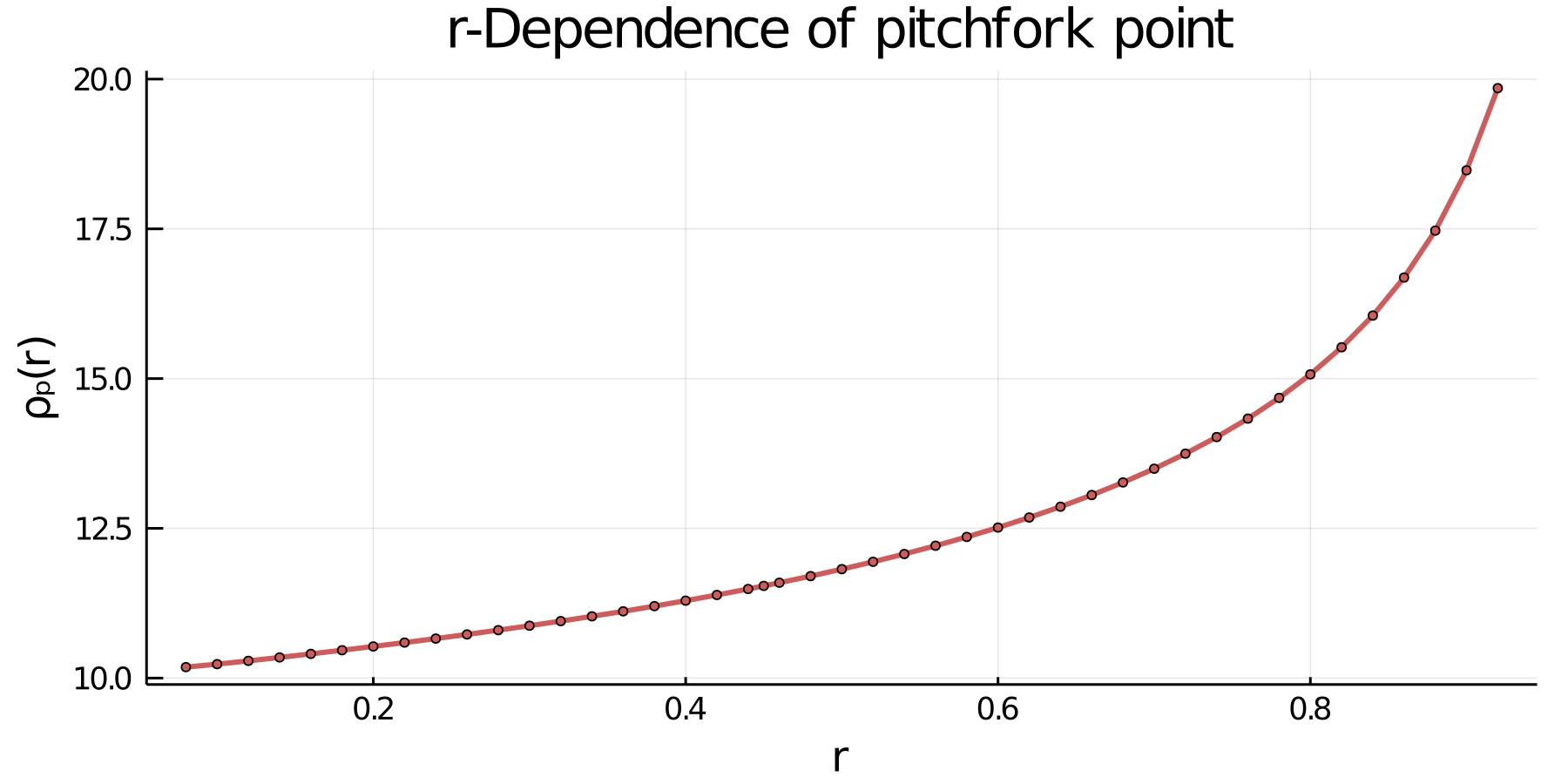}
  \caption{Left: 3-D Bifurcation diagram, computed from
    (\ref{gm:2cells_all}), showing symmetric and asymmetric
    steady-states of a two-cell ring pattern with ring radius $r=0.5$
    and GM kinetics (\ref{cell:GM}). Asymmetric steady-states emerge
    at the supercritical pitchfork bifurcation point
    $\rho=\rho_p\approx 9.79168$ along the symmetric branch. Right:
    The pitchfork bifurcation value of $\rho$ increases rapidly as the
    ring radius $r$, and consequently the distance between the cells,
    increases. The dots are the values computed from
    (\ref{gm:2cells_symmb}), while the curve is the interpolation by
    the plotting function in Julia \cite{julia}.  Parameters:
    $D_u=D_v=5, \sigma_u =\sigma_v=0.6, d_u=0.09$, and
    $\varepsilon=0.03$.}
    \label{fig:GMnonhystdu0p09}
\end{figure}

\begin{figure}[]
    \centering
	\begin{subfigure}[b]{.45\textwidth}
	    \begin{subfigure}[b]{1.\textwidth}
	        \centering
                \def\svgwidth{1\textwidth}
                    \def\svgheight{4.2cm}
	        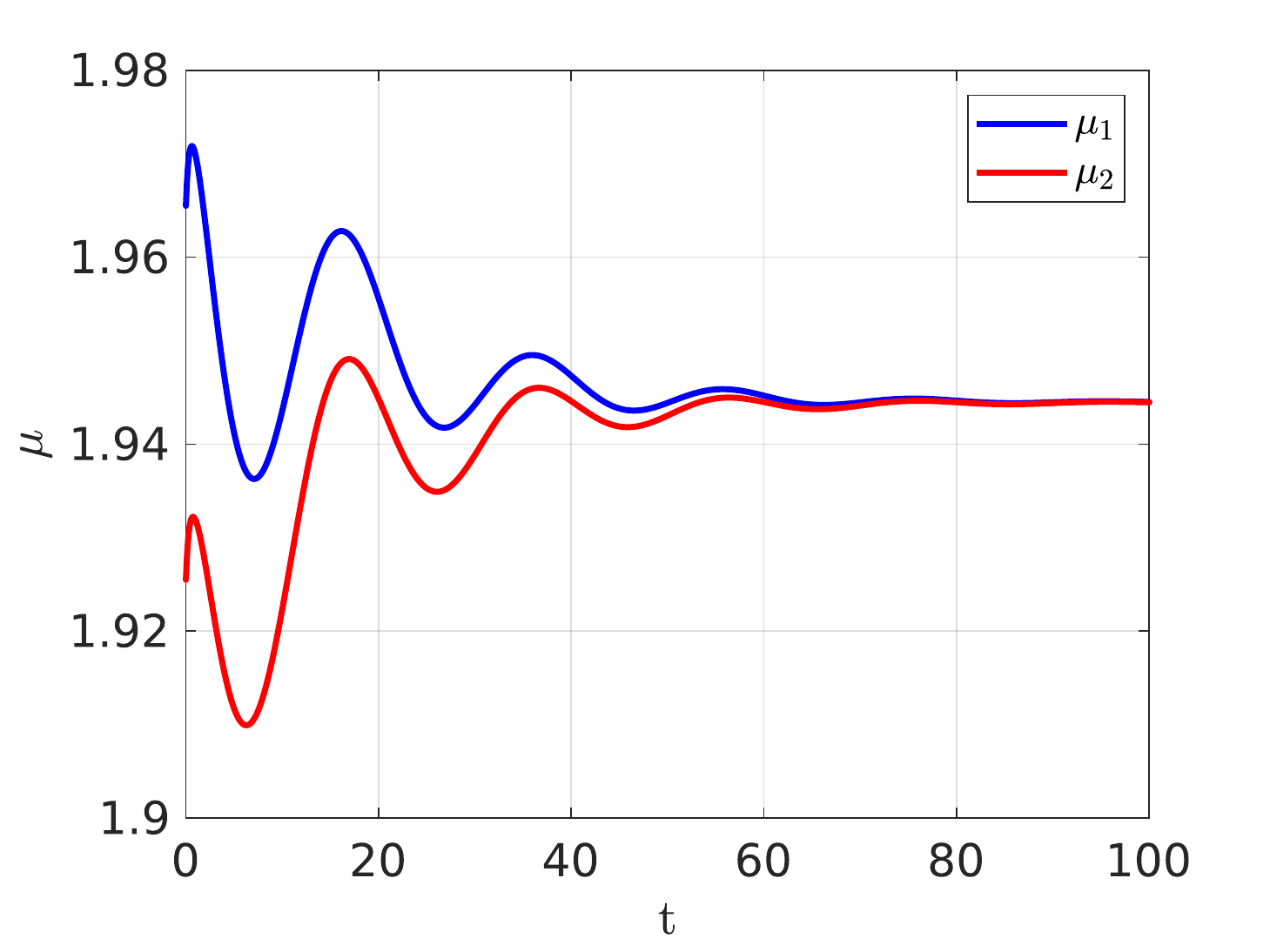
	    \end{subfigure}
	    \begin{subfigure}[b]{1.\textwidth}
	        \centering
                \def\svgwidth{1\textwidth}
                    \def\svgheight{4.2cm}
	        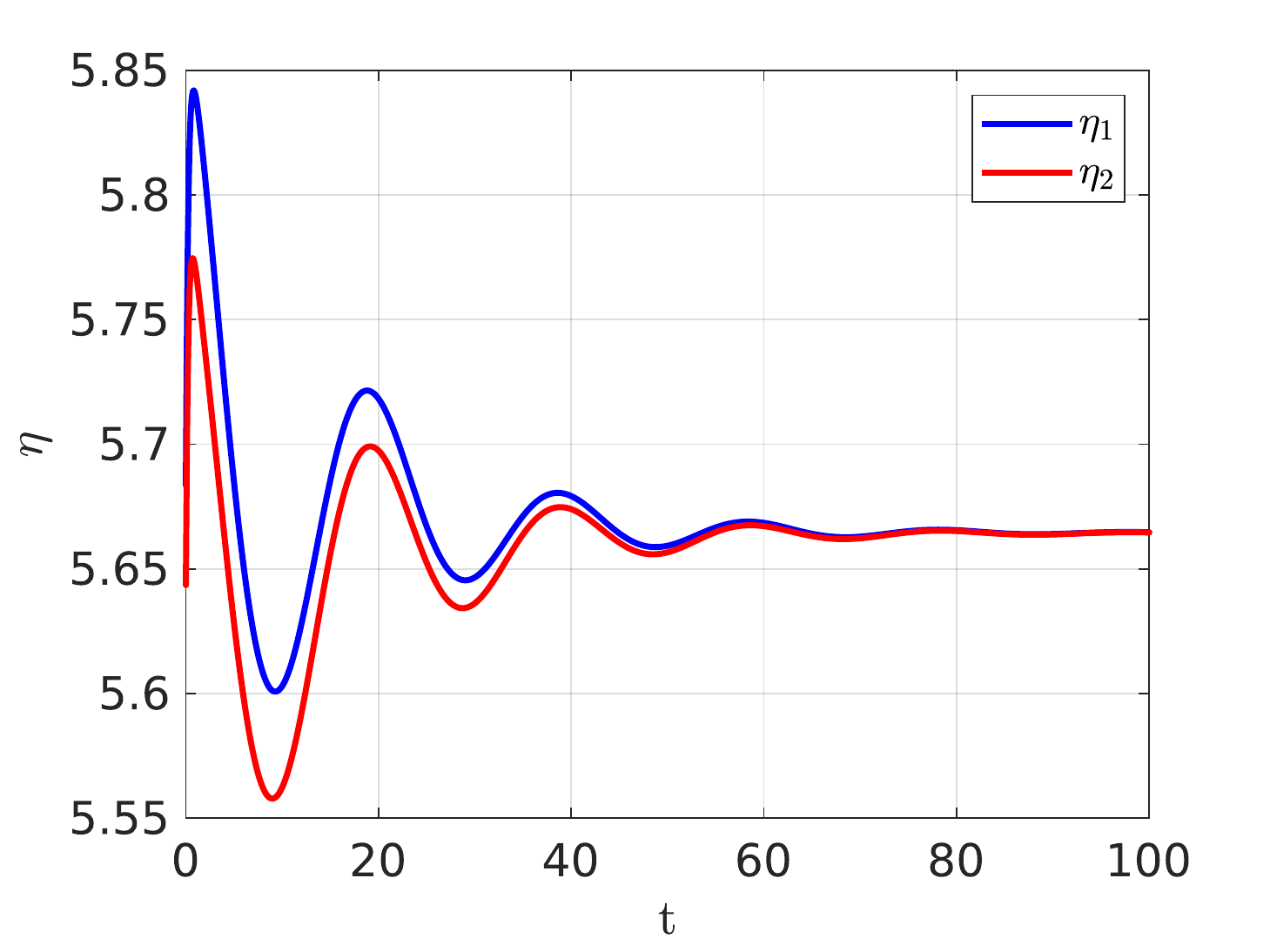
	    \end{subfigure}
	    \begin{subfigure}[b]{1.\textwidth}
    		\centering
                \def\svgwidth{1\textwidth}
                    \def\svgheight{4.2cm}
			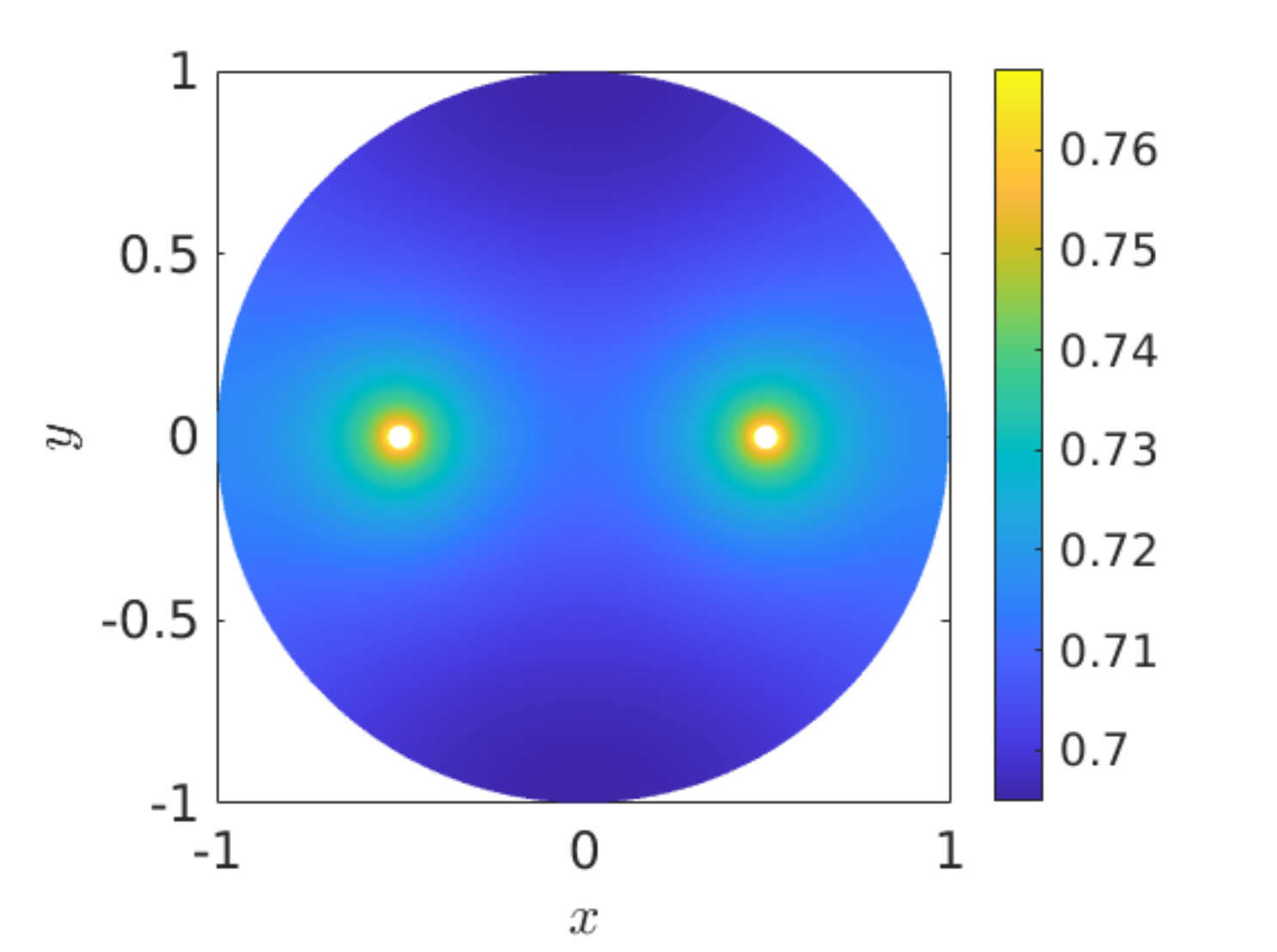  
		\end{subfigure}
	\end{subfigure}
	\begin{subfigure}[b]{.45\textwidth}
	    \begin{subfigure}[b]{1.\textwidth}
	        \centering
                \def\svgwidth{1\textwidth}
                    \def\svgheight{4.2cm}
	        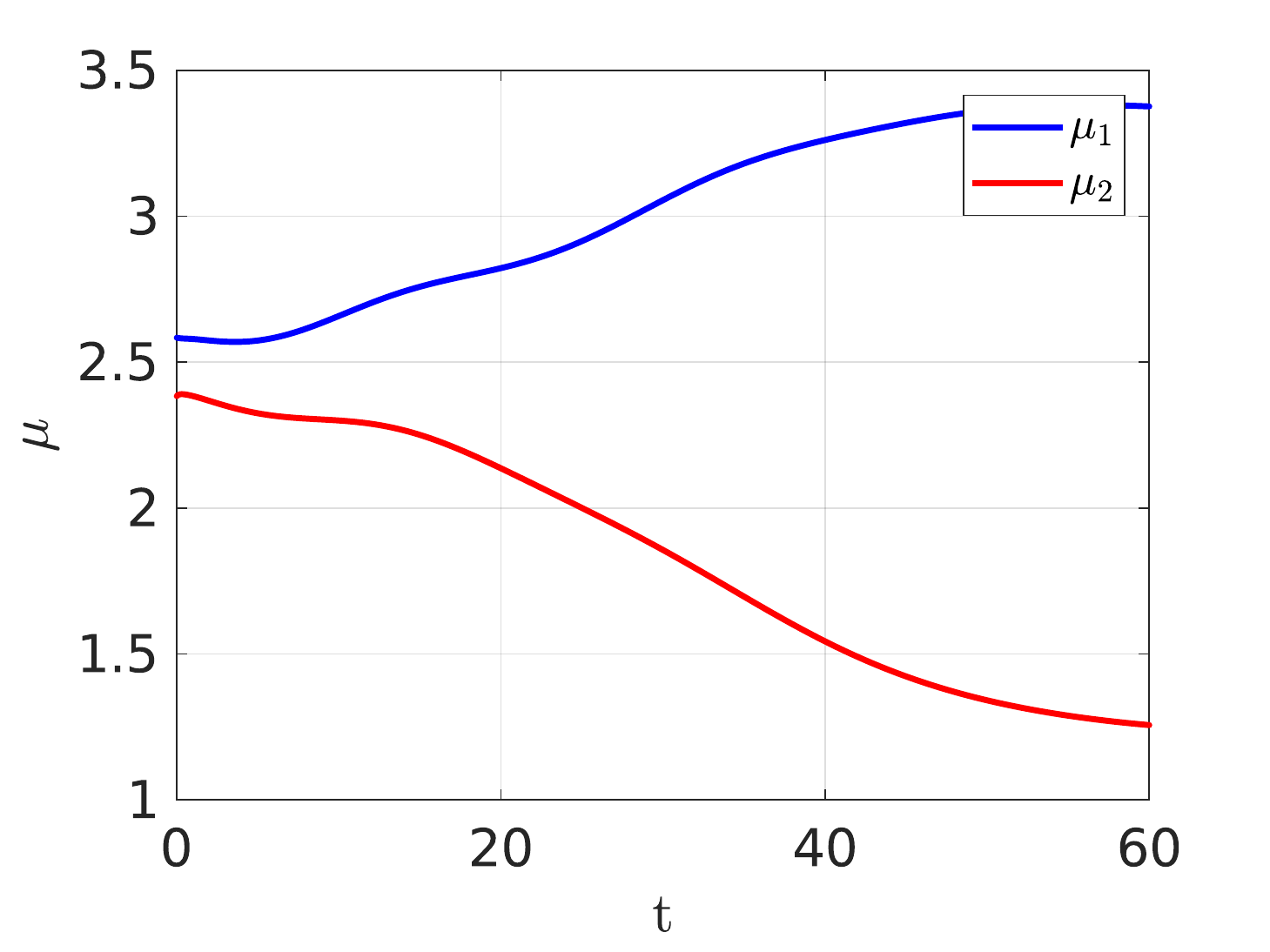
	    \end{subfigure}
	    \begin{subfigure}[b]{1.\textwidth}
	        \centering
                \def\svgwidth{1\textwidth}
                    \def\svgheight{4.2cm}
	        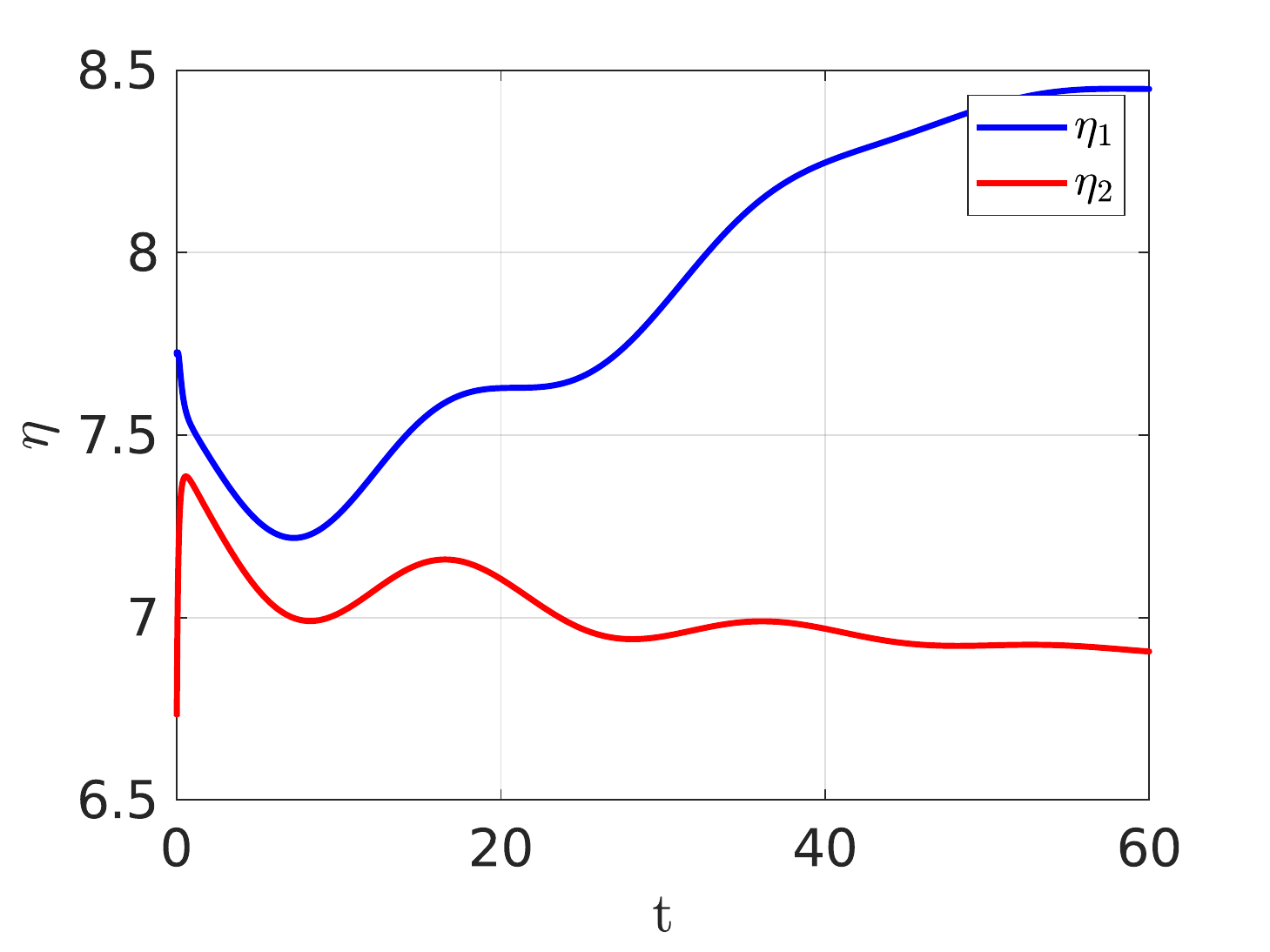
	    \end{subfigure}
	    \begin{subfigure}[b]{1.\textwidth}
    		\centering
                \def\svgwidth{1\textwidth}
                    \def\svgheight{4.2cm}
			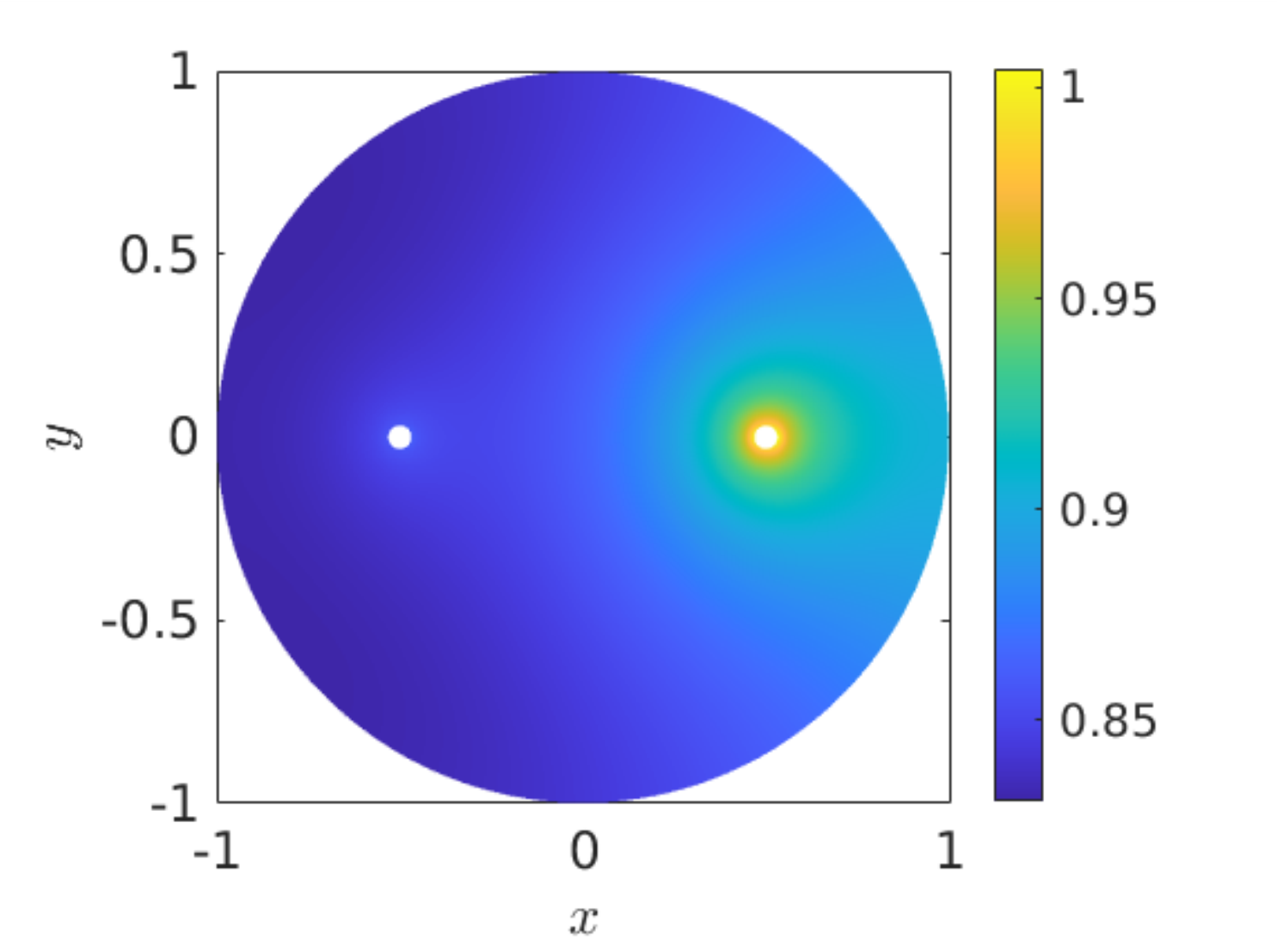  
		\end{subfigure}
	\end{subfigure}
	\caption{Full numerical PDE simulation results of
          \eqref{eqsys:full} with FlexPDE \cite{flexpde} for GM
          kinetics (\ref{cell:GM}). The bottom panels show
          the concentration of $u$. Left: convergence to the symmetric
          branch for $\rho=5$ before the supercritical pitchfork point
          $\rho_p=9.79168$, for an initial condition close to the
          symmetric branch. Right: convergence to the asymmetric
          branch selected by the eigenperturbation direction
          ${\bf q}_2=(1,-1)^T$ for $\rho=15$ and starting near the
          symmetric branch. Parameters:
          $D_u=D_v=5, \sigma_u =\sigma_v=0.6, d_u=0.09,
          \varepsilon=0.03$, and $r=0.5$.}
	\label{fig:GMsuperpitch}
\end{figure}

In the left panel of Figure \ref{fig:GMnonhystdu0p09} we plot the
bifurcation diagram of solutions to (\ref{gm:2cells_all}) for a
parameter set where $D_v=D_u$ and with the other parameter values as
in the figure caption. We observe that a supercritical
symmetry-breaking pitchfork bifurcation from the symmetric branch
occurs at the critical value $\rho=\rho_p\approx 9.79168$. In Figure
\ref{fig:GMsuperpitch} we show full PDE results for (\ref{eqsys:full})
computed with FlexPDE \cite{flexpde} for values of $\rho$ on either
side of this theoretically predicted bifurcation value.  In the left
panels of this figure, we observe that when $\rho = 5<\rho_p$, an
initial perturbation of the symmetric steady-state converges back to
the symmetric steady-state as time increases. In contrast, when
$\rho=15>\rho_p$, we observe from the right panels of Figure
\ref{fig:GMsuperpitch} that, for an initial condition near the
symmetric steady-state, the time-dependent PDE solution converges as
time increases to the asymmetric steady-state predicted in the left
panel of Figure \ref{fig:GMnonhystdu0p09}.

In the right panel of Figure \ref{fig:GMnonhystdu0p09} we show that
the pitchfork bifurcation value for the emergence of asymmetric
steady-states increases substantially when the ring radius for the
two-cell pattern increases. As a result, we conclude that for cells
that are farther apart, a larger value of $\rho$ is needed to create a
stable asymmetric pattern.

\begin{figure}[]
\centering
    \begin{subfigure}[H]{0.48\textwidth}
      \def\svgwidth{1.12\textwidth}
          \def\svgheight{4.5cm}
    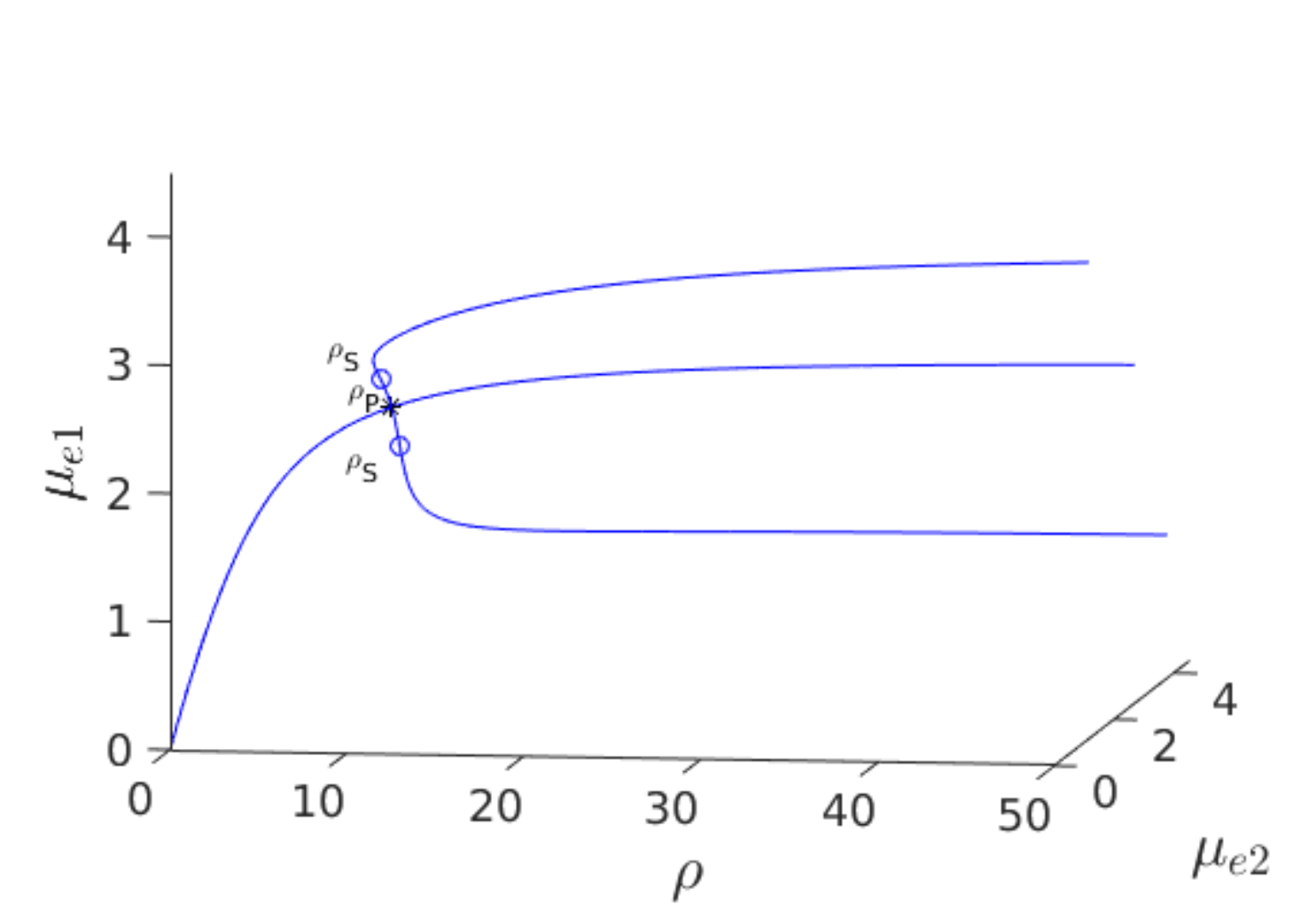
    \label{fig:GMnonhystdu0p08}
    \end{subfigure}
    \begin{subfigure}[H]{0.48\textwidth}
      \def\svgwidth{1.12\textwidth}
          \def\svgheight{4.5cm}
    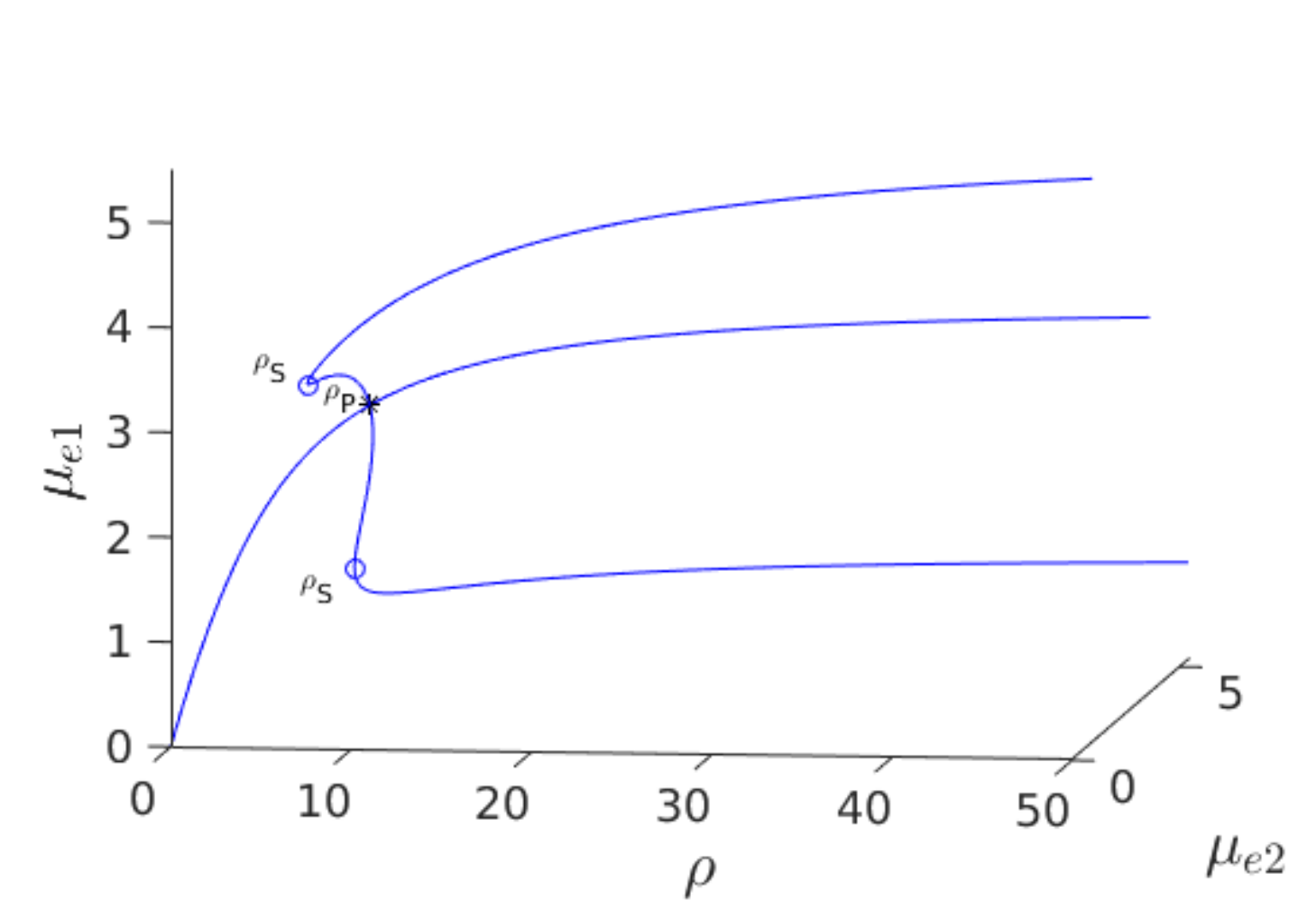
    \label{fig:GMnonhystdu0p05}
    \end{subfigure}
    \caption{3-D Bifurcation diagram, computed from (\ref{gm:2cells_all}),
      for symmetric and asymmetric steady-states of a two-cell ring
      pattern with ring radius $r=0.5$ and two different values of
      $d_u$ with GM kinetics (\ref{cell:GM}). Left panel:
      $d_u=0.08$. Right panel: $d_u=0.05$. For these values of $d_u$,
      the steady-states exhibit hysteresis, i.e., a subcritical
      pitchfork bifurcation occurs from the symmetric equilibrium
      branch, with the emerging unstable asymmetric equilibrium
      branches regaining stability at a secondary fold point. Observe
      that the extent of the hysteresis increases when $d_u$
      decreases. Parameters:
      $D_u=D_v=5, \sigma_u =\sigma_v=0.6, \varepsilon=0.03$, and
      $r=0.5$.}
    \label{fig:GMhystbifdiag}
\end{figure}

We now show that by varying the membrane reaction rate $d_u$, which
necessarily varies the membrane reaction rate to the $v$-species
according to $d_v=\rho\,d_u$, the steady-state solution branches with
GM kinetics (\ref{cell:GM}) computed from (\ref{gm:2cells_all}) can
exhibit a hysteresis structure for low $d_u$. The numerical results of
Figure \ref{fig:GMhystbifdiag} show such a hysteretic bifurcation
structure between the asymmetric and symmetric solution branches for
two values of $d_u$. We observe that as $d_u$ decreases the extent of
the hysteresis increases. The range where hysteresis occurs is given
by the separation $\rho_p-\rho_s$ between the pitchfork point $\rho_p$
and the secondary fold bifurcation point $\rho_s$ along each asymmetric
branch. Numerical results for this range for a parameter set where
hysteresis occurs when $d_u < 0.09$ is given in Table \ref{tab:2-cell
  GM hysteresis du}.

\begin{table}[ht]
\centering
\footnotesize
\begin{tabular}{|c||c|c|c|c|c|c|c|c|c|c|}
    \hline
    $d_u$             & 0.05      & 0.06       & 0.07      & 0.08       & 0.09      & 0.1       & 0.11 & 0.12  & 0.13 & 0.135\\
    \hline\hline
    $\rho_p$            & 7.70971  & 7.66508  & 11.42015  & 8.62258  & 9.79168  & 11.81838  & 15.65552 & 24.82347 & 70.62460 & $>$ 1000 or $\nexists$\\
    \hline
    $\mu_e$             & 2.78094   & 2.54938   & 3.21380   & 2.28060   & 2.19994   & 2.14061   & 2.09668 & 2.06422 & 2.04050 &\\
    \hline\hline
    $\rho_s$            & 6.27944   & 6.93251   & 6.82631   & 8.60260   & -  & -   & - & - & - & - \\
    $\mu_{e1}$           & 3.27136   & 3.09001   & 3.45845   & 2.56189   & -   & -   & - & - & - & - \\
    $\mu_{e2}$           & 1.12375   & 1.34489   & 0.87895   & 1.93057   & -   & -  & - & - & - & -\\
    \hline
\end{tabular}
\caption{Numerical values (rounded to 5th decimal place) of the
  subcritical (or supercritical) pitchfork bifurcation point $\rho_p$,
  the fold bifurcation point $\rho_s$, and the associated
  values for the symmetric $\mu_e$ and one of the asymmetric
  ($\mu_{e1}$, $\mu_{e2}$) solution branches. As $d_u$ increases from
  $0.05$, the range of $\rho$ where hysteresis occurs decreases, until
  a supercritical pitchfork bifurcation occurs when $d_u\approx
  0.85$. Parameters:
  $D_u=D_v=5, \sigma_u =\sigma_v=0.6,\varepsilon=0.03, r=0.5$.}
\label{tab:2-cell GM hysteresis du}
\end{table}

\begin{figure}[]
    \centering
	\begin{subfigure}[b]{.45\textwidth}
	    \begin{subfigure}[b]{1.\textwidth}
	        \centering
                \def\svgwidth{1\textwidth}
                            \def\svgheight{4.2cm}
	        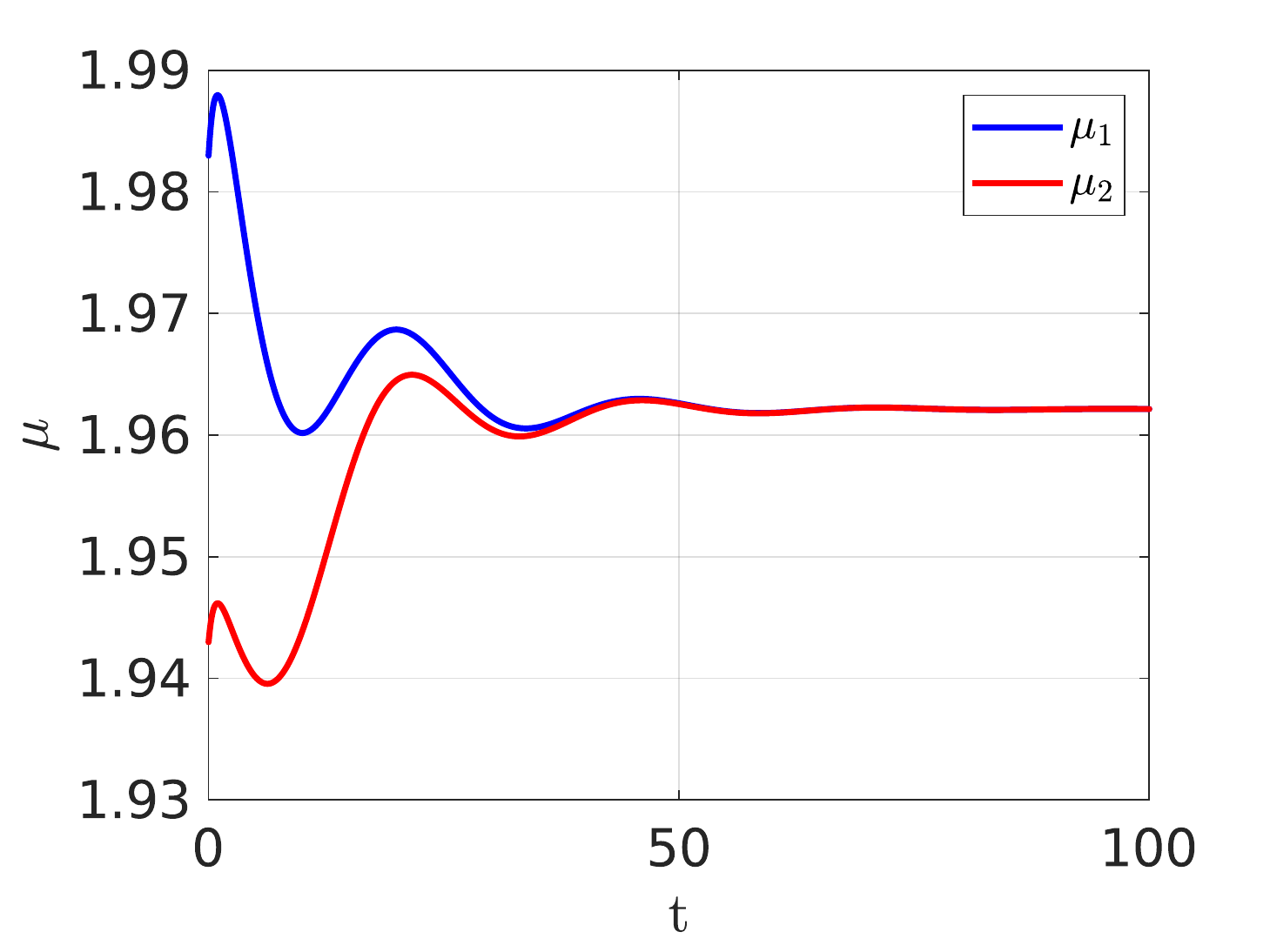
	    \end{subfigure}
	    \begin{subfigure}[b]{1.\textwidth}
	        \centering
                \def\svgwidth{1\textwidth}
                                            \def\svgheight{4.2cm}
	        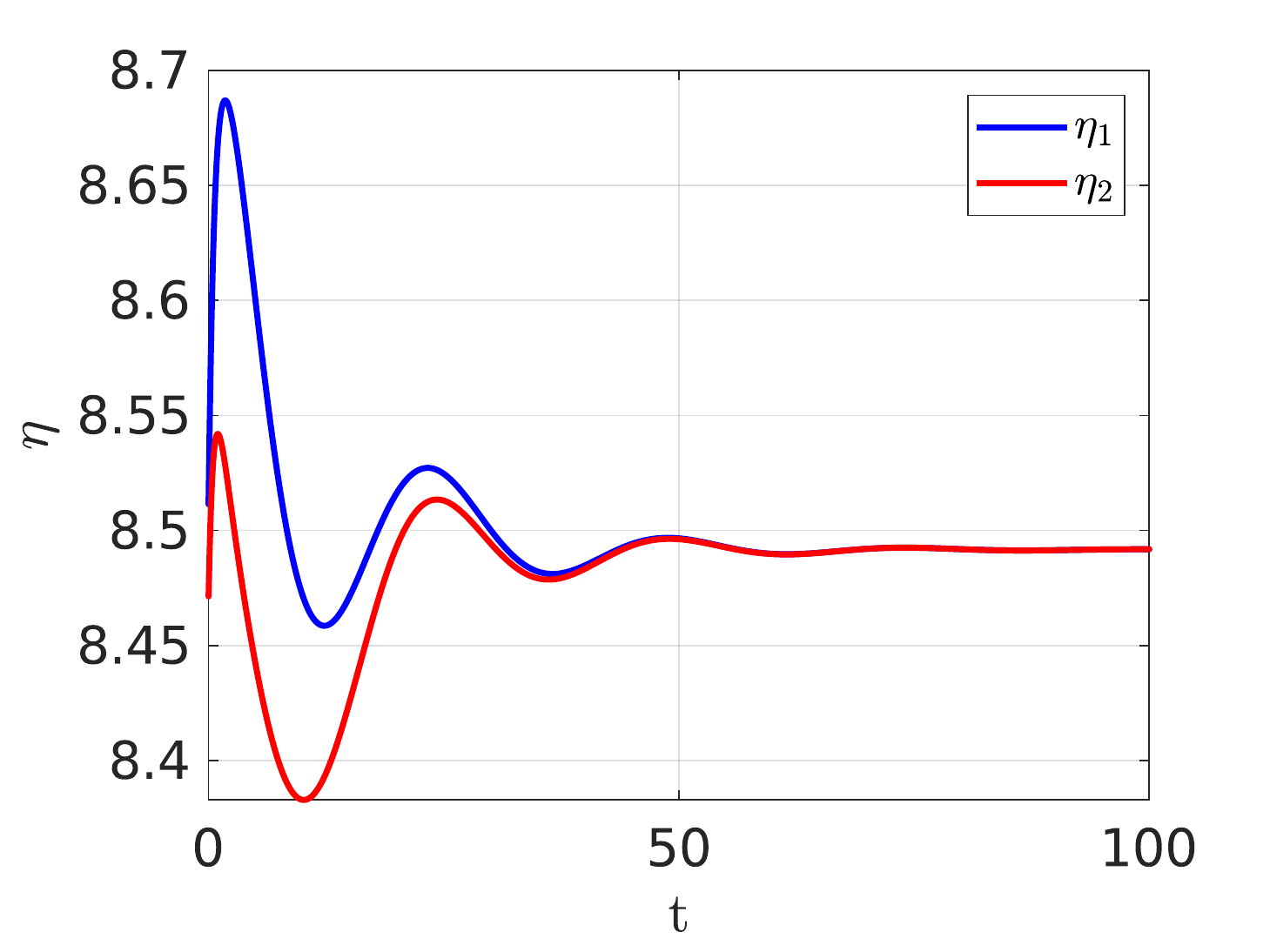
	    \end{subfigure}
	    \begin{subfigure}[b]{1.\textwidth}
    		\centering
                \def\svgwidth{1\textwidth}
                                            \def\svgheight{4.2cm}
			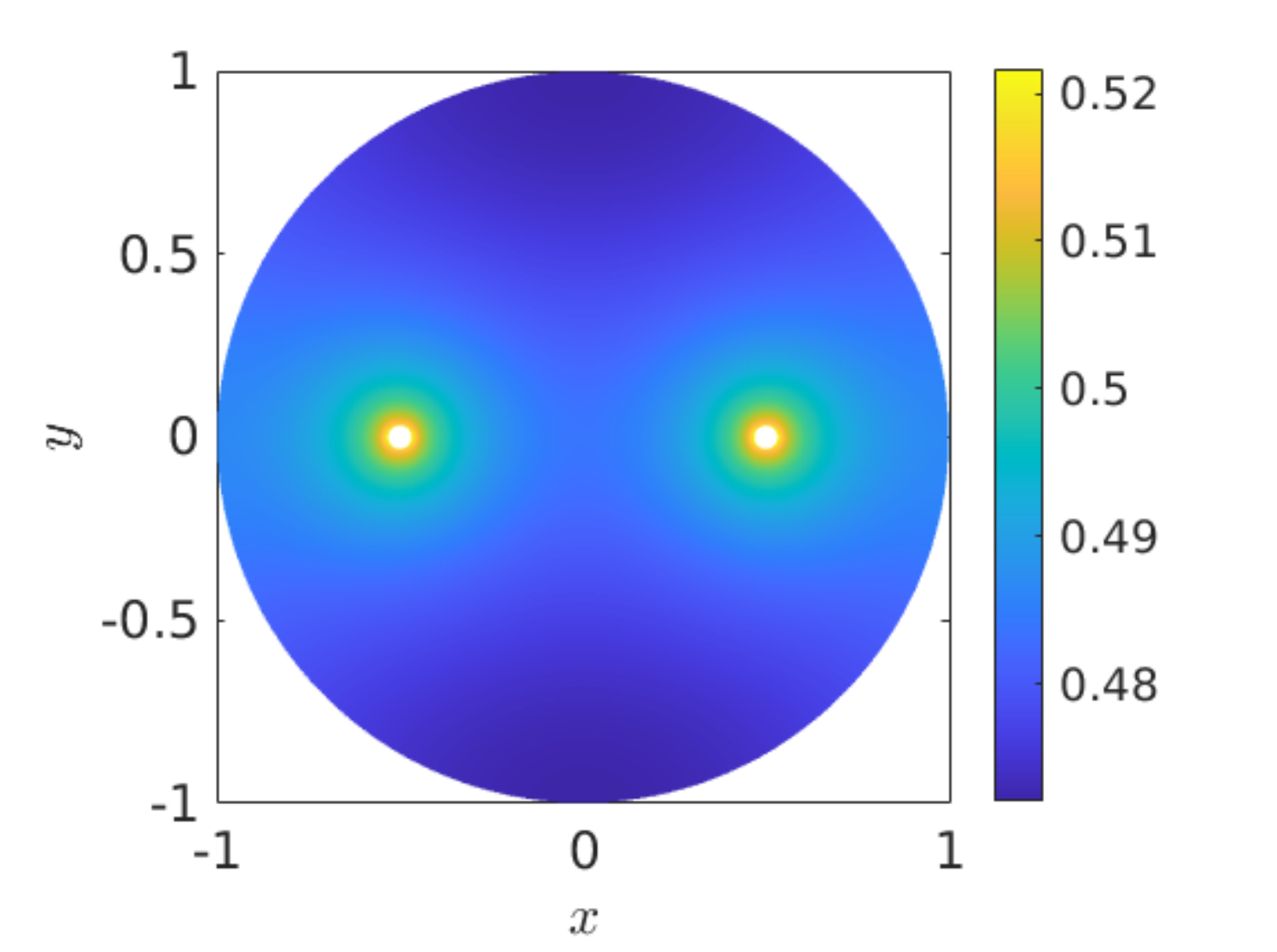  
		\end{subfigure}
	\end{subfigure}
	\begin{subfigure}[b]{.45\textwidth}
	    \begin{subfigure}[b]{1.\textwidth}
              \centering
              \def\svgwidth{1\textwidth}
                                          \def\svgheight{4.2cm}
	        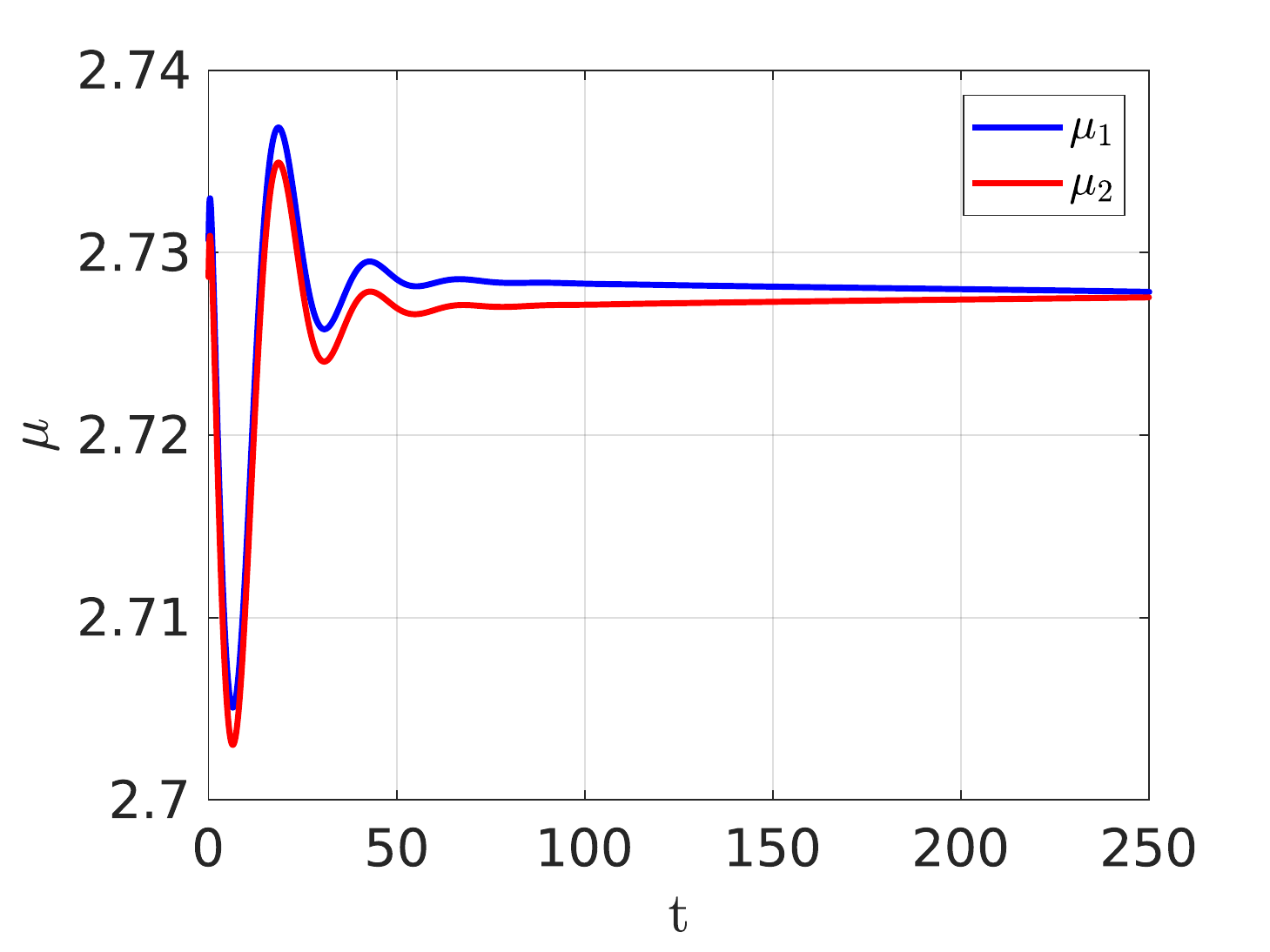
	    \end{subfigure}
	    \begin{subfigure}[b]{1.\textwidth}
	        \centering
                \def\svgwidth{1\textwidth}
                                            \def\svgheight{4.2cm}
	        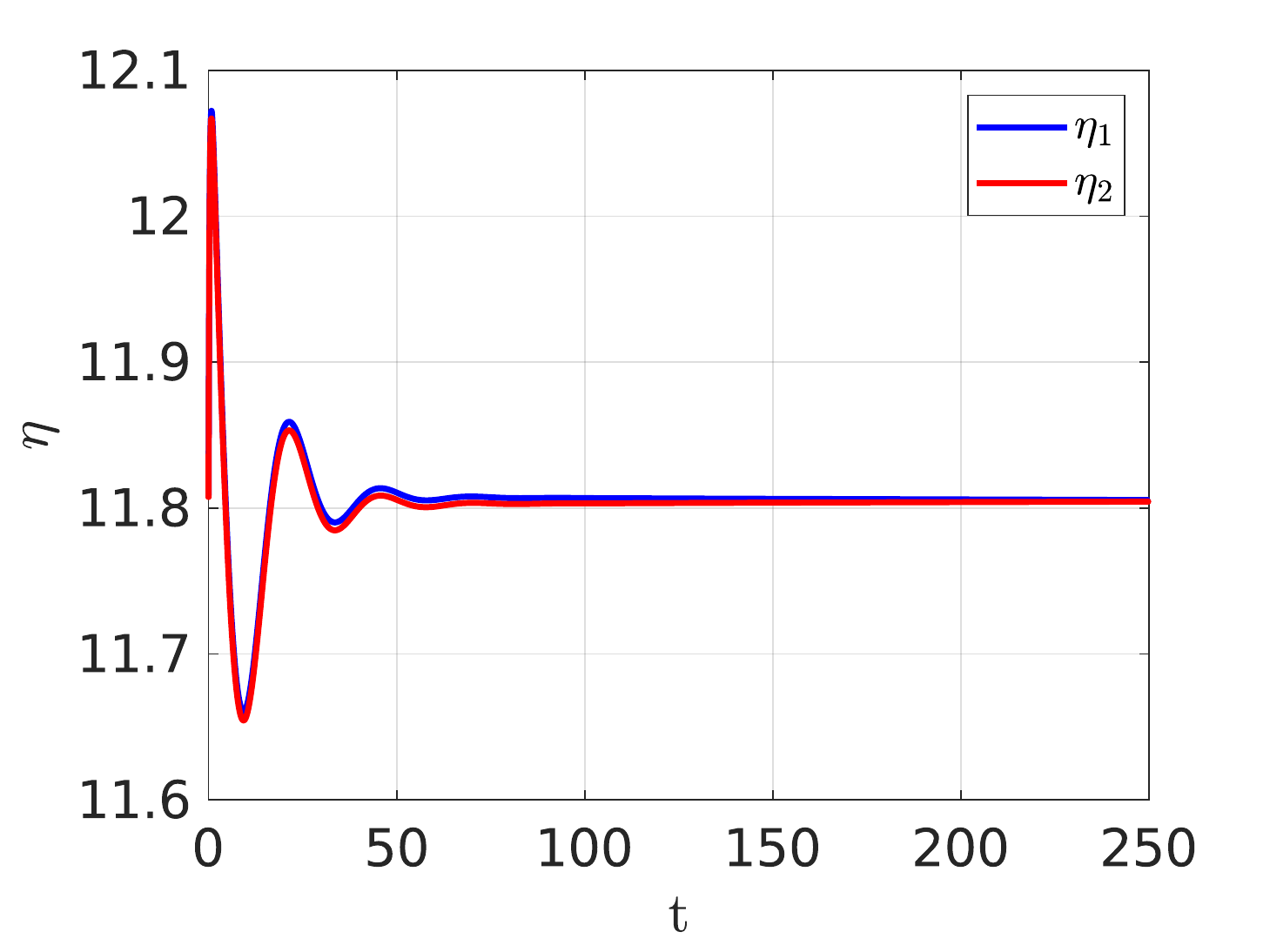
	    \end{subfigure}
	    \begin{subfigure}[b]{1.\textwidth}
    		\centering
                \def\svgwidth{1\textwidth}
                                            \def\svgheight{4.2cm}
			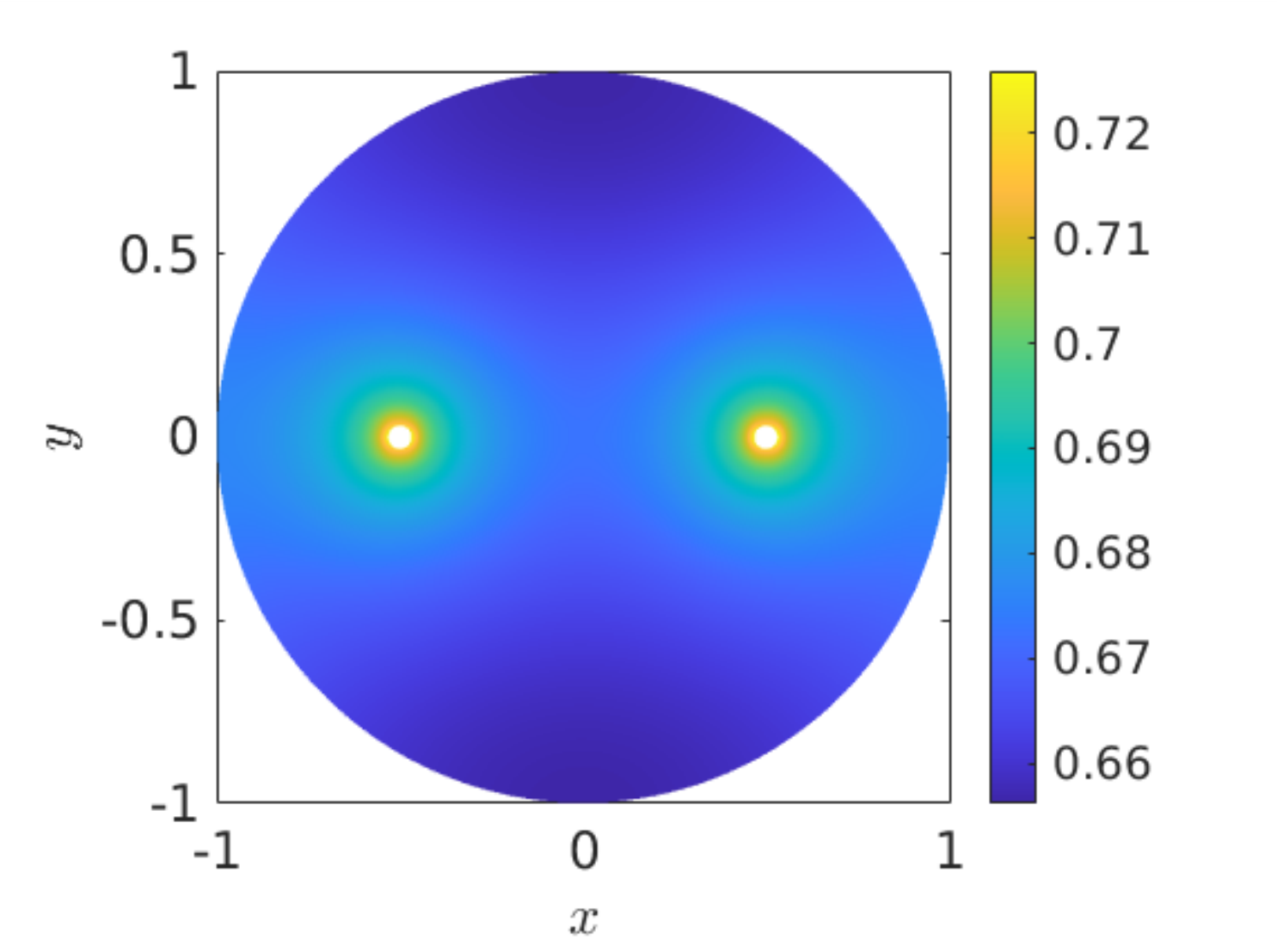  
		\end{subfigure}
	\end{subfigure}
	\caption{Full numerical simulation results of
          \eqref{eqsys:full} with FlexPDE \cite{flexpde} for GM
          kinetics (\ref{cell:GM}). Left: for an initial condition
          near the symmetric branch we observe convergence to the
          symmetric branch when $\rho=4$, which is before the
          hysteresis region bounded by the fold point
          $\rho_s \approx 6.27945$ and the subcritical pitchfork point
          $\rho_p \approx 7.70971$. Right: convergence to the
          symmetric branch for $\rho = 7.2$, which lies on the range
          $\rho_s<\rho<\rho_p$, when starting near the symmetric
          branch. The bottom panels show the concentration of
            $u$. Parameters:
          $D_u=D_v=5, \sigma_u =\sigma_v=0.6, d_u=0.05,
          \varepsilon=0.03, r=0.5$.}
	\label{fig:GMhystsym}
\end{figure}

\begin{figure}[]
    \centering
	\begin{subfigure}[b]{.45\textwidth}
	    \begin{subfigure}[b]{1.\textwidth}
	        \centering
                \def\svgwidth{1\textwidth}
                                            \def\svgheight{4.2cm}
	        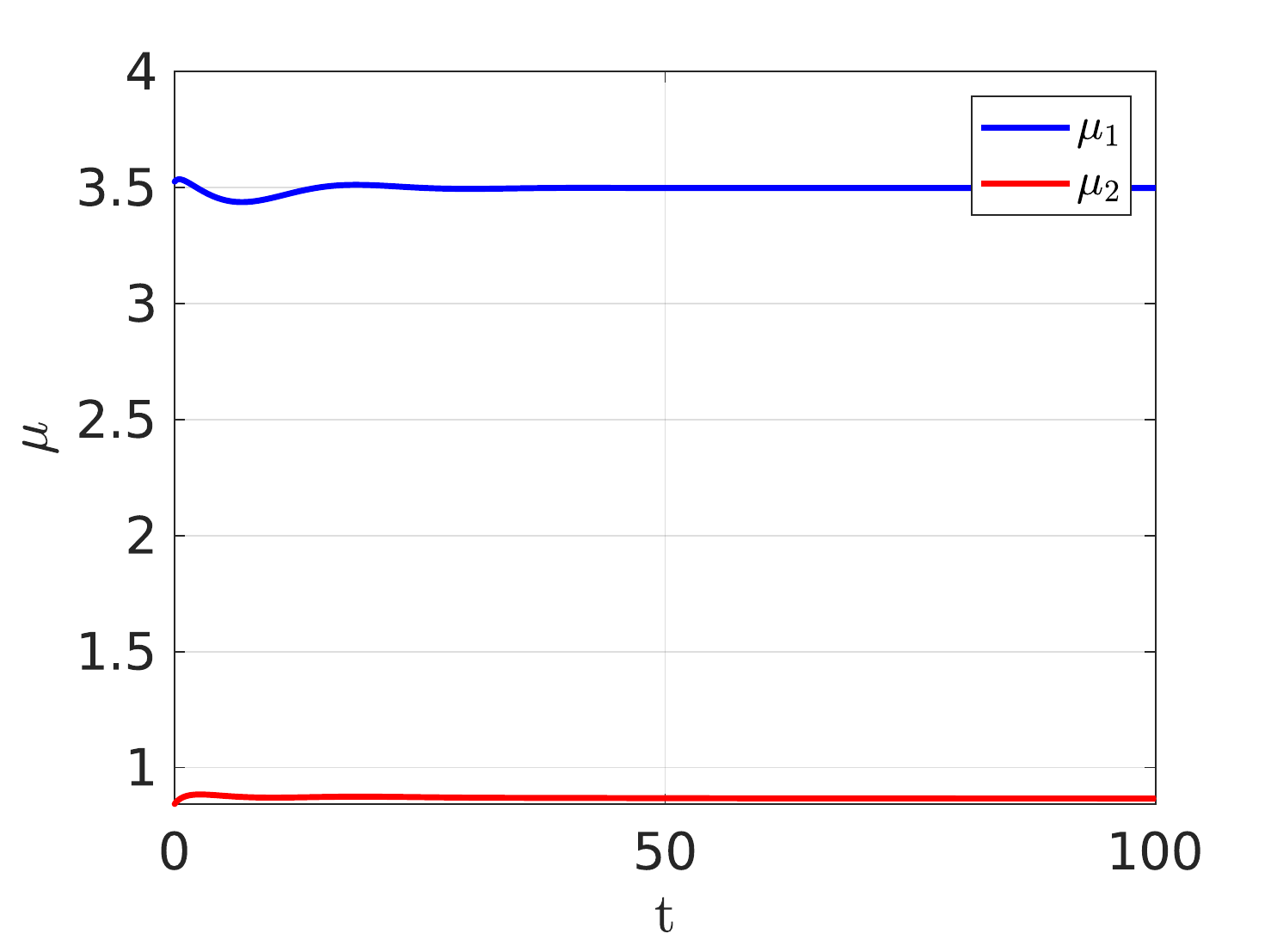
	    \end{subfigure}
	    \begin{subfigure}[b]{1.\textwidth}
	        \centering
                \def\svgwidth{1\textwidth}
                                            \def\svgheight{4.2cm}
	        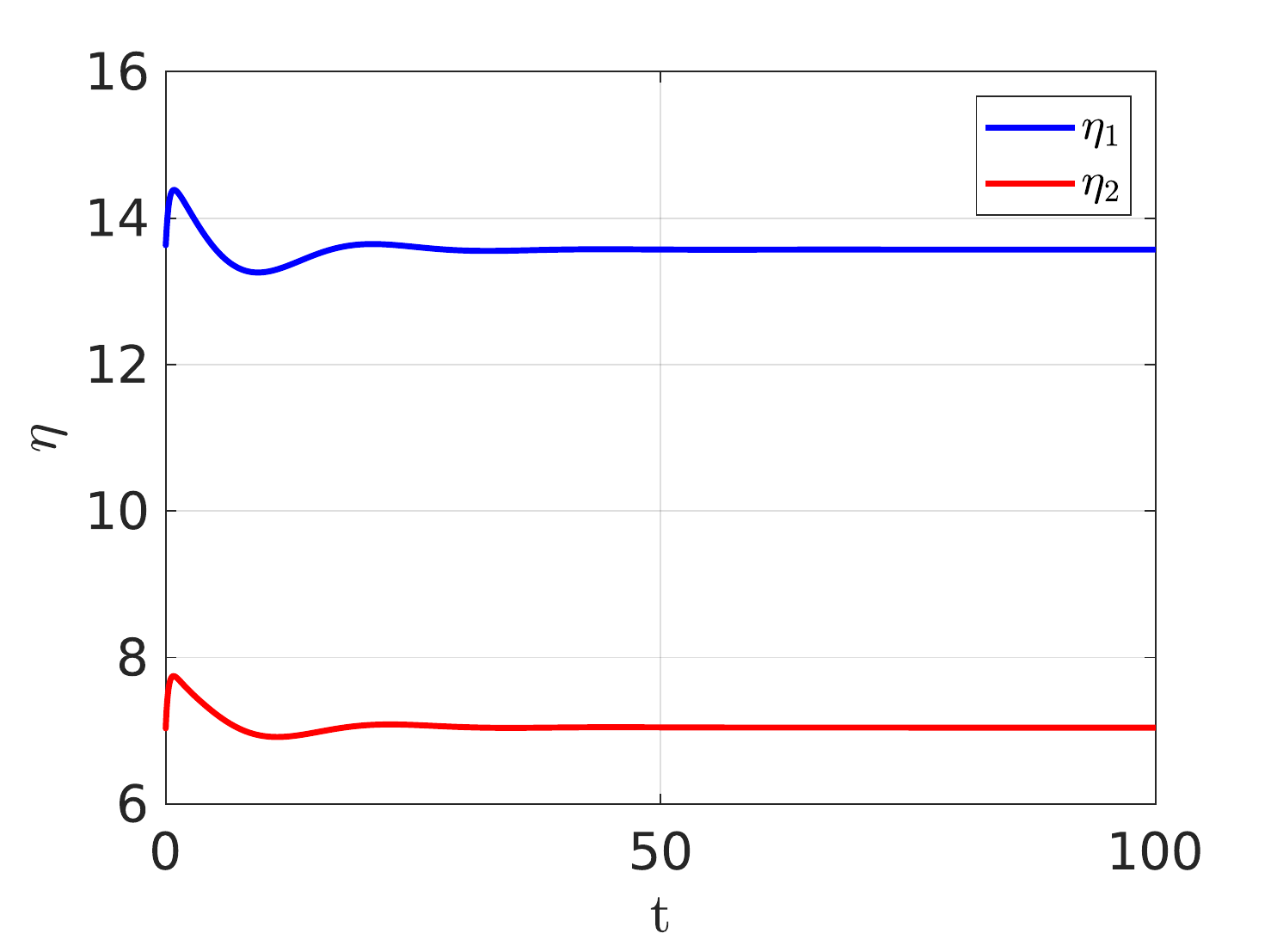
	    \end{subfigure}
	    \begin{subfigure}[b]{1.\textwidth}
    		\centering
                \def\svgwidth{1\textwidth}
                                            \def\svgheight{4.2cm}
			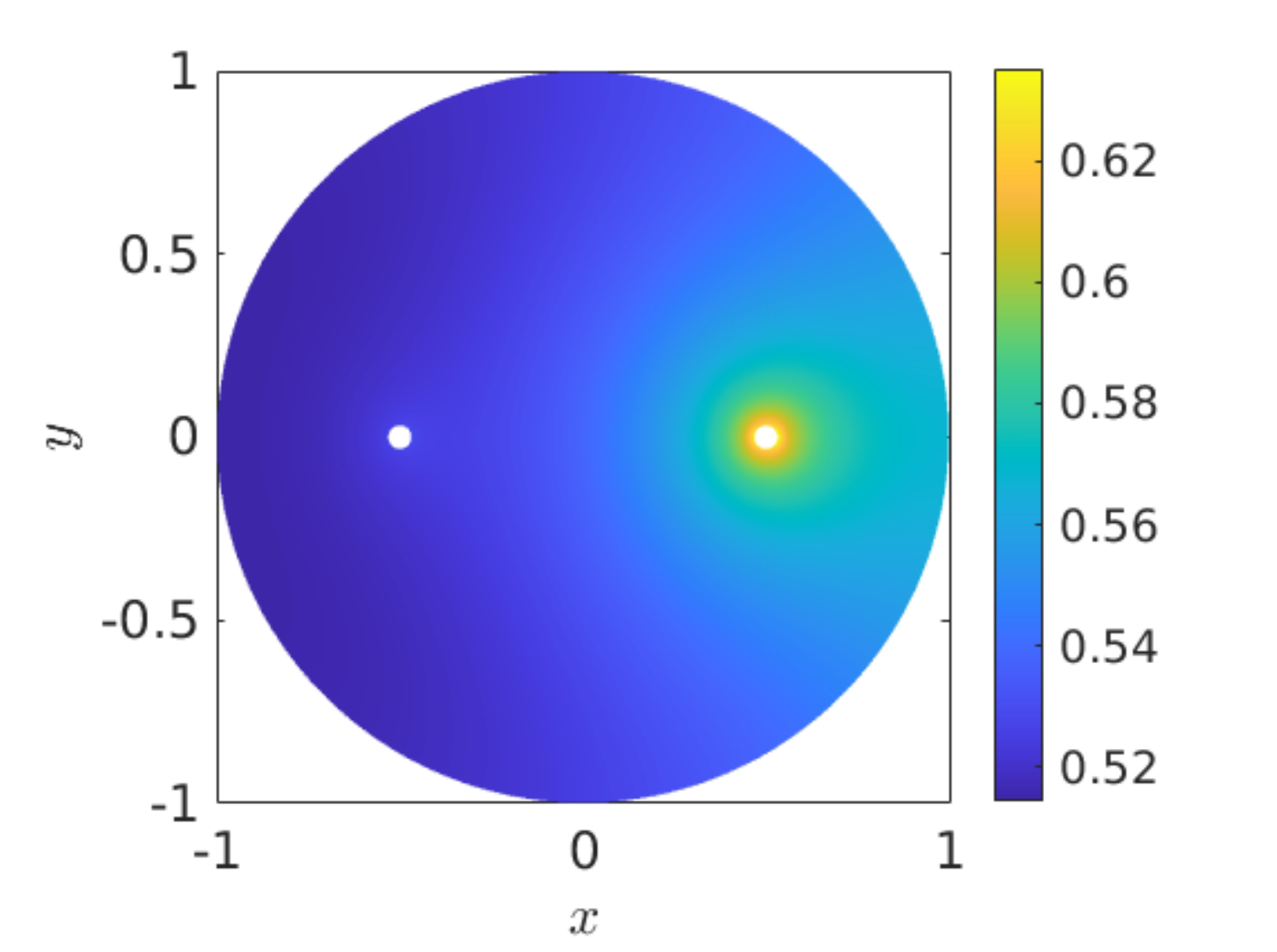  
		\end{subfigure}
	\end{subfigure}
	\begin{subfigure}[b]{.45\textwidth}
	    \begin{subfigure}[b]{1.\textwidth}
	        \centering
                \def\svgwidth{1\textwidth}
                                            \def\svgheight{4.2cm}
	        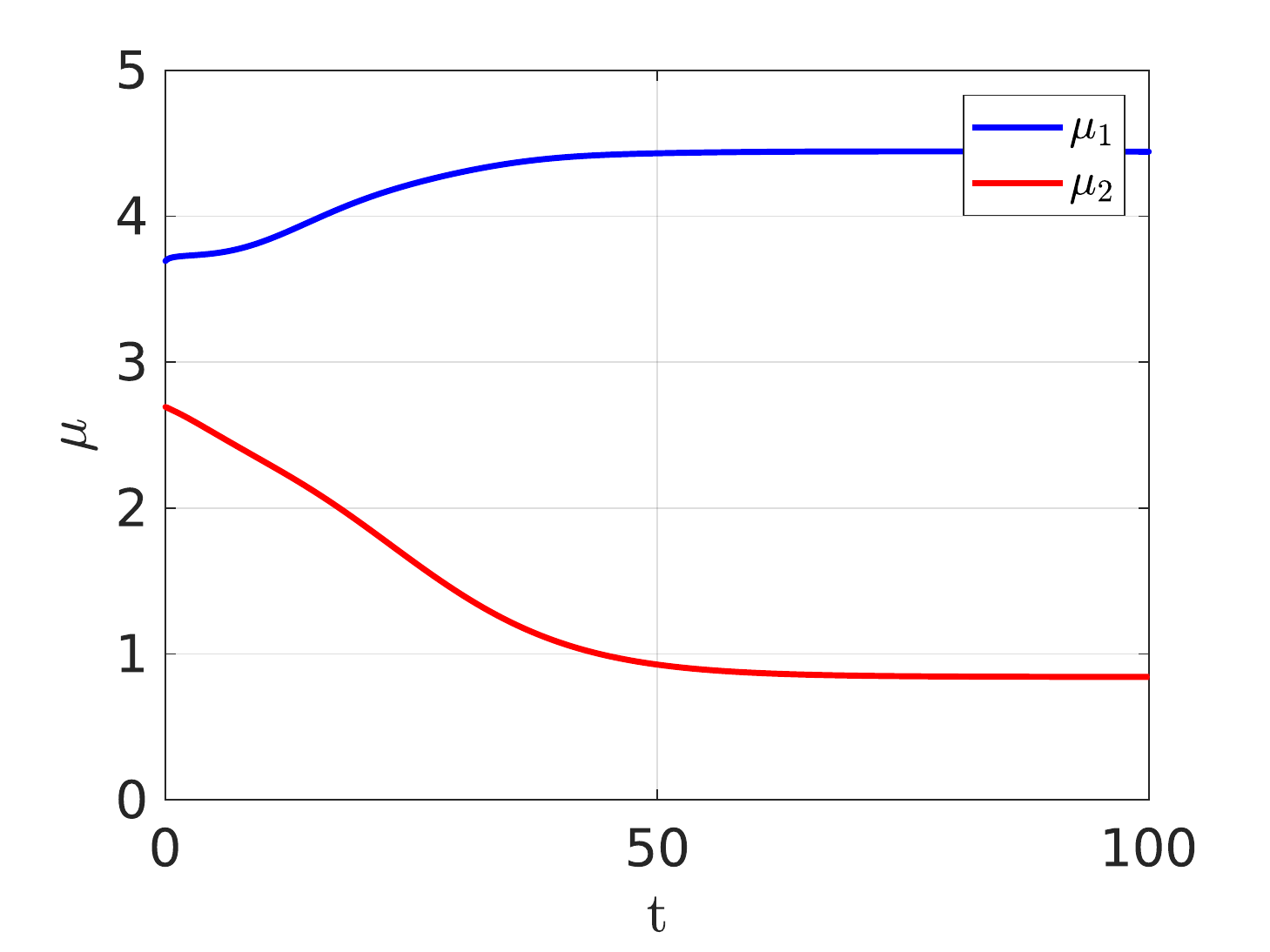
	    \end{subfigure}
	    \begin{subfigure}[b]{1.\textwidth}
	        \centering
                \def\svgwidth{1\textwidth}
                                            \def\svgheight{4.2cm}
	        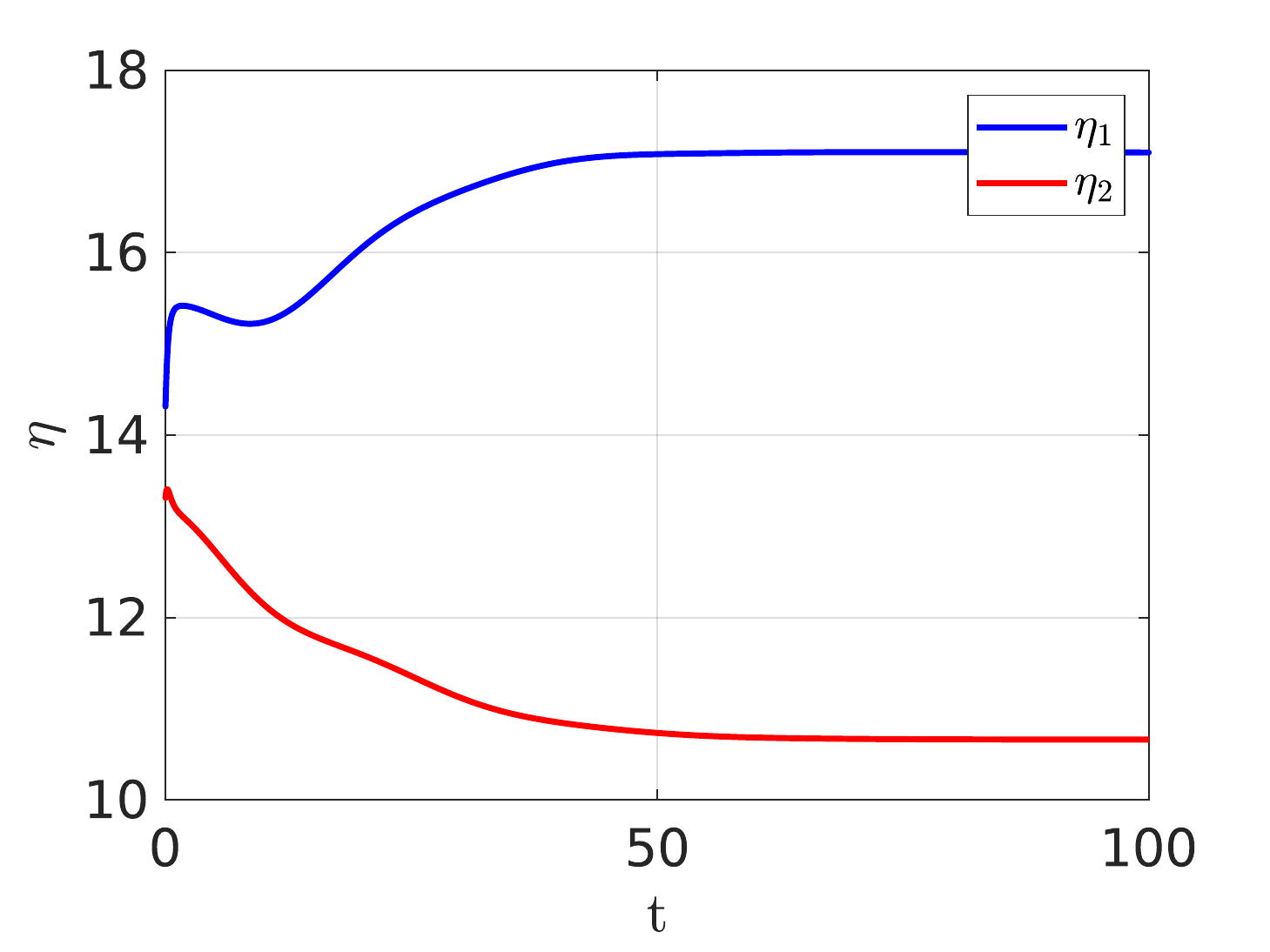
	    \end{subfigure}
	    \begin{subfigure}[b]{1.\textwidth}
    		\centering
                \def\svgwidth{1\textwidth}
                                            \def\svgheight{4.2cm}
			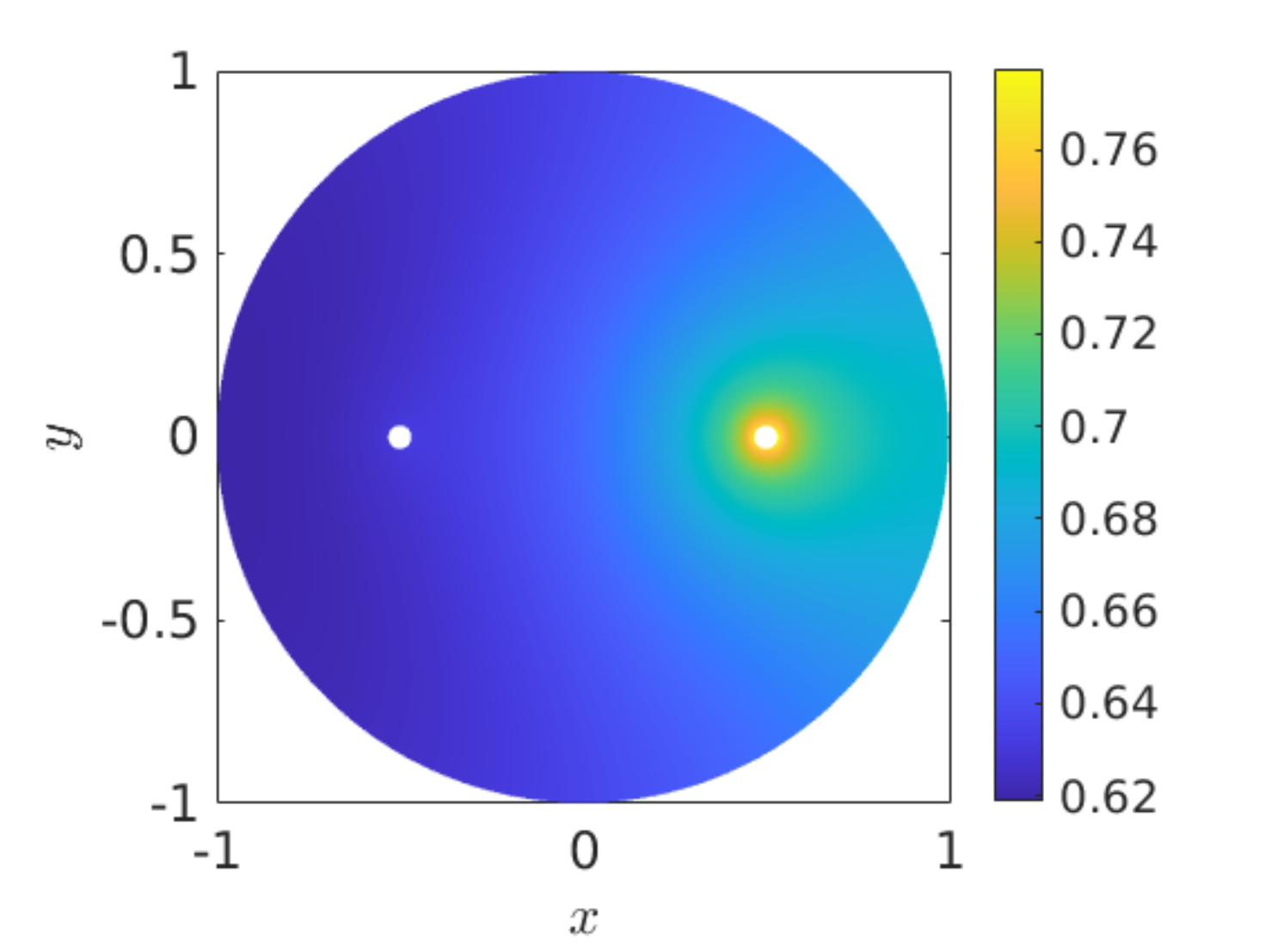  
		\end{subfigure}
	\end{subfigure}
	\caption{Full numerical simulation results of
          \eqref{eqsys:full} with FlexPDE \cite{flexpde} for GM
          kinetics (\ref{cell:GM}). Left: convergence to the
          asymmetric branch for an initial condition near this branch
          when $\rho=7.2$ lies in the hysteresis region between the
          fold point $\rho_s \approx 6.27945$ and the subcritical
          pitchfork point $\rho_p \approx 7.70971$. Right: convergence
          to an asymmetric steady-state as selected by a small initial
          perturbation of the symmetric solution in the direction
          ${\bf q}_2=(1,-1)$ when $\rho = 15>\rho_p$. The bottom
            panels show the concentration of $u$. Parameters:
          $D_u=D_v=5, \sigma_u =\sigma_v=0.6,
          d_u=0.05,\varepsilon=0.03$, and $r=0.5$.}
	\label{fig:GMhystasym}
\end{figure}

For the parameter set with $d_u=0.05$, which corresponds to the
bifurcation diagram shown in the right panel of Figure
\ref{fig:GMhystbifdiag}, the full time-dependent computations of
(\ref{eqsys:full}) with FlexPDE \cite{flexpde}, as shown in Figure
\ref{fig:GMhystsym}, illustrate that for an initial condition near the
symmetric steady-state branch, and with $\rho$ either satisfying
$\rho<\rho_s$ or $\rho_{s}<\rho<\rho_p$, the time-dependent solution
converges to the stable symmetric steady-state solution. However, as
shown in the left panel of Figure \ref{fig:GMhystasym}, for an
initial condition near the asymmetric branch when $\rho$ is in the
hysteresis region $\rho_s<\rho<\rho_p$, the time-dependent solution
converges to the asymmetric steady-state. Moreover, if $\rho>\rho_p$, the
right panel of Figure \ref{fig:GMhystasym} shows that for an initial
condition near the unstable symmetric steady-state the time-dependent
solution converges to the asymmetric steady-state solution.

\begin{figure}[]
	 \begin{subfigure}[b]{0.49\textwidth}
	        \centering
            \includegraphics[width=\textwidth]{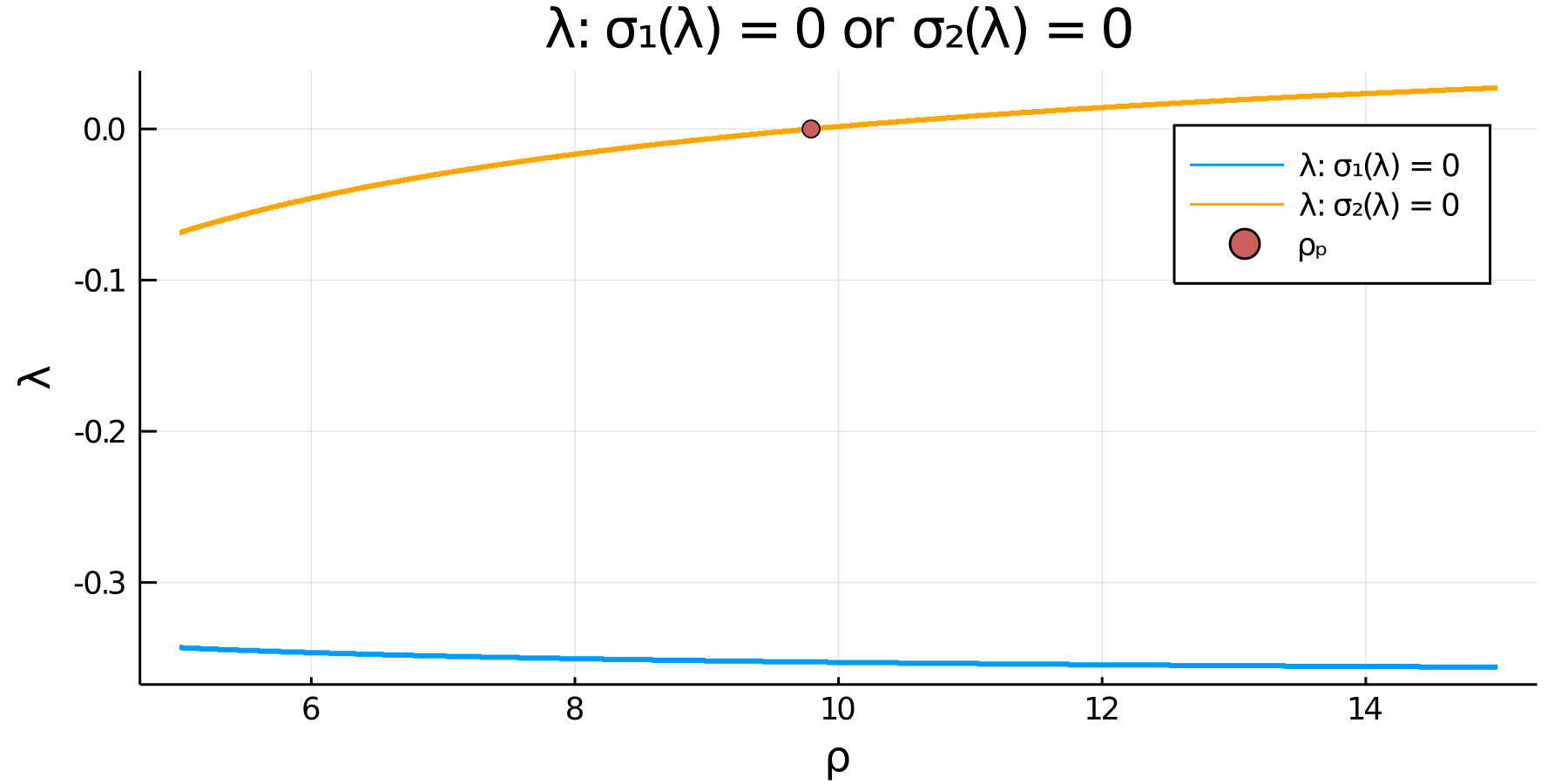}
	    \end{subfigure}
	    \begin{subfigure}[b]{0.49\textwidth}
	        \centering
            \includegraphics[width=\textwidth]{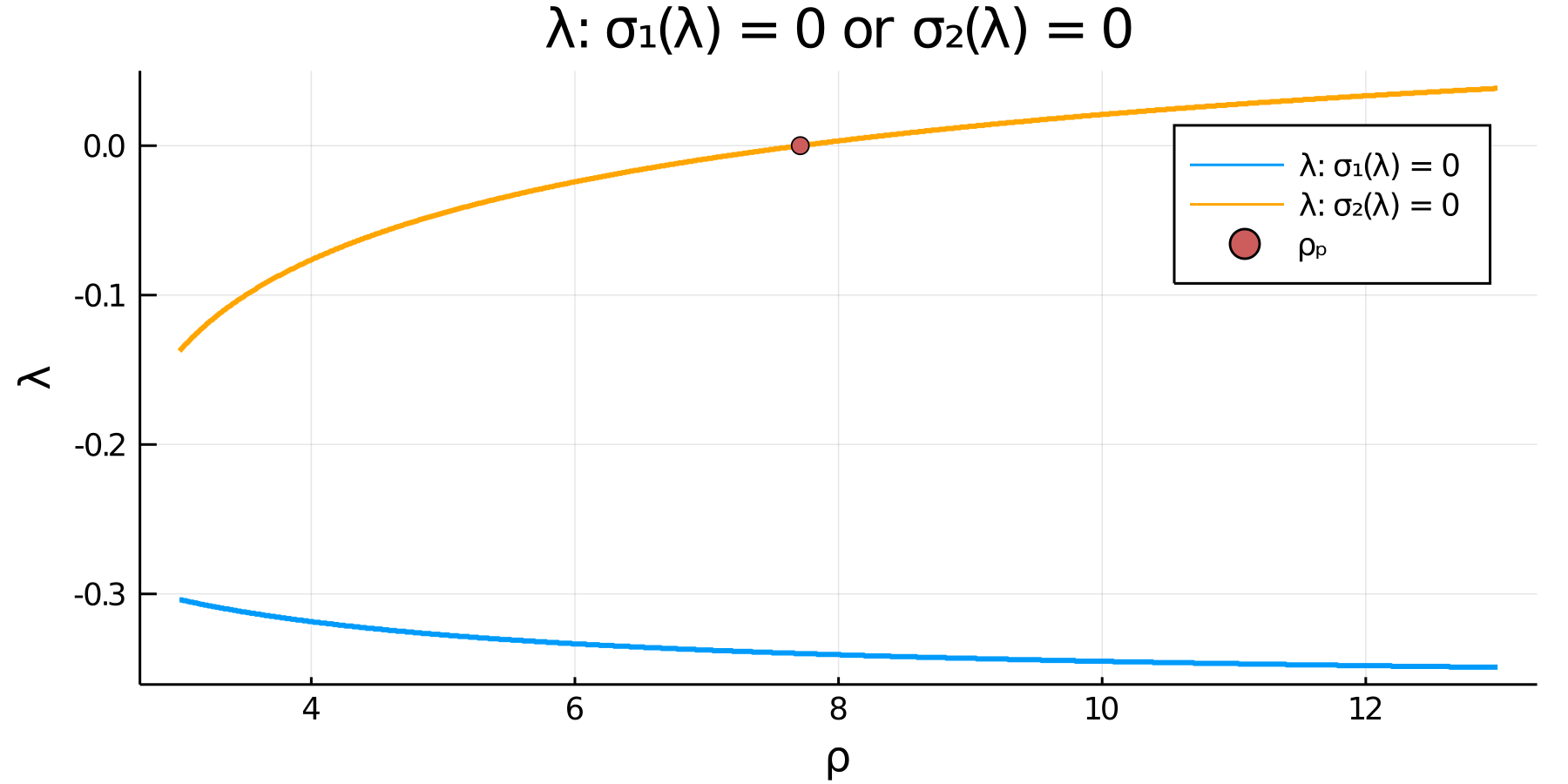}
	    \end{subfigure}
	    \caption{Plots of the numerically computed largest roots
              of $\sigma_1(\lambda)=0$ and $\sigma_2(\lambda)=0$
              versus $\rho$, as defined in (\ref{lin:root}), that
              determine the linear stability properties to either
              in-phase ${\bf e}=(1,1)^T$ or anti-phase
              ${\bf q}_2=(1,-1)^T$ eigenperturbations of the symmetric
              steady-state solution, respectively. Left panel: for
              $d_u=0.09$ we have $\rho_p\approx 9.79168$. Right panel:
              for $d_u=0.05$ we have $\rho_p\approx 7.70971$. Observe
              that in-phase eigenperturbations are always linearly
              stable, whereas anti-phase eigenperturbations are
              linearly stable only on the range $\rho<\rho_p$ before
              the pitchfork point $\rho_p$. Parameters:
              $D_u=D_v=5, \sigma_u =\sigma_v=0.6,\varepsilon=0.03,
              r=0.5$.}
	    \label{fig:eigcrossingGM}
\end{figure}

To determine the linear stability properties of the symmetric
steady-state solution branch as $\rho$ is varied in Figures
\ref{fig:GMnonhystdu0p09} and \ref{fig:GMhystbifdiag} we must
determine the eigenvalues $\lambda$ in the set
(\ref{new:TransDent}). This is done by numerically computing the
largest roots to $\sigma_1(\lambda)=0$ and to $\sigma_2(\lambda)=0$,
as defined in (\ref{lin:root}). In Figure \ref{fig:eigcrossingGM} we
plot these roots versus $\rho$ for two values of $d_u$. From this
figure, we observe that in-phase perturbations of the symmetric
steady-state solution branch, as determined by the roots of
$\sigma_1(\lambda)=0$, are always linearly stable. In contrast,
anti-phase perturbations of the symmetric steady-state, as
characterized by the roots of $\sigma_2(\lambda)=0$, are linearly
stable only for $\rho<\rho_p$, where $\rho_p$ is the symmetry-breaking
threshold. For $\rho>\rho_p$, the symmetric steady-state solution
branch is unstable to an anti-phase eigenperturbation
${\bf q}_2=(1, -1)^T$.

\begin{figure}[]
	\centering
    \includegraphics[width=0.9\textwidth]{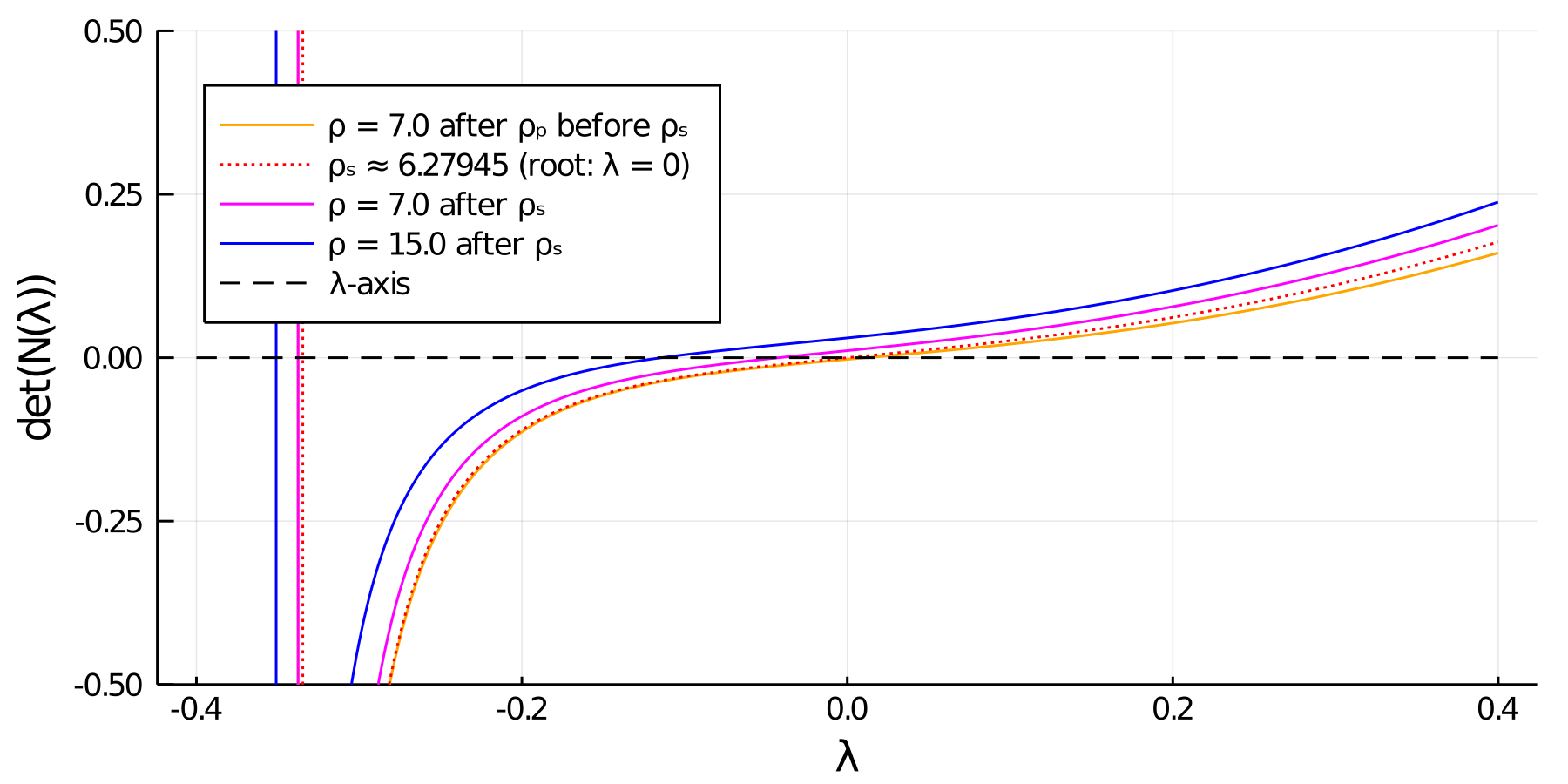}
    \caption{Zero-crossings of $\det(\mathcal{N}(\lambda))=0$, as
      defined in (\ref{lin:glin_non}), determine the linear stability
      properties of an asymmetric steady-state solution with two
      cells.  On the range $\rho_p<\rho<\rho_s$, before the secondary
      fold point along the asymmetric branch, we observe that
      $\lambda>0$. This implies that the subcritical portion of the
      asymmetric steady-state branch between the pitchfork point and
      the fold point is unstable. Further along past the fold point
      the asymmetric branch regains stability. Parameters:
          $D_u=D_v=5, \sigma_u =\sigma_v=0.6,d_u=0.05, \varepsilon=0.03,
          r=0.5$.
}        \label{fig:lin_asymm}
\end{figure}

Next, to study the linear stability properties of the asymmetric
steady-state solution branches we must determine whether the
root-finding condition $\det(\mathcal{N}(\lambda))=0$ in
(\ref{lin:glin_non}) yields an eigenvalue with
$\mbox{Re}\lambda>0$. The numerical results shown in Figure
\ref{fig:lin_asymm} for $d_u=0.05$ (corresponding to the right panel
in Figure \ref{fig:GMhystbifdiag}), establishes that the asymmetric
branch on the subcritical range $\rho_s<\rho<\rho_p$, which emanates
from the symmetric steady-state branch, is unstable. However, as
observed from Figure \ref{fig:lin_asymm}, the upper portion of the
asymmetric branch for $\rho>\rho_s$ is linearly stable.
          
We now explore how the pitchfork bifurcation point depends on
decreasing values of the diffusion coefficient ratio ${D_v/D_u}$ when
$d_u=0.09$. When this ratio is unity, there was no hysteresis between
the symmetric and asymmetric steady-state solution branches (see Table
\ref{tab:2-cell GM hysteresis du}). We remark that a similar numerical
experiment was performed in \S 2.3 of \cite{1dturing} for a 1-D
compartmental-reaction diffusion model with GM kinetics with
dynamically reactive boundaries.  In our 2-D setting, we observe from
the numerical results in Table \ref{tab:2-cell GM hysteresis DvDu
  du=0.09} that a symmetry-breaking bifurcation can occur on the range
${D_v/D_u}<1$, but only up until some minimum diffusion ratio
threshold at which the pitchfork bifurcation point given by the root
of (\ref{gm:2cells_symmb}) no longer exists. In addition, we observe
that reducing the diffusion ratio threshold ${D_v/D_u}$ below unity
for fixed $d_u=0.09$ does not introduce new hysteresis behavior, and
the symmetry-breaking bifurcation remains supercritical.

\begin{figure}[H]
    \centering
    \begin{subfigure}{0.48\textwidth}
        \includegraphics[width=\textwidth,height=4.5cm]{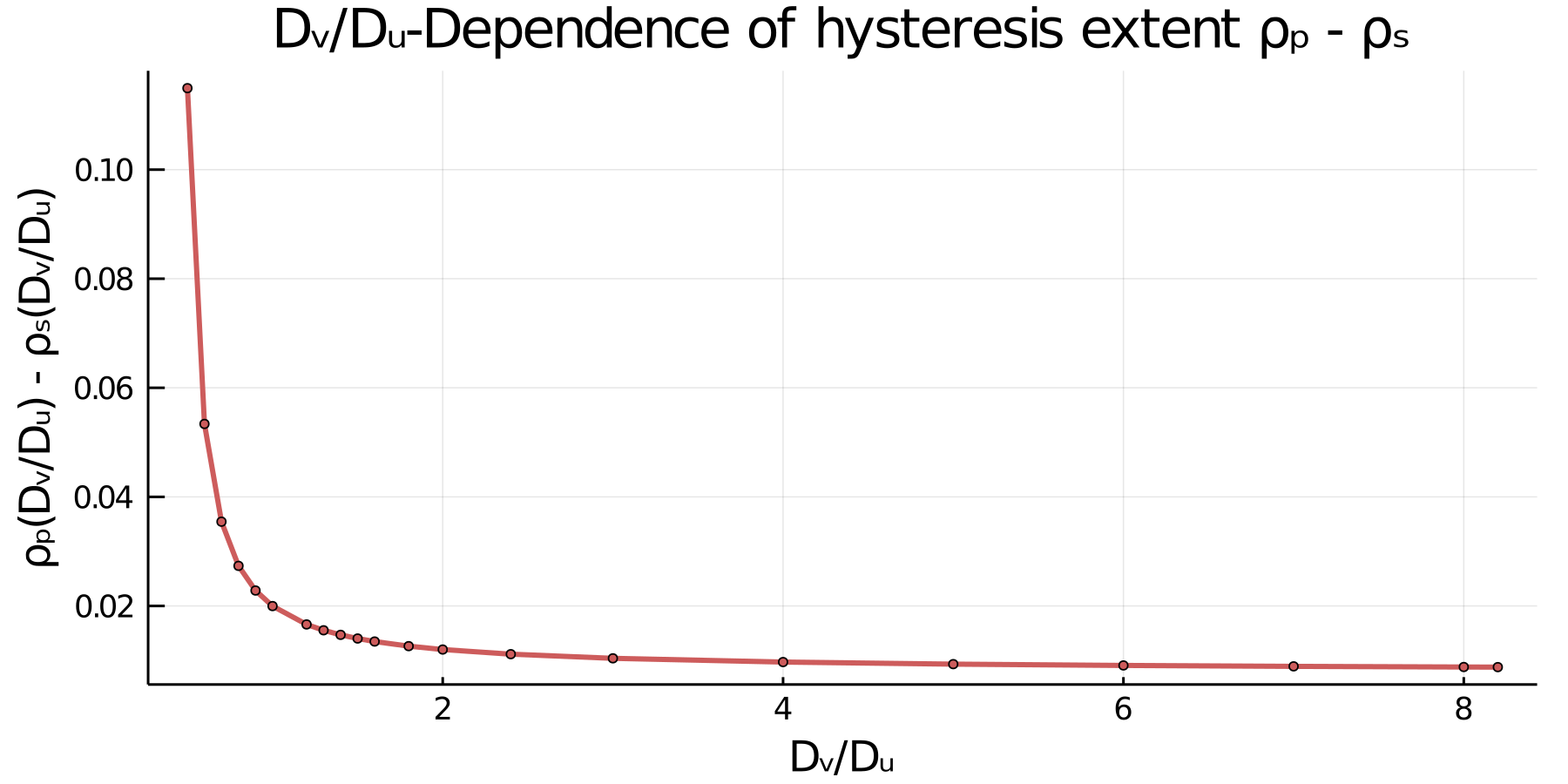}
    \end{subfigure}
    \begin{subfigure}{0.48\textwidth}
        \includegraphics[width=\textwidth,height=4.5cm]{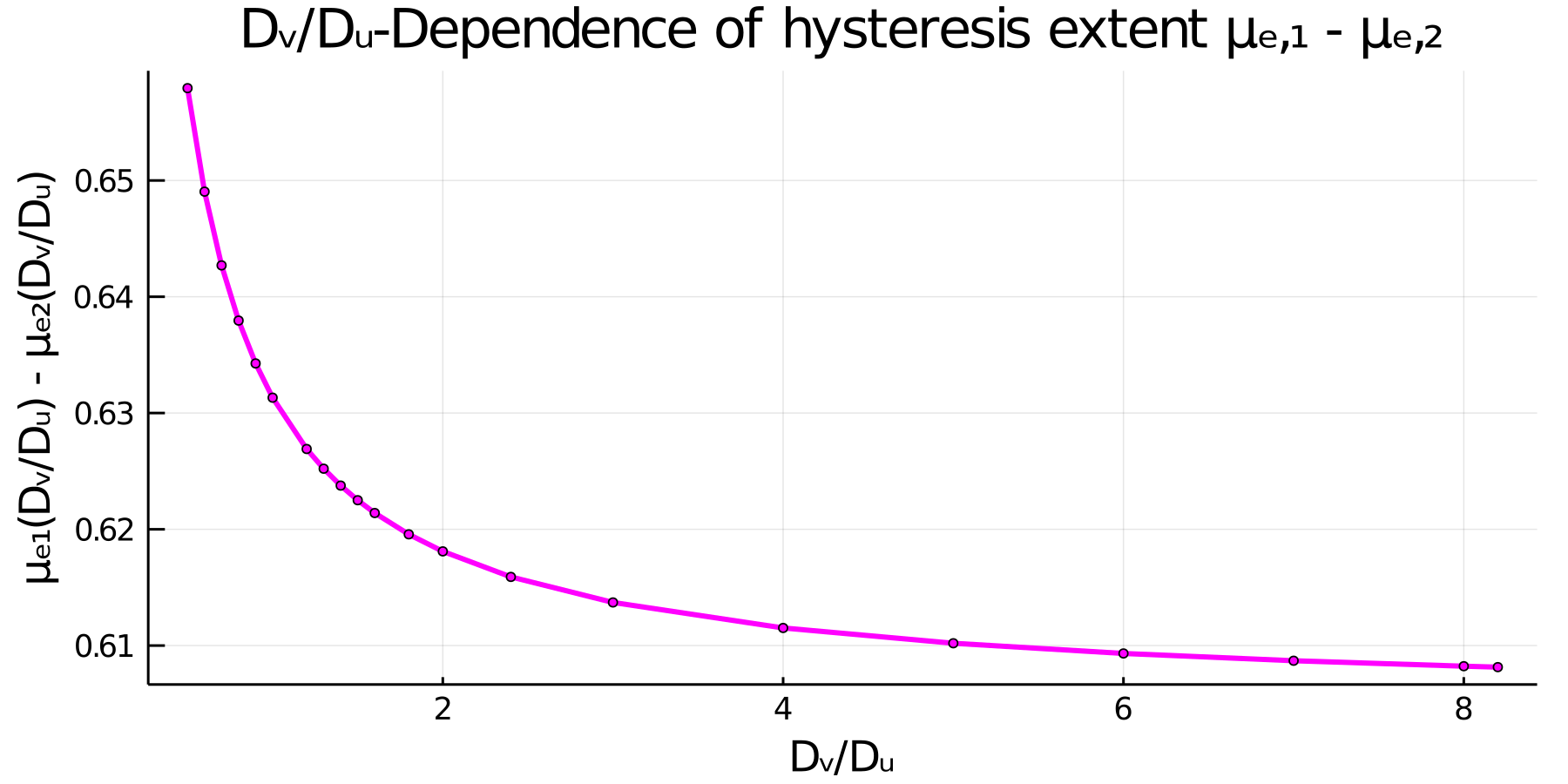}
    \end{subfigure}
    \caption{Effect of the diffusivity ratio ${D_v/D_u}$ of the two
      bulk species on the extent of the hysteresis when $d_u=0.08$, as
      measured by the distance between the fold bifurcation points and
      the subcritical pitchfork bifurcation point (left) and by the
      distance of the two asymmetric equilibria $\mu_{e1}$ and
      $\mu_{e2}$ from each other (right). The diffusivity $D_u=5$ is
      fixed and the remaining parameters are as in Table
      \ref{tab:2-cell GM hysteresis DvDu du=0.08}. The dots are the
      numerically computed values using MatCont \cite{matcont} that
      are interpolated by the plotting function in Julia
      \cite{julia}.}
    \label{fig:DvDueffectonrhop}
\end{figure}

Next, we set $d_u=0.08$ where hysteresis occurs when ${D_v/D_u}=1$,
and we vary this diffusion ratio to determine whether hysteresis can
be eliminated. Our numerical results in Figure
\ref{fig:DvDueffectonrhop} and Table \ref{tab:2-cell GM hysteresis
  DvDu du=0.08} indicates that varying ${D_v/D_u}$ does not eliminate
the hysteresis between the symmetric and asymmetric steady-state
branches. However, the extent of the hysteresis decreases as the ratio
${D_v/D_u}$ increases.

\begin{table}[htbp]
\small
\centering
\begin{tabular}{|c||c|c|c|c|c|c|c|c|}
    \hline
    ${D_v/D_u}$         & 0.42 & 0.43   & 0.5       & 0.6       & 0.8       & 1       & 1.2       & 1.4\\
    \hline\hline
    $\rho_p$            & $>1000$ or $\nexists$ & 573.56743 & 38.45836  & 19.56926  & 12.06861  & 9.79168  & 8.69082  & 8.04185 \\
    \hline
    $\mu_e$             &       & 2.32310 & 2.29271   & 2.26170   & 2.22305   & 2.19994   & 2.18456   & 2.17360\\
    \hline\hline
    $\rho_f$            &-       & -   & -   & -   & -   & -   & - & - \\
    $\mu_{e1}$           &-       & -   & -   & -   & -   & -   & - & - \\
    $\mu_{e2}$           &-       & -   & -   & -   & -   & -   & - & - \\
    \hline
\end{tabular}
\caption{Decreasing the ratio ${D_v/D_u}$ does not trigger hysteresis when
  $d_u=0.09$, but rather there is a minimum threshold of the
  diffusivity ratio where the symmetry-breaking pitchfork bifurcation
  point exists. The numerical values for the pitchfork point
  $(\rho_p,\mu_e)$ on the symmetric steady-state branch are again
  rounded to the 5th decimal place. Parameters:
  $D_u=5, \sigma_u =\sigma_v=0.6, d_u=0.09, \varepsilon=0.03, r=0.5$.}
\label{tab:2-cell GM hysteresis DvDu du=0.09}
\end{table}

\begin{table}[htbp]
\footnotesize
\centering
\begin{tabular}{|c||c|c|c|c|c|c|c|c|c|}
    \hline
    ${D_v/D_u}$   & 0.37       & 0.38       & 0.4       & 0.6       & 0.8       & 1 & 3 & 5 & 8 \\
    \hline\hline
    $\rho_p$      & $>$ 1000 or $\nexists$  & 197.98732  & 72.56533  & 14.30013  & 10.14347  & 8.62258 & 6.13144 & 5.79198 & 5.61640 \\
    \hline
    $\mu_e$       &      & 2.43797   & 2.42519   & 2.34462   & 2.30457   & 2.28061 & 2.21699 & 2.20432 & 2.19719 \\
    \hline\hline
    $\rho_f$      &      & 188.58078   & 71.25577   & 14.24676   & 10.11612   & 8.60260 & 6.12105 & 5.78264 & 5.60759\\
    $\mu_{e1}$     &      & 2.7386611   & 2.72431   & 2.63380   & 2.58881   & 2.56189 & 2.49043 & 2.47619 & 2.46819 \\
    $\mu_{e2}$     &      & 2.063782   & 2.05297   & 1.98476   & 1.95085   &  1.93057 & 1.87672 & 1.86599 & 1.85996 \\
    \hline
\end{tabular}
\caption{Increasing the diffusivity ratio ${D_v/D_u}$ when $d_u=0.08$
  does not eliminate hysteresis, as the symmetry-breaking bifurcation
  point remains subcritical. The numerical values for the pitchfork
  point $(\rho_p,\mu_e)$ on the symmetric equilibrium branch and for
  one of the fold points $(\rho_s, \mu_{e1},\mu_{e2})$ on the
  asymmetric branch are again rounded to the 5th decimal
  place. Parameters:
  $D_u=5, \sigma_u =\sigma_v=0.6, d_u=0.08, \varepsilon=0.03, r=0.5$.}
\label{tab:2-cell GM hysteresis DvDu du=0.08}
\end{table}

\begin{figure}[H]
    \centering
    \begin{subfigure}{0.48\textwidth}
        \includegraphics[width=\textwidth,height=4.5cm]{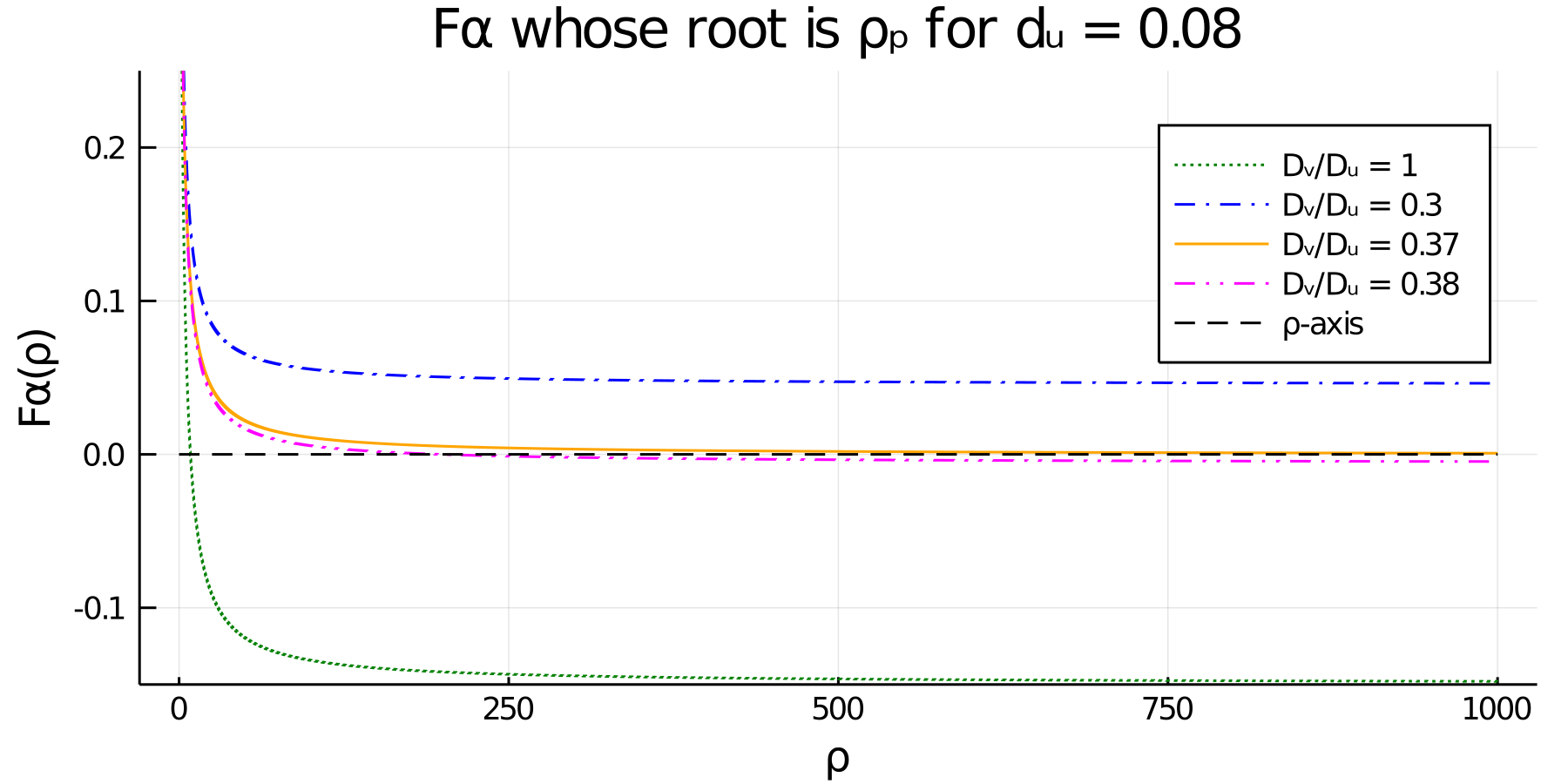}
    \end{subfigure}
    \begin{subfigure}{0.48\textwidth}
        \includegraphics[width=\textwidth,height=4.5cm]{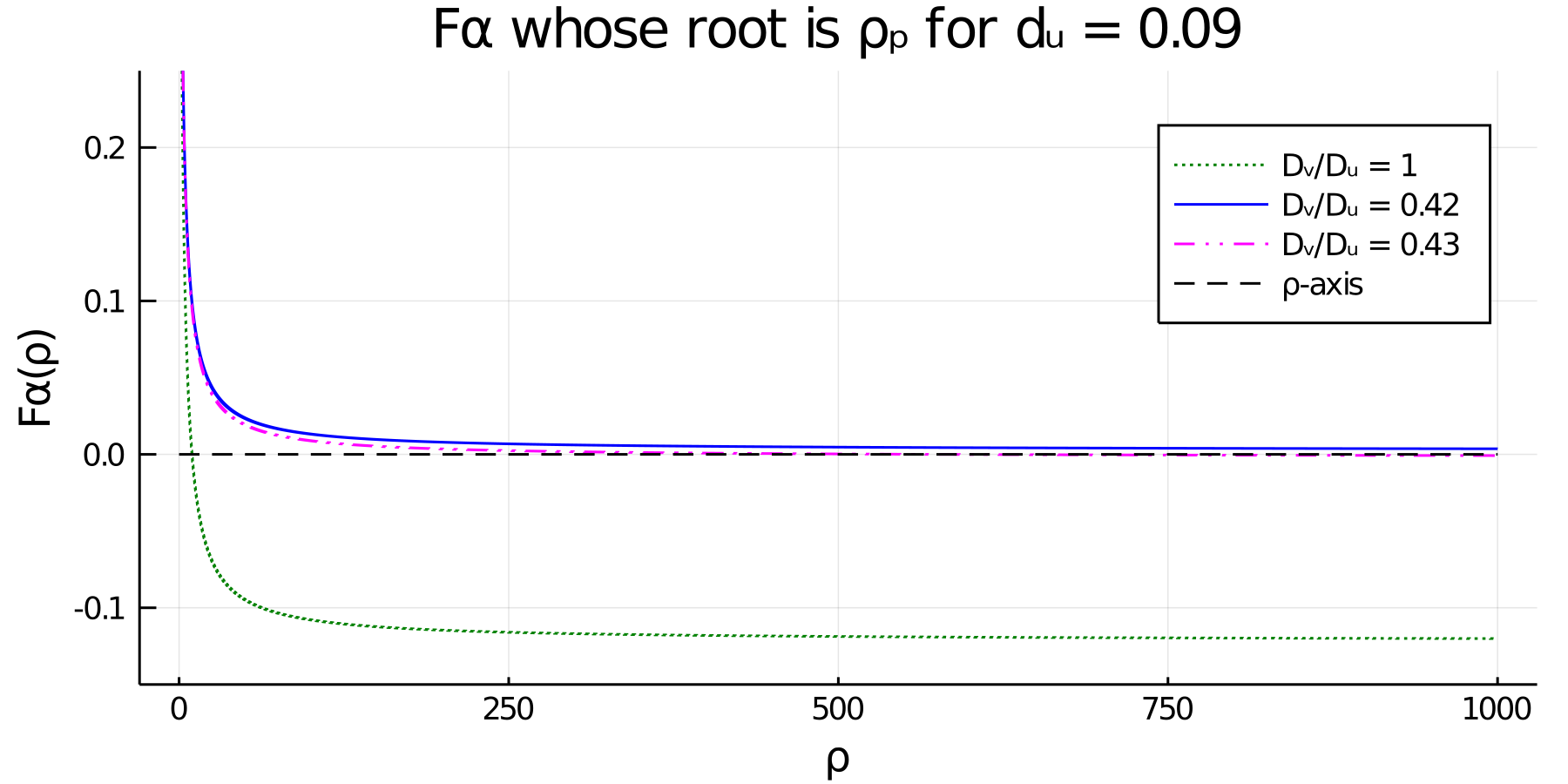}
    \end{subfigure}
    \caption{Effect of the diffusivity ratio ${D_v/D_u}$ on the
      existence of the pitchfork point when $d_u=0.08$ (left) and
      $d_u=0.09$ (right). The numerical results show that the
      asymptote of
      $F_\alpha(\rho) \equiv \alpha_v/\alpha_{v,2}^{\perp} +
      \alpha_{u,2}^{\perp}/(2\alpha_u) - 1$ is positive for smaller
      values of $D_v/D_u$, as suggested by Tables \ref{tab:2-cell GM
        hysteresis DvDu du=0.09} and \ref{tab:2-cell GM hysteresis
        DvDu du=0.08}. Therefore, when ${D_v/D_u}$ falls below a
      threshold, the pitchfork bifurcation condition $F_\alpha(\rho) = 0$,
      which is equivalent to \eqref{gm:2cells_symmb}, no longer holds
      for any $\rho>0$.}
    \label{fig:vanishingpitch}
\end{figure}

To obtain some analytical insight into the disappearance of the
pitchfork point as shown in Tables \ref{tab:2-cell GM hysteresis DvDu
  du=0.09} and \ref{tab:2-cell GM hysteresis DvDu du=0.08} when the
diffusivity ratio ${D_v/D_u}$ decreases below a threshold, in
Figure \ref{fig:vanishingpitch} we plot the function
$F_\alpha(\rho) \equiv \alpha_v/\alpha_{v,2}^{\perp} +
\alpha_{u,2}^{\perp}/(2\alpha_u) - 1$ versus $\rho$, representing the
left-hand side of the pitchfork bifurcation condition
(\ref{gm:2cells_symmb}), for several values of ${D_v/D_u}$, and for
either $d_u=0.08$ (left panel) or $d_u=0.09$ (right panel). From
(\ref{gm:2cells_symmb}) a root of $F_{\alpha}(\rho)=0$ corresponds to
a symmetry-breaking bifurcation point along the symmetric solution
branch.  From Figure \ref{fig:vanishingpitch} we observe that the
asymptote of $F_\alpha(\rho)$ as $\rho\to \infty$ is positive when
${D_v/D_u}$ is below a threshold, which eliminates the possibility of
a pitchfork bifurcation point.

\subsection{Rauch-Millonas reaction kinetics}

Next, we consider the activator-inhibitor system proposed in
\cite{rauch} to universally model two-species signal transduction
reaction kinetics between cells. This Rauch-Millonas (RM)
intracellular kinetics of \cite{rauch} is given by
\begin{equation}\label{cell:RM}
    \begin{array}{rclcl}
      \dot{\mu} &=& c_u - q_u \mu + \frac{a_1^u \mu}{b_1^u+\mu}
          - \frac{a_2^u \mu\eta}{b_2^u+\mu} &\equiv& f(\mu,\eta) \\
         \dot{\eta} &=& c_v + w_v\mu - q_v\eta &\equiv& g(\mu,\eta)\,.
    \end{array}
\end{equation}
Since $g$ has the form in (\ref{g:linear}), we identify that
$g_1(\mu)=c_v + w_v\mu$ and $g_2=q_v$.  We will choose a parameter set
for which the reaction kinetics when uncoupled from the bulk has a
unique linearly stable steady-state.

From (\ref{g:mu_solve}), all steady-states of the bulk-cell model for
a two-cell ring pattern are associated with the nonlinear algebraic
system
\begin{equation} \label{rm:2cells_all}
    \begin{array}{rcl}
      f(\mu_{e1}, {\bf e}_1^T\left(q_v I + \Theta_v\right)^{-1}
      (c_v + w_v\mu_{e1}, c_v + w_v\mu_{e2})^T) -
      {\bf e}_1^T \Theta_u(\mu_{e1},\mu_{e2})^T &=& 0 \\ 
      f(\mu_{e2}, {\bf e}_2^T \left(q_v I + \Theta_v\right)^{-1}
      (c_v + w_v\mu_{e1}, c_v + w_v\mu_{e2})^T) -
      {\bf e}_2^T \Theta_u(\mu_{e1},\mu_{e2})^T &=& 0\,.
    \end{array}
\end{equation}
By using (\ref{symm:scalar}), the symmetric steady-state solution branch is
obtained from the solution $\mu_e$ to
\begin{equation} \label{rm:symeqRM}
  c_u - q_u \mu_e + \frac{a_1^u \mu_e}{b_1^u+\mu_e} -
  \frac{a_2^u \mu_e}{b_2^u+\mu_e} \frac{(c_v + w_v\mu_e)}{q_v+\alpha_v} -
  \alpha_u\mu_e = 0\,,
\end{equation}
where $\alpha_u$ and $\alpha_v$ are given in (\ref{symm:alpkap}).
Symmetry-breaking bifurcation points are obtained by solving the
zero-eigenvalue crossing condition (\ref{two:red_simp}) together with
(\ref{rm:symeqRM}). By solving for $w_v=w_v(\mu_e)$ in
(\ref{two:red_simp}), we calculate
\begin{equation}\label{rm:wv_symm}
  w_v(\mu_e) = \frac{-q_u+\frac{a_1^u}{b_1^u+\mu_e} -
    \frac{ a_1^u \mu_e }{ (b_1^u+\mu_e)^2 } -
    \frac{ a_2^u }{ b_2^u+\mu_e } \frac{ c_v }{ q_v+\alpha_v } +
    \frac{ a_2^u\mu_e }{ (b_2^u+\mu_e)^2 } \frac{c_v}{q_v+\alpha_v} -
    \alpha_{u,2}^{\perp}}
  {\frac{ a_2^u }{ b_2^u+\mu_e }\frac{ \mu_e }{ q_v+\alpha_v } -
    \frac{ a_2^u\mu_e }{ (b_2^u + \mu_e)^2 }\frac{ \mu_e }{ q_v + \alpha_v }
    + \frac{ a_2^u\mu_e }{ b_2^u+\mu_e }\frac{1}{ q_v+\alpha_{v,2}^{\perp} } }\,,
\end{equation}
where $\alpha_{u,2}^{\perp}$ and $\alpha_{v,2}^{\perp}$ are defined in
(\ref{two:alpha_def}). By using (\ref{rm:wv_symm}) to eliminate
$\omega_v$ in (\ref{rm:symeqRM}) we obtain a nonlinear algebraic
equation that determines any symmetry-breaking bifurcation value for
$\mu_e$ along the symmetric solution branch. The corresponding
bifurcation value for $w_v$ is obtained from (\ref{rm:wv_symm}).

\begin{figure}[htbp]
  \centering
  \begin{subfigure}[b]{0.48\textwidth}
    		\centering
         \def\svgwidth{1\textwidth}
		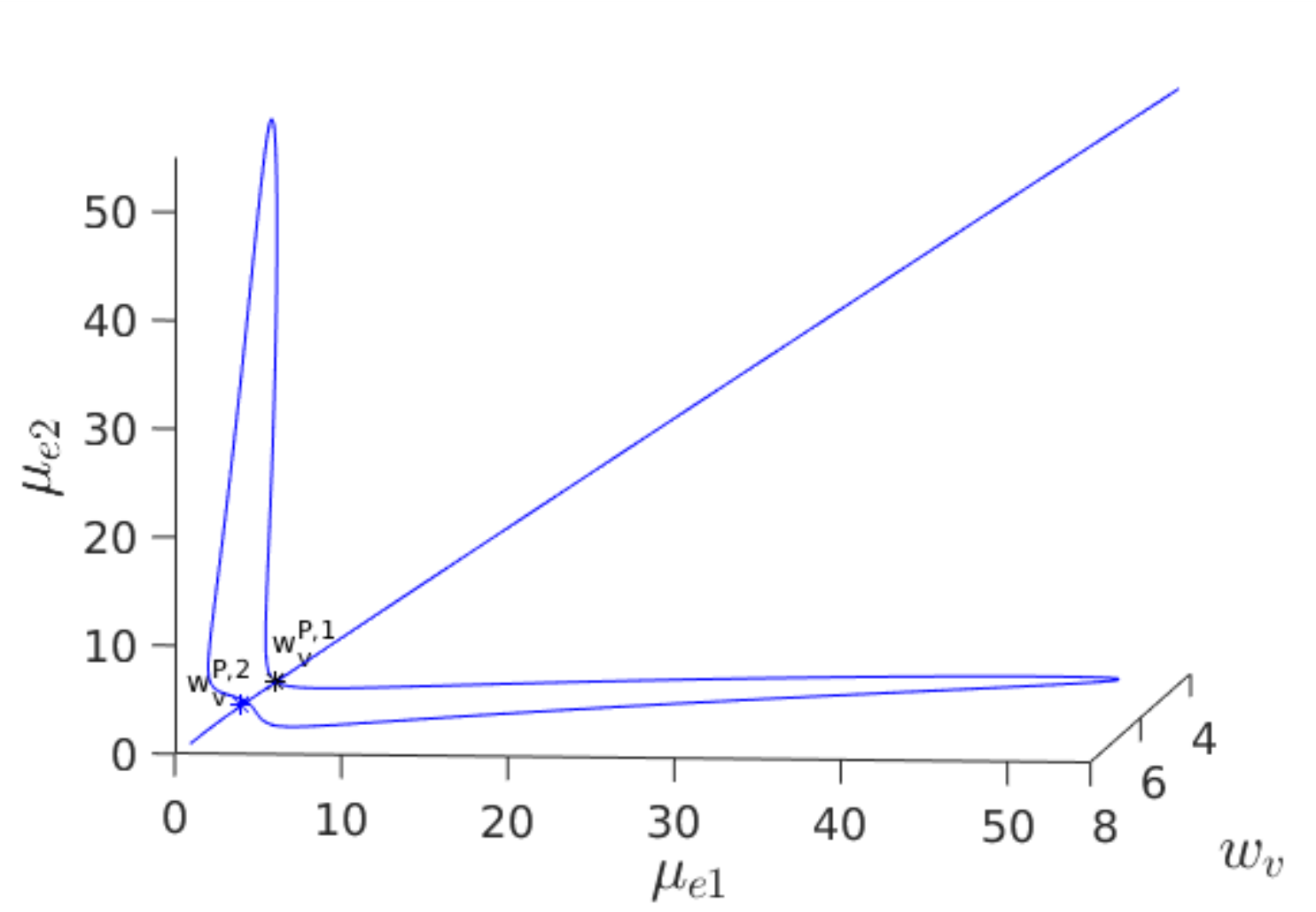  
	\end{subfigure}
  \begin{subfigure}[b]{0.48\textwidth}
    		\centering
         \def\svgwidth{1\textwidth}
		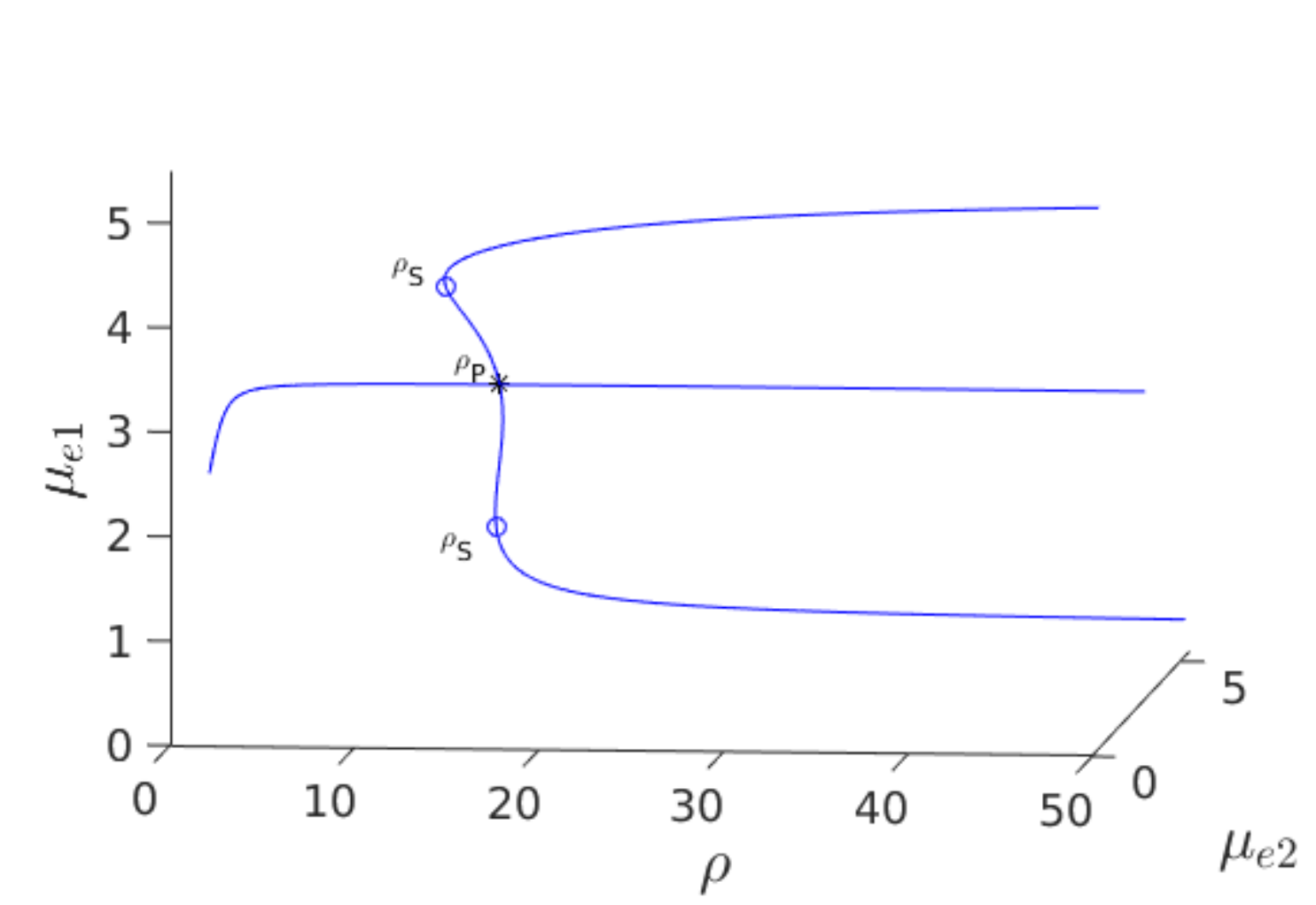  
	\end{subfigure}
        \caption{3-D Bifurcation diagram, computed from
          (\ref{rm:2cells_all}) using MatCont \cite{matcont}, for
          symmetric and asymmetric steady-states of a two-cell ring
          pattern with ring radius $r=0.5$ and with RM kinetics
          (\ref{cell:RM}). Left: 3-D Plot of $(\mu_{e1}, \mu_{e2})$
          versus the kinetic parameter $w_v$ in (\ref{cell:RM}) at a
          fixed $\rho={d_v/d_u}=15$, showing that asymmetric
          steady-states occur inside the degenerate pitchfork bubble
          delimited by $w_v^{P,1}\approx 6.88285$ and
          $w_v^{P,2}\approx 7.08723$.  Note that the bubble lobes are
          stretching into decreasing $w_v$ and that there exists
          hysteresis at $w_v^{P,2}$. Right: In terms of $\rho$, a
          supercritical pitchfork bifurcation from the symmetric
          branch occurs when $w_v=w_v^{P,2}\approx 7.08723$.  The
          asymmetric branches are linearly stable past this
          bifurcation threshold in $\rho$.  Parameters:
            $D_u = D_v= 1, \sigma_u=\sigma_v =0.1, \varepsilon = 0.03,
            c_u=c_v=1, q_u={1/100}, q_v=7, a_1^u=600, a_2^u=6,
            b_1^u=100, b_2^u={1/10}$, and
            $d_u=0.14$.} \label{fig:pbubbleandrhobifRM}
\end{figure}

\begin{figure}[htbp]
  \centering
   \includegraphics[width=0.7\textwidth]{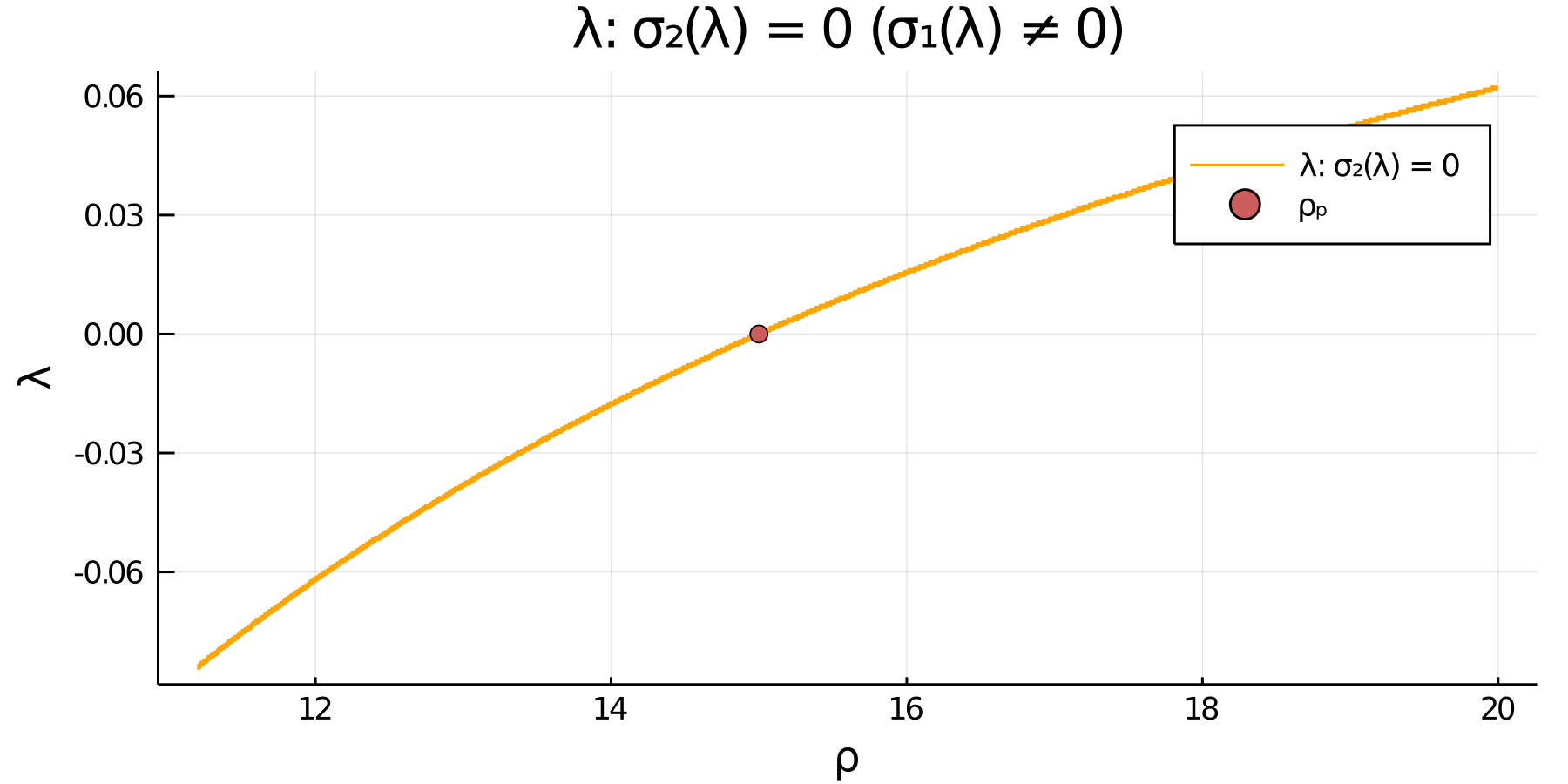}
   \caption{For the two-cell system with RM kinetics (\ref{cell:RM})
     and parameters as in the caption of Figure
     \ref{fig:pbubbleandrhobifRM} with $w_v=w_v^{P,2}$, we plot the
     eigenvalue $\lambda$, satisfying $\sigma_2(\lambda)=0$ in
     (\ref{lin:root}), versus $\rho$ that determines the linear
     stability of the symmetric steady-state solution branch to
     eigenperturbations of the form ${\bf q}_2=(1,-1)^T$. We observe
     that the symmetric steady-state branch is unstable only for
     $\rho>\rho_p=15$. There is no root to $\sigma_1(\lambda)=0$ on
     this range.} \label{fig:eigcrossingsRM}
\end{figure}

For the parameter set given in the figure caption, we show in the left
panel of Figure \ref{fig:pbubbleandrhobifRM} that, for the fixed value
$\rho=15$, there is a degenerate $w_v$-pitchfork bubble, which is
characterized by the emergence of asymmetric steady-state solutions at
two values of $w_v$. From the right panel of Figure
\ref{fig:pbubbleandrhobifRM} we observe that in terms of $\rho$, and
at a fixed $w_v$, the symmetry-breaking bifurcation is supercritical
in $\rho$.  For the parameter set in the right panel of Figure
\ref{fig:pbubbleandrhobifRM}, we observe from Figure
\ref{fig:eigcrossingsRM} that the eigenvalue $\lambda$ determined by
the root-finding condition $\sigma_2(\lambda)=0$, with $\sigma_2$
given in (\ref{lin:root}), crosses through zero at the
$\rho$-pitchfork bifurcation point along the symmetric steady-state
branch. As a result, when $w_v=w_v^{P,2}\approx 7.08723$, the
symmetric steady-state solution is linearly stable for
$\rho<\rho_p=15$, and is unstable on the range $\rho>\rho_p=15$ to
eigenperturbations in the direction of ${\bf q}_2=(1,-1)^T$.

\subsection{FitzHugh-Nagumo reaction kinetics}

Finally, we consider a ring pattern for the bulk-cell system
(\ref{eqsys:full}) with two cells and with FitzHugh-Nagumo (FN)
intracellular reaction kinetics \cite{gomez2007}. The uncoupled
intracellular kinetics are
\begin{equation}\label{cell:FN}
  \dot{\mu}(t) =  \mu - q(\mu-2)^3 + 4 - \eta \equiv f(\mu,\eta)\,,
  \qquad \dot{\eta}(t) =  \delta z \mu - \delta \eta \equiv
  g(\mu,\eta) \,,
\end{equation}
with $q>0$, $\delta>0$ and $z>0$. Since $g$ has the form in
(\ref{g:linear}), we identify that $g_1(\mu)=\delta z$ and
$g_2=\delta$.

\begin{figure}[htbp]
  \centering
  \begin{subfigure}[b]{0.48\textwidth}
    		\centering
         \def\svgwidth{1\textwidth}
		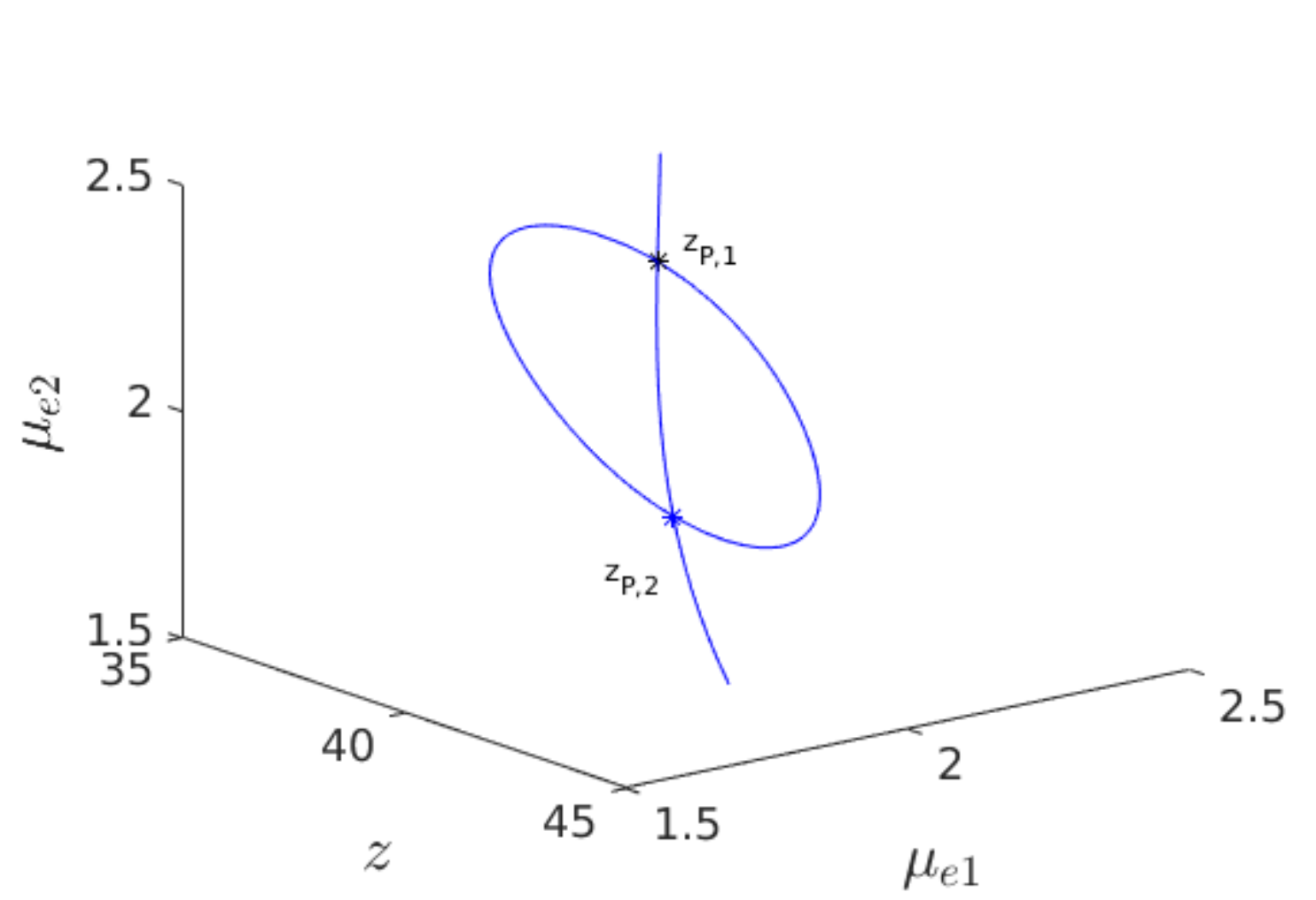  
	\end{subfigure}
  \begin{subfigure}[b]{0.48\textwidth}
    		\centering
         \def\svgwidth{1\textwidth}
		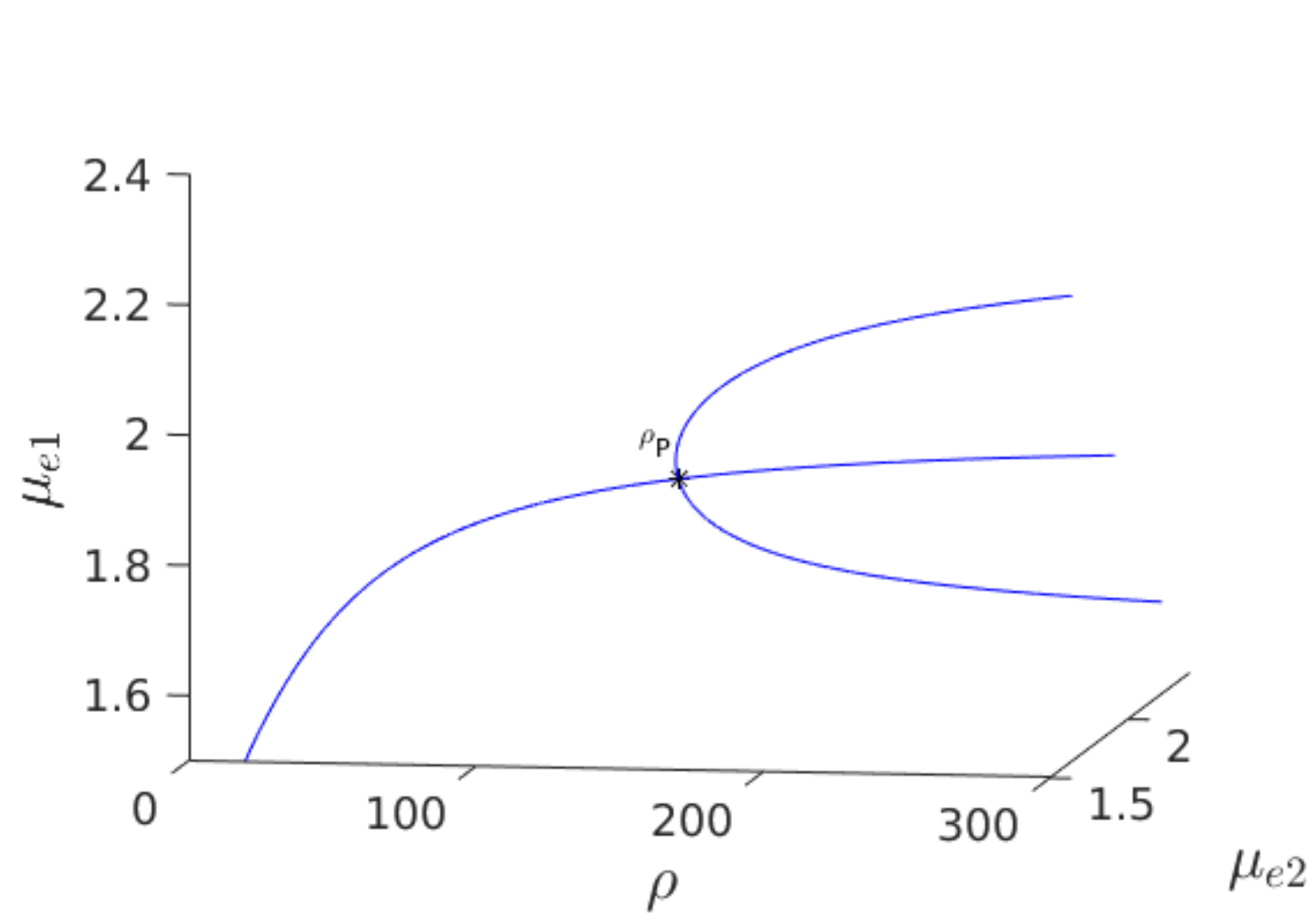  
              \end{subfigure}
              \caption{3-D Bifurcation diagram, computed from
                (\ref{fn:2cells_all}) using MatCont \cite{matcont},
                for symmetric and asymmetric steady-states of a
                two-cell ring pattern with ring radius $r=0.5$ and
                with FN kinetics (\ref{cell:FN}). Left: 3-D Plot of
                $(\mu_{e1}, \mu_{e2})$ showing that asymmetric
                steady-states occur inside the supercritical pitchfork
                bubble delimited by $z_{P,1} \approx 36.75458$ and
                $z_{P,2}\approx 41.26889$ when
                $\rho={d_v/d_u}=150$. Right: Supercritical pitchfork
                bifurcation from the symmetric branch occurs at
                $\rho_p=150$ when $z = z_{P,2}$.  Linearly stable
                asymmetric branches exist past this threshold in
                $\rho$.  Parameters:
                $D_u = 1, D_v= 4, \sigma_u=\sigma_v =1, \varepsilon =
                0.03, r=0.5, q = 1, \delta = 0.1$, and
                $d_u=0.04$.} \label{fig:pbubbleandrhobifFN}
\end{figure}

\begin{figure}[htbp]
  \centering
      \includegraphics[width=0.6\textwidth,height=5.3cm]{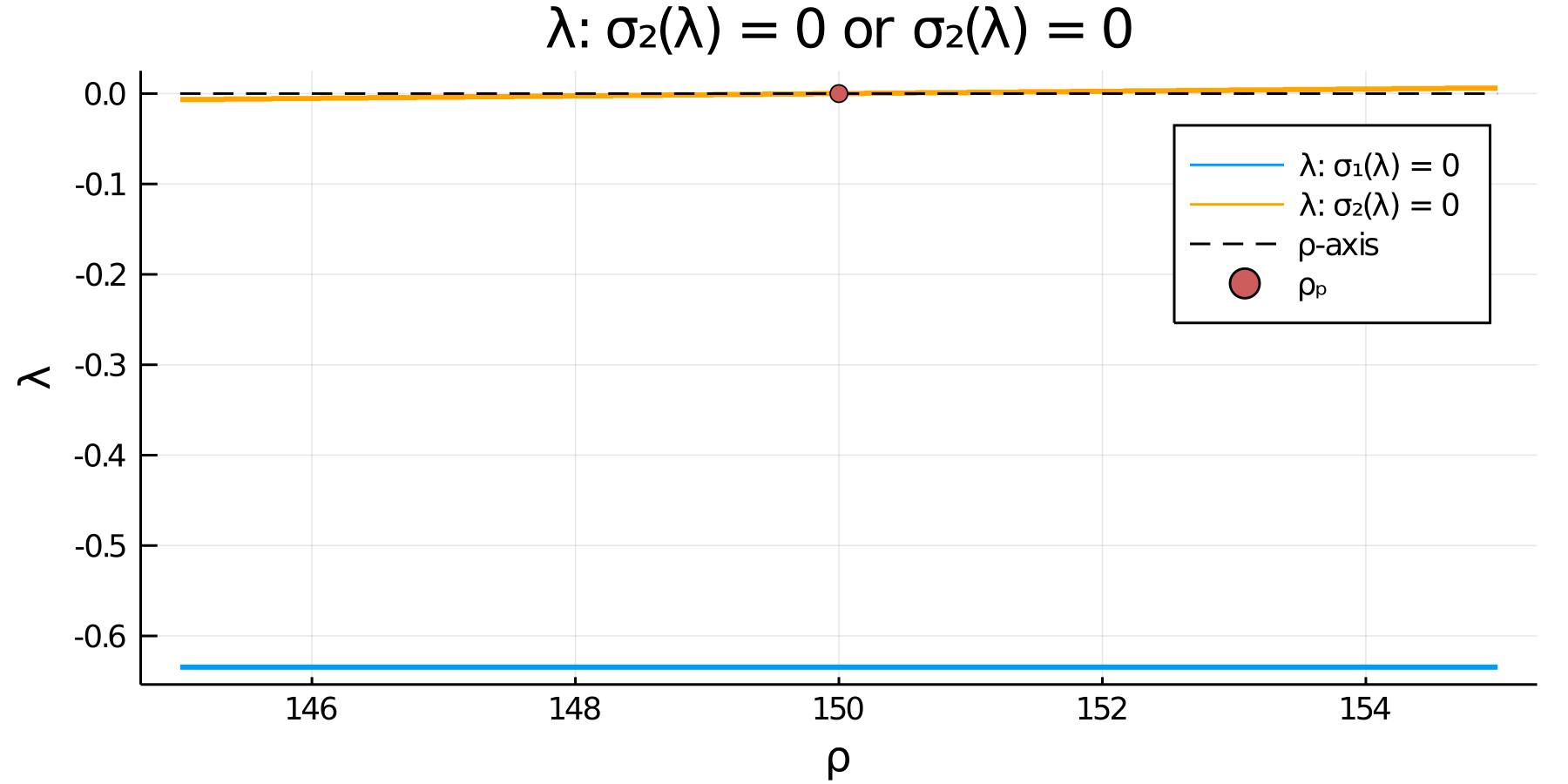}
      \caption{For the two-cell system with FN kinetics
        (\ref{cell:FN}) and parameters as in the caption of Figure
        \ref{fig:pbubbleandrhobifFN} with $z=z_{P,2}$, we plot the
        numerically computed eigenvalue $\lambda$, satisfying
        $\sigma_2(\lambda)=0$ in (\ref{lin:root}), versus
        $\rho$. Since $\lambda>0$ only on the range $\rho>\rho_p=150$,
        we conclude that the symmetric steady-state solution is
        linearly stable to anti-phase eigenperturbations of the form
        ${\bf q}_2=(1,-1)^T$ only when $\rho<\rho_p=150$. Moreover,
        since the root to $\sigma_1(\lambda)=0$ satisfies $\lambda<0$,
        we conclude that the symmetric steady-state branch is always
        linearly stable to in-phase eigenperturbations of the
        symmetric steady-state.}
	    \label{fig:eigcrossingsFN}
\end{figure}

We will choose a parameter set for which there is a unique linearly stable
steady-state of the intra-compartmental dynamics (\ref{cell:FN}).
From (\ref{g:mu_solve}), all steady-states of the bulk-cell model for
a two-cell ring pattern are obtained from the nonlinear algebraic
problem
\begin{equation} \label{fn:2cells_all}
    \begin{array}{rcl}
      f(\mu_{e1}, \delta z {\bf e}_1^T(\delta I +
      \Theta_v)^{-1}(\mu_{e1},
      \mu_{e2})^T) - {\bf e}_1^T \Theta_u(\mu_{e1},\mu_{e2})^T &=& 0 \\ 
      f(\mu_{e2}, \delta z {\bf e}_2^T(\delta I +
      \Theta_v)^{-1}(\mu_{e1},
      \mu_{e2})^T) -
      {\bf e}_2^T \Theta_u(\mu_{e1},\mu_{e2})^T &=& 0\,.
    \end{array}
\end{equation}
The symmetric steady-state solution branch, as characterized by
(\ref{symm:scalar}), is obtained from the root $\mu_e$ of the cubic equation
\begin{equation} \label{eq:symFN}
  \mu_e - q(\mu_e-2)^3 + 4 - \frac{\delta z \mu_e}{\delta +
    \alpha_v}
  - \alpha_u \mu_e = 0\,,
\end{equation}
where $\alpha_u$ and $\alpha_v$ are given in (\ref{symm:alpkap}).  The
symmetry-breaking bifurcation condition (\ref{two:red_simp}) along
the symmetric steady-state solution branch yields that
\begin{equation*}
  1-3q(\mu_e-2)^2 - \frac{\delta z}{\delta+\alpha_{v,2}^\perp} -
  \alpha_{u,2}^\perp = 0  \qquad \Leftrightarrow \qquad z(\mu_e) =
  \frac{\delta+\alpha_{v,2}^\perp}{\delta}
  \left(1-3q(\mu_e-2)^2-\alpha_{u,2}^\perp\right)\,,
\end{equation*}
where $\alpha_{u,2}^{\perp}$ and $\alpha_{v,2}^{\perp}$ are defined in
(\ref{two:alpha_def}).  We substitute $z(\mu_e)$ into
(\ref{eq:symFN}), and solve the resulting equation numerically for
$\mu_e$. For $\rho=150$, and with the parameters as in the caption of
Figure \ref{fig:pbubbleandrhobifFN}, we obtain that there are two
supercritical pitchfork bifurcation points $z_{P,1}$ and $z_{P,2}$ on
the symmetric steady-state branch. The linearly stable asymmetric
steady-state branches that exist on the range $z_{P,1}<z<z_{P,2}$
between the two pitchfork points are shown in the left panel of Figure
\ref{fig:pbubbleandrhobifFN}. When $z=z_{P,2}$, we observe from the
bifurcation diagram in the right panel of Figure
\ref{fig:pbubbleandrhobifFN}, together with the eigenvalue
computations in Figure \ref{fig:eigcrossingsFN}, that the
symmetry-breaking bifurcation is supercritical in terms of $\rho$.

\begin{figure}[htbp]
  \centering
      \begin{subfigure}{0.48\textwidth}
        \includegraphics[width=\textwidth,height=4.8cm]{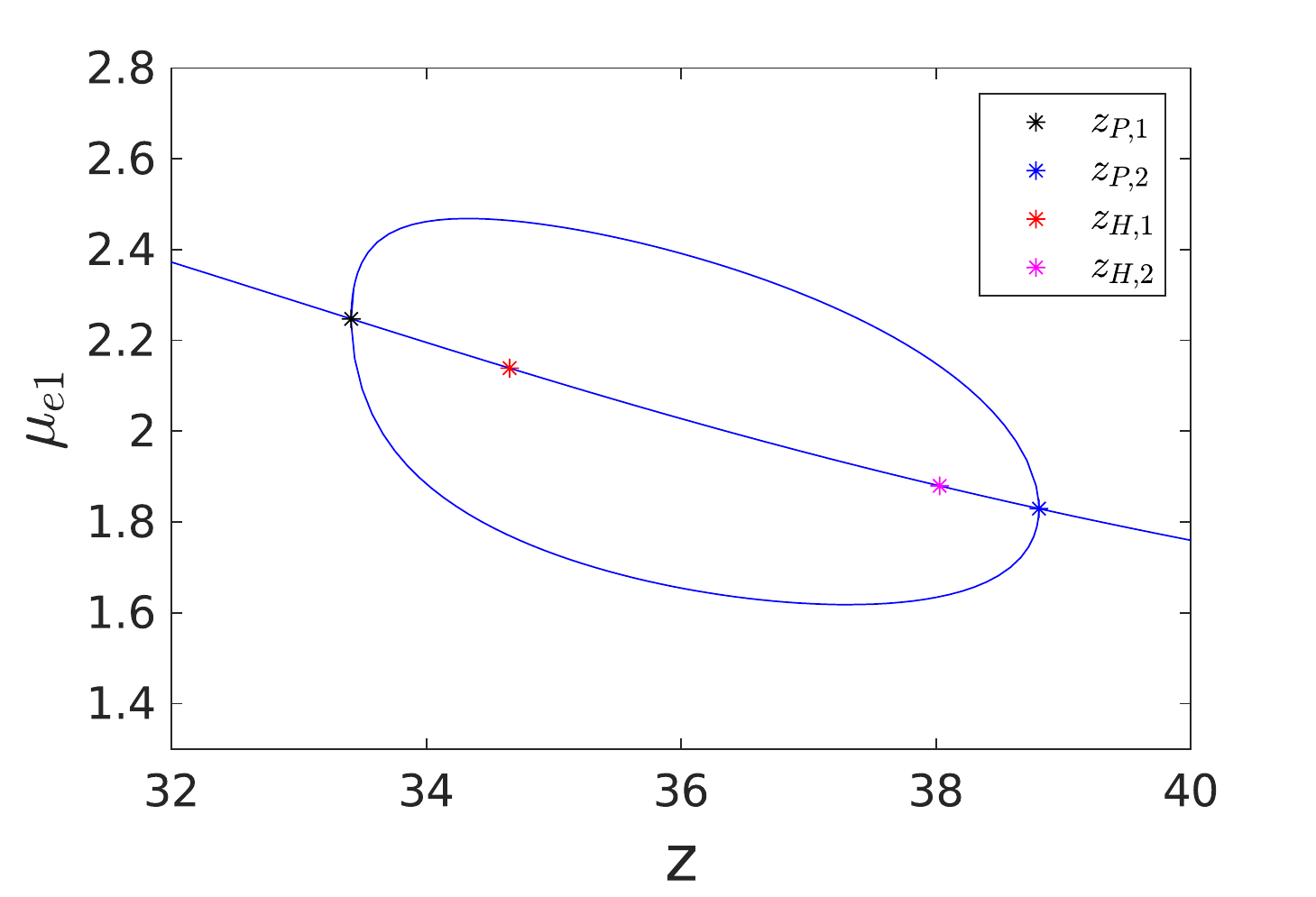}
    \end{subfigure}
    \begin{subfigure}{0.48\textwidth}
        \includegraphics[width=\textwidth,height=4.8cm]{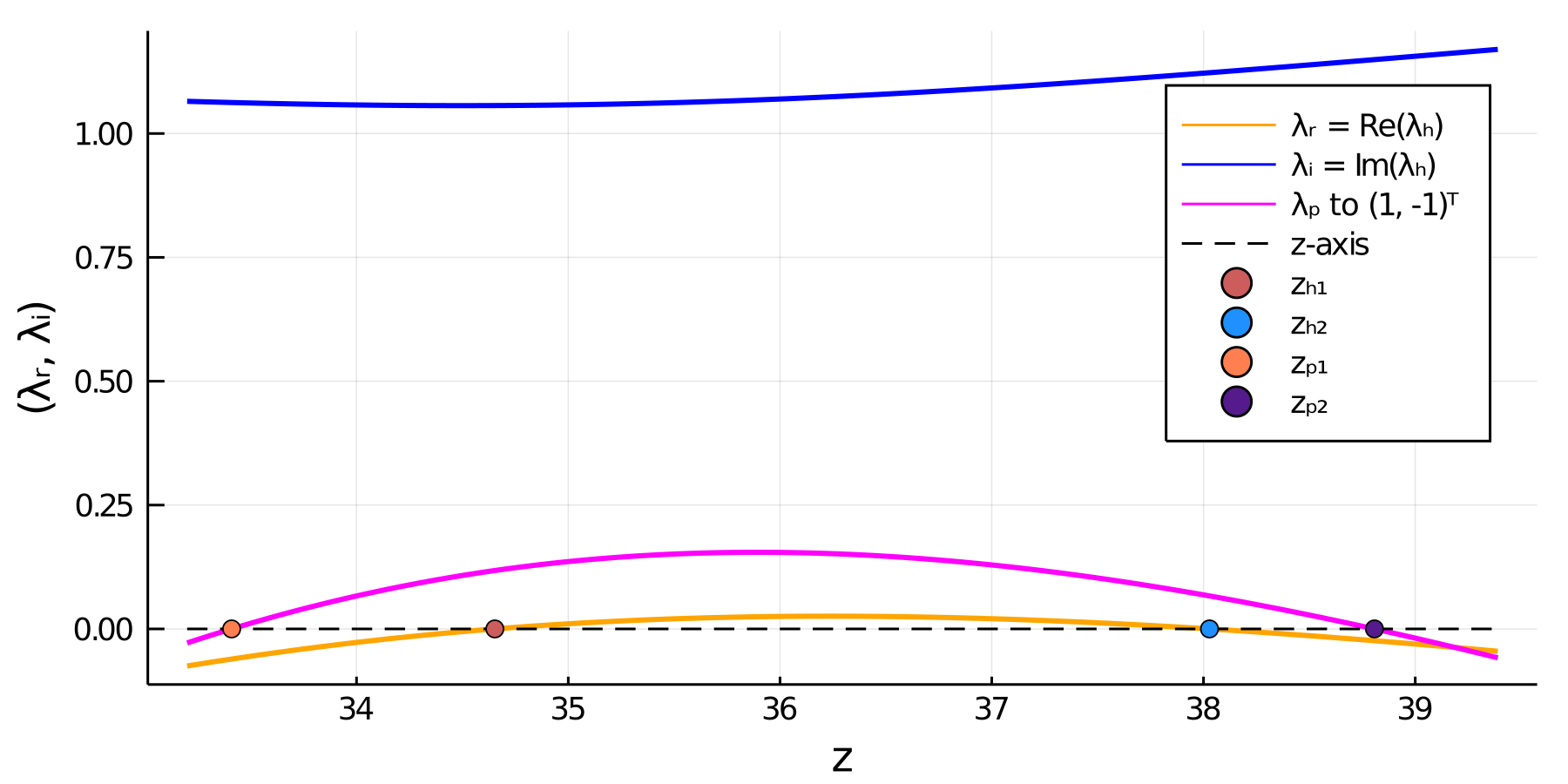}
    \end{subfigure}
    \caption{Left panel: Bifurcation diagram of $\mu_{e1}$ versus $z$,
      computed from (\ref{fn:2cells_all}) using MatCont
      \cite{matcont}, for symmetric and asymmetric steady-states of a
      two-cell ring pattern with the same parameters as in the left
      panel of Figure \ref{fig:pbubbleandrhobifFN} except that now the
      degradation rates are decreased slightly to
      $\sigma_u=\sigma_v=0.9$. For this parameter set, there are Hopf
      bifurcation points associated with in-phase perturbations of the
      symmetric steady-state that emerge at
      $z=z_{H,1}\approx 34.65328$ and $z=z_{H,2}\approx 38.02834$
      between the two symmetry-breaking pitchfork bifurcation points
      located at $z=z_{P,1}\approx 33.41022$ and
      $z=z_{P,2}\approx 38.80742$. Right panel: the root $\lambda$ of
      $\sigma_2(\lambda)=0$ versus $z$ (pink curve), as computed from
      (\ref{lin:root}), shows that the symmetric steady-state solution
      branch is unstable to anti-phase perturbations on the range
      $z_{P,1}<z<z_{P,2}$. The plotted real and imaginary parts of the
      complex-valued root $\lambda_h\equiv\lambda_r+i\lambda_i$ to
      $\sigma_1(\lambda_h)=0$, from (\ref{lin:root}), shows that
      $\mbox{Re}(\lambda_h)>0$ on the range $z_{H,1}<z<z_{H,2}$. On
      this range of $z$, a synchronous oscillatory instability of the
      symmetric steady-state solution can also occur, but it has a
      smaller growth rate than that for the anti-phase
      mode.} \label{fig:hopfFN}
\end{figure}

Next, we illustrate that the bulk-cell model with FN kinetics can also
exhibit oscillatory instabilities for in-phase perturbations of the
symmetric steady-state. In Figure \ref{fig:hopfFN} we plot the
bifurcation diagram of $\mu_{e1}$ versus $z$ for the same parameter
set as in the caption of Figure \ref{fig:pbubbleandrhobifFN} except
that the bulk degradation rates have been decreased slightly to
$\sigma_u=\sigma_v=0.9$. We observe that there are now two Hopf
bifurcation values $z_{H,1}$ and $z_{H,2}$ of $z$ along the symmetric
steady-state branch for the in-phase mode that lie within the interval
delimited by the two pitchfork bifurcation points. In the right panel of
Figure \ref{fig:hopfFN}, we plot the real and imaginary parts of the
complex-valued root of $\sigma_1(\lambda)=0$, as computed from
(\ref{lin:root}), which shows that $\mbox{Re}(\lambda)>0$ and
$\mbox{Im}(\lambda)\neq 0$ when $z_{H,1}<z<z_{H,2}$. This leads to the
possibility of a synchronous oscillatory instability. As a result, on
the range $z_{H,1}<z<z_{H,2}$, the symmetric steady-state solution
branch is unstable to both anti-phase and in-phase
perturbations. However, as seen from the right-panel of Figure
\ref{fig:hopfFN}, where we also plot the growth rate $\lambda$ for the
anti-phase mode as obtained by setting $\sigma_2(\lambda)=0$ in
(\ref{lin:root}), the anti-phase instability has a larger growth rate
than the in-phase instability.

\subsection{Numerical experiments with closely-spaced
  cells: GM kinetics}\label{sec:gm_close}

We now briefly explore, from full PDE simulations of
(\ref{eqsys:full}), symmetry-breaking behavior leading to stable
asymmetric patterns that can occur for closely spaced cells when the
ratio $\rho={d_v/d_u}$ is increased. For realistic modeling of pattern
formation properties of biological tissues one needs to consider the
situation where cells are closely spaced in the sense that the cell
radii are either comparable to the distance between the cells, or that
there are only narrow gaps between cells. Although the asymptotic
theory of \S \ref{sec:2d} and \S \ref{sec:stab} is no longer valid for
such closely spaced cell arrangements, the FlexPDE simulations of
(\ref{eqsys:full}) shown below reveal a similar qualitative solution
behavior as we have analyzed for spatially segregated cells. More
specifically, although we no longer have an analytical theory to
predict a bifurcation diagram of all steady-state solutions, our full
PDE numerical results suggest that stable symmetric steady-states
occur only when $\rho$ is below some threshold. When $\rho$ exceeds
some symmetry-breaking threshold, stable asymmetric steady-states will
be the preferred state. Our numerical results suggest that the
critical threshold of $\rho$ that is needed to establish this
symmetry-breaking behavior for closely spaced cells is smaller than
that needed for spatially segregated cells, if in fact such a
threshold exists.

\begin{figure}[]
    \centering
	\begin{subfigure}[b]{.45\textwidth}
	    \begin{subfigure}[b]{1.\textwidth}
	        \centering
                \def\svgwidth{1\textwidth}
                                    \def\svgheight{4.2cm}
	        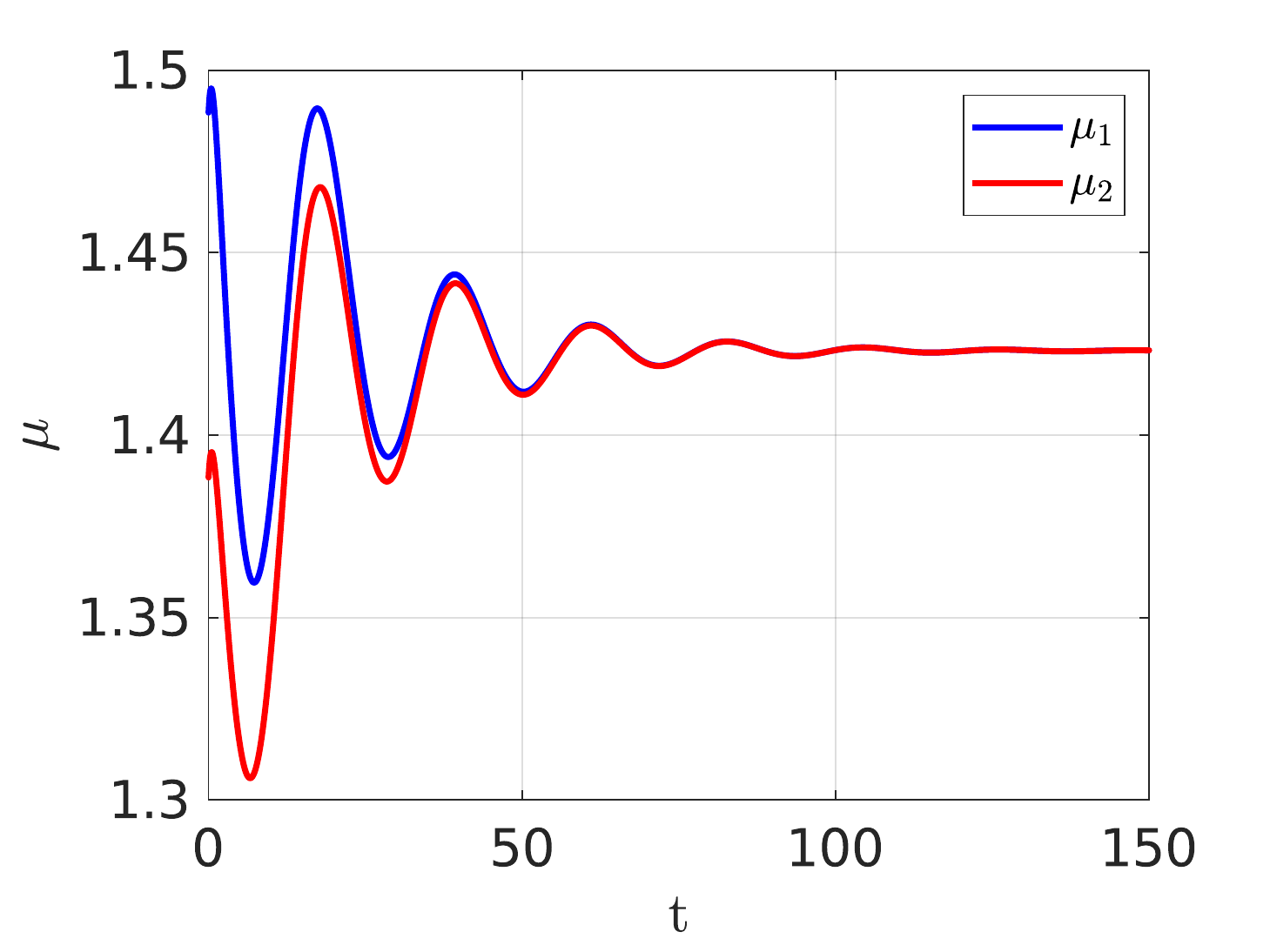
	    \end{subfigure}
	    \begin{subfigure}[b]{1.\textwidth}
	        \centering
                \def\svgwidth{1\textwidth}
                                    \def\svgheight{4.2cm}
	        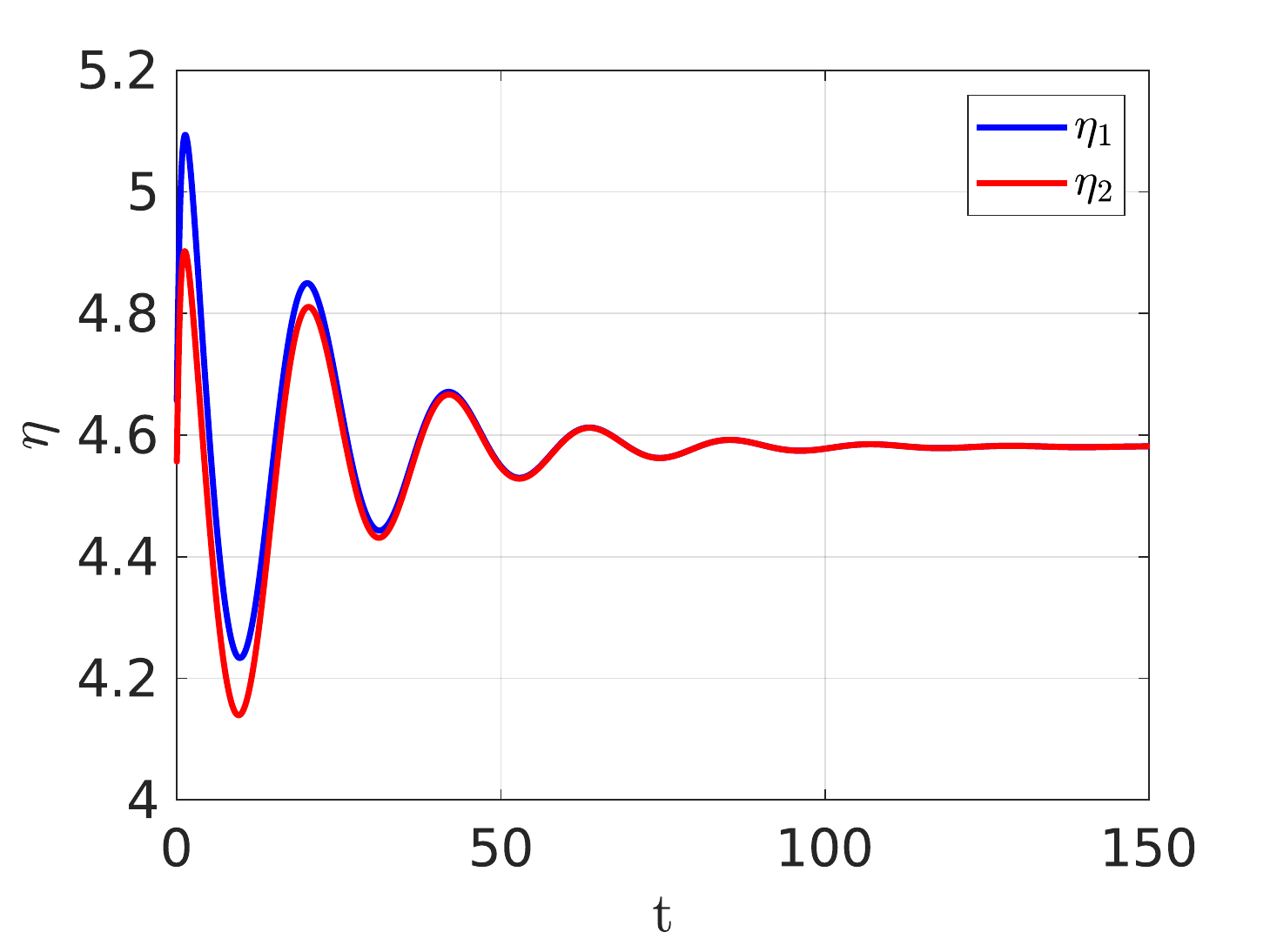
	    \end{subfigure}
	    \begin{subfigure}[b]{1.\textwidth}
    		\centering
                \def\svgwidth{1\textwidth}
                                    \def\svgheight{4.2cm}
			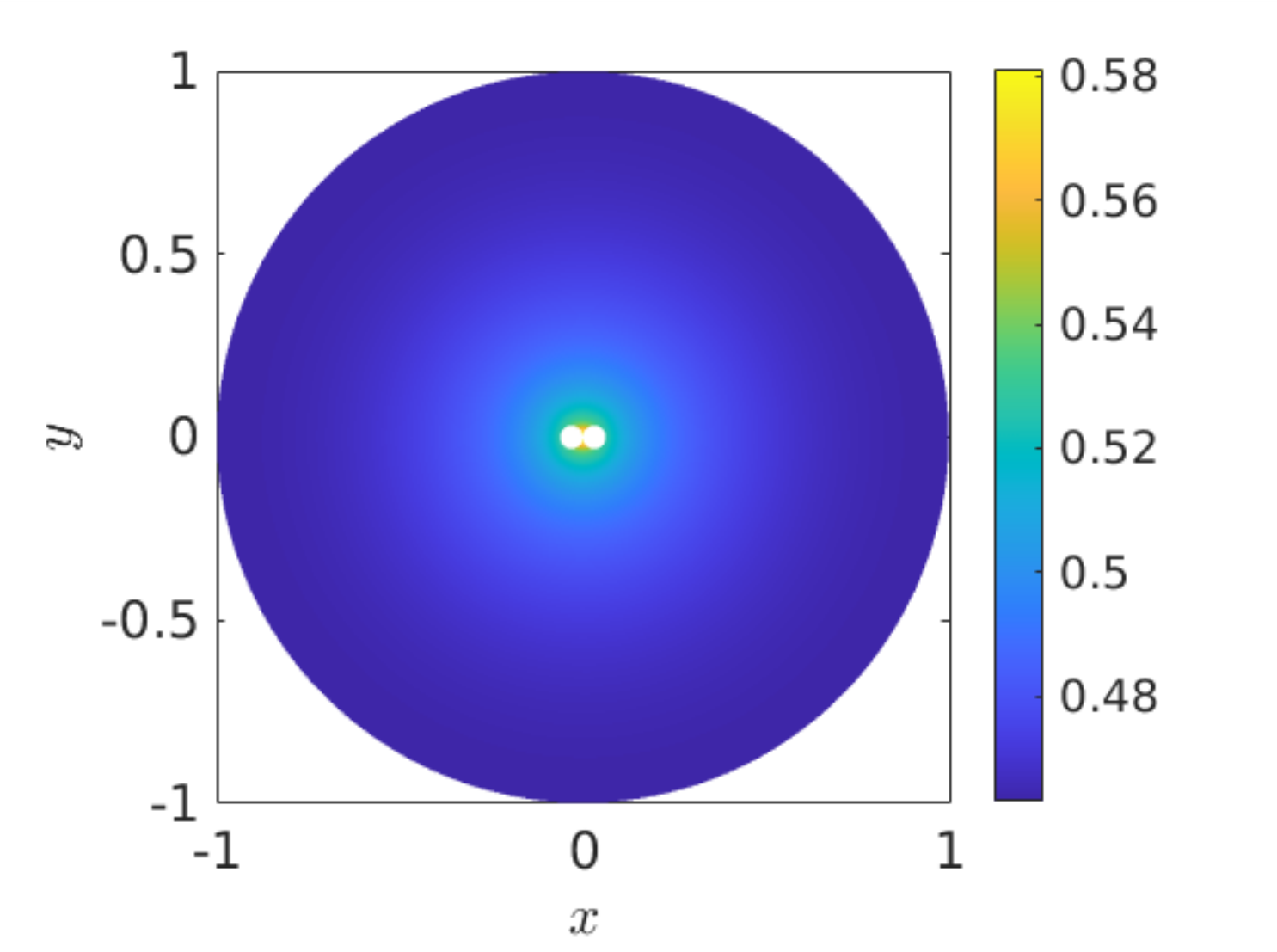  
		\end{subfigure}
	\end{subfigure}
	\begin{subfigure}[b]{.45\textwidth}
	    \begin{subfigure}[b]{1.\textwidth}
	        \centering
                \def\svgwidth{1\textwidth}
                                    \def\svgheight{4.2cm}
	        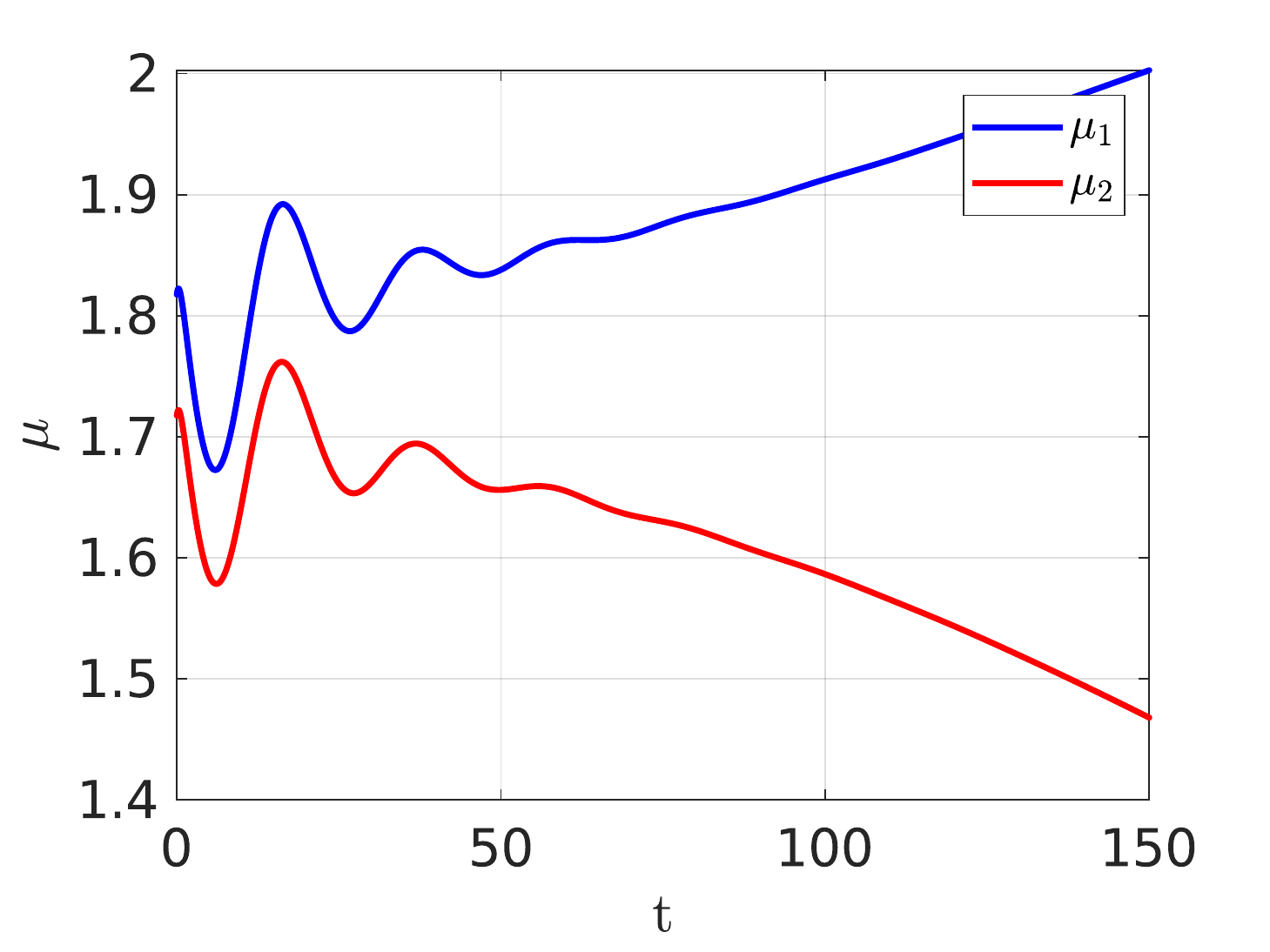
	    \end{subfigure}
	    \begin{subfigure}[b]{1.\textwidth}
	        \centering
                \def\svgwidth{1\textwidth}
                                    \def\svgheight{4.2cm}
	        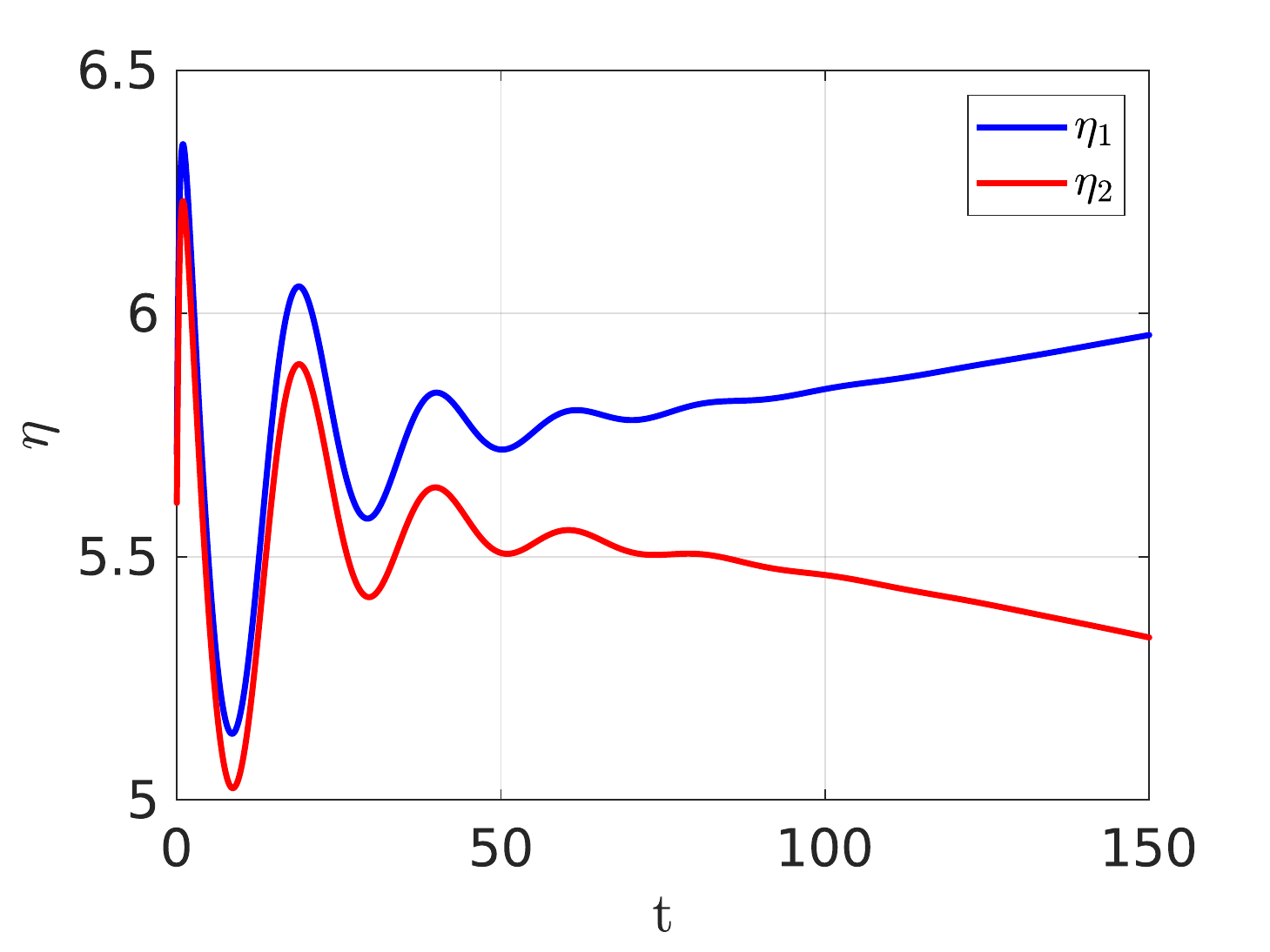
	    \end{subfigure}
	    \begin{subfigure}[b]{1.\textwidth}
    		\centering
                \def\svgwidth{1\textwidth}
                                    \def\svgheight{4.2cm}
			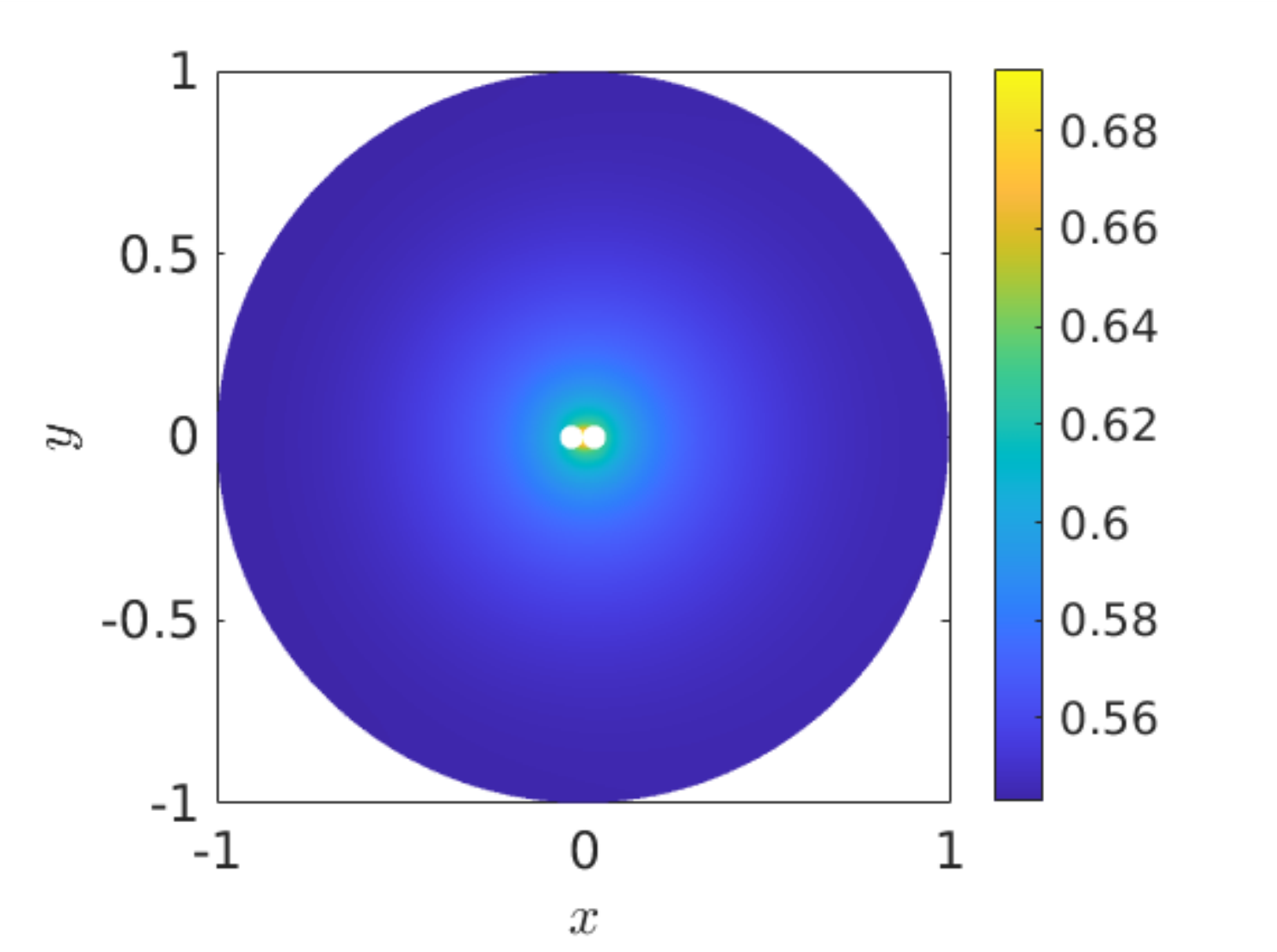  
		\end{subfigure}
	\end{subfigure}
	\caption{Full numerical simulation results of
          \eqref{eqsys:full} with FlexPDE \cite{flexpde} for GM
          kinetics (\ref{cell:GM}) with two closely spaced cells centered
          on a ring of radius $r=0.031$ and with minimum cell separation of
          $0.002$. The other parameters are the same as in Table
          \ref{tab:2-cell GM hysteresis DvDu du=0.08}. The only difference
          here is that the cells are now much more closely spaced.
          The bottom two panels show the concentration of $u$.
          Left: convergence to a stable symmetric steady-state
          solution when $\rho=3$. Right: convergence to a stable
          asymmetric steady-state soluton for $\rho = 8$ when
          starting with a symmetric initial condition. Parameters:
          $D_u=5$, $D_v=1.5$, $\sigma_u =\sigma_v=0.6$, $d_u=0.08$,
          and $\varepsilon=0.03$.}
	\label{fig:GM2closeDvDu0p3}
\end{figure}

To illustrate this, in Figure \ref{fig:GM2closeDvDu0p3} we take two
closely spaced cells centered near the origin that have a minimum
separation of $0.002$. The degradation rates, cell radius, and the
value of $d_u$ used for Figure \ref{fig:GM2closeDvDu0p3} are the same
as in Table \ref{tab:2-cell GM hysteresis DvDu du=0.08}, where
bifurcation values were given for the two-cell arrangement at
different ratios of ${D_v/D_u}$ with $D_u=5$ and for a ring radius
$r=0.5$.  In the cell arrangement in Figure \ref{fig:GM2closeDvDu0p3},
the only difference is that the two cells are now much more closely
spaced than in Table \ref{tab:2-cell GM hysteresis DvDu du=0.08} and
we fix ${D_v/D_u}=0.3$ and $D_u=5$. For these parameter values, we
observe from Table \ref{tab:2-cell GM hysteresis DvDu du=0.08} that no
symmetry-breaking bifurcations are possible for this diffusivity ratio
when the ring radius is $r=0.5$. However, as suggested from the
results shown in Figure \ref{fig:GM2closeDvDu0p3}, when the cells are
closely spaced there is a symmetry-breaking pitchfork bifurcation
point that occurs on the range $3<\rho<8$. We remark that, rather
surprisingly, if we use the symmetry-breaking bifurcation condition
(\ref{gm:2cells_symmb}) from the asymptotic theory for this case of
two-closely spaced cells it predicts that $\rho_p\approx 6.53$, which
lies within the range $3<\rho<8$. However, we emphasize that the
asymptotic theory is not valid for closely-spaced cells.

\begin{figure}[]
    \centering
	\begin{subfigure}[b]{.45\textwidth}
	    \begin{subfigure}[b]{1.\textwidth}
	        \centering
                \def\svgwidth{1\textwidth}
                                    \def\svgheight{4.2cm}
	        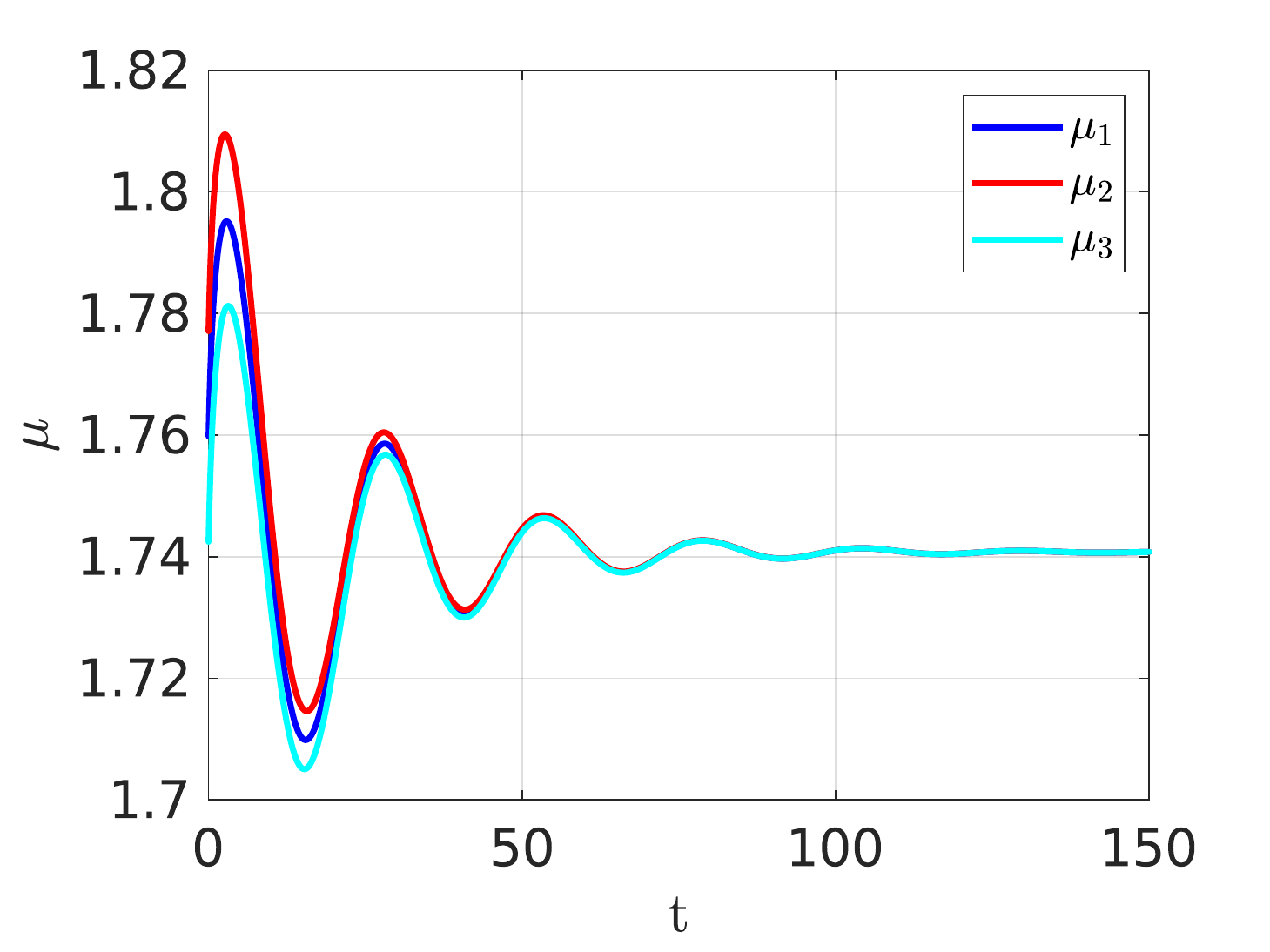
	    \end{subfigure}
	    \begin{subfigure}[b]{1.\textwidth}
	        \centering
                \def\svgwidth{1\textwidth}
                                    \def\svgheight{4.2cm}
	        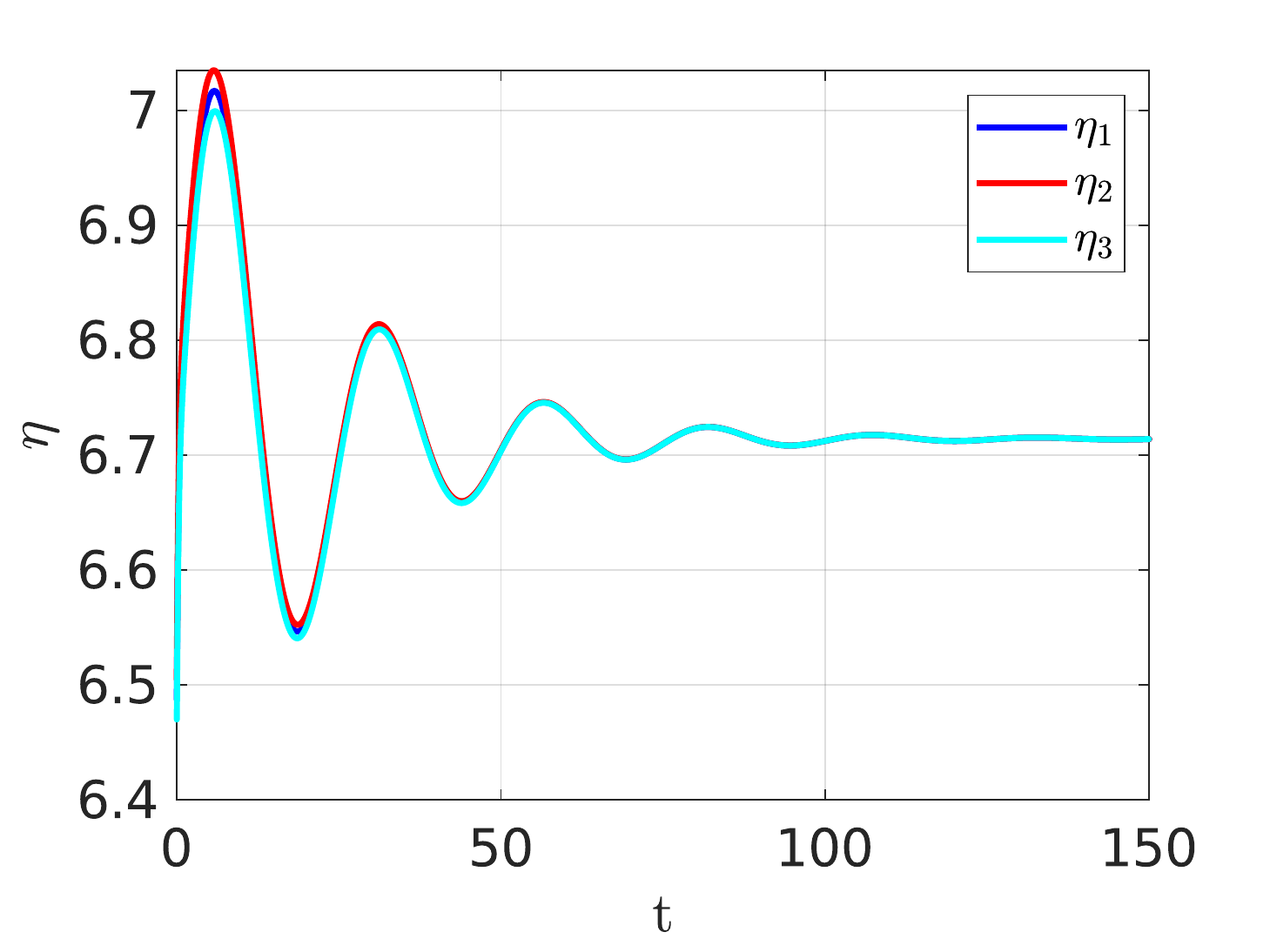
	    \end{subfigure}
	    \begin{subfigure}[b]{1.\textwidth}
    		\centering
                \def\svgwidth{1\textwidth}
                                    \def\svgheight{4.2cm}
			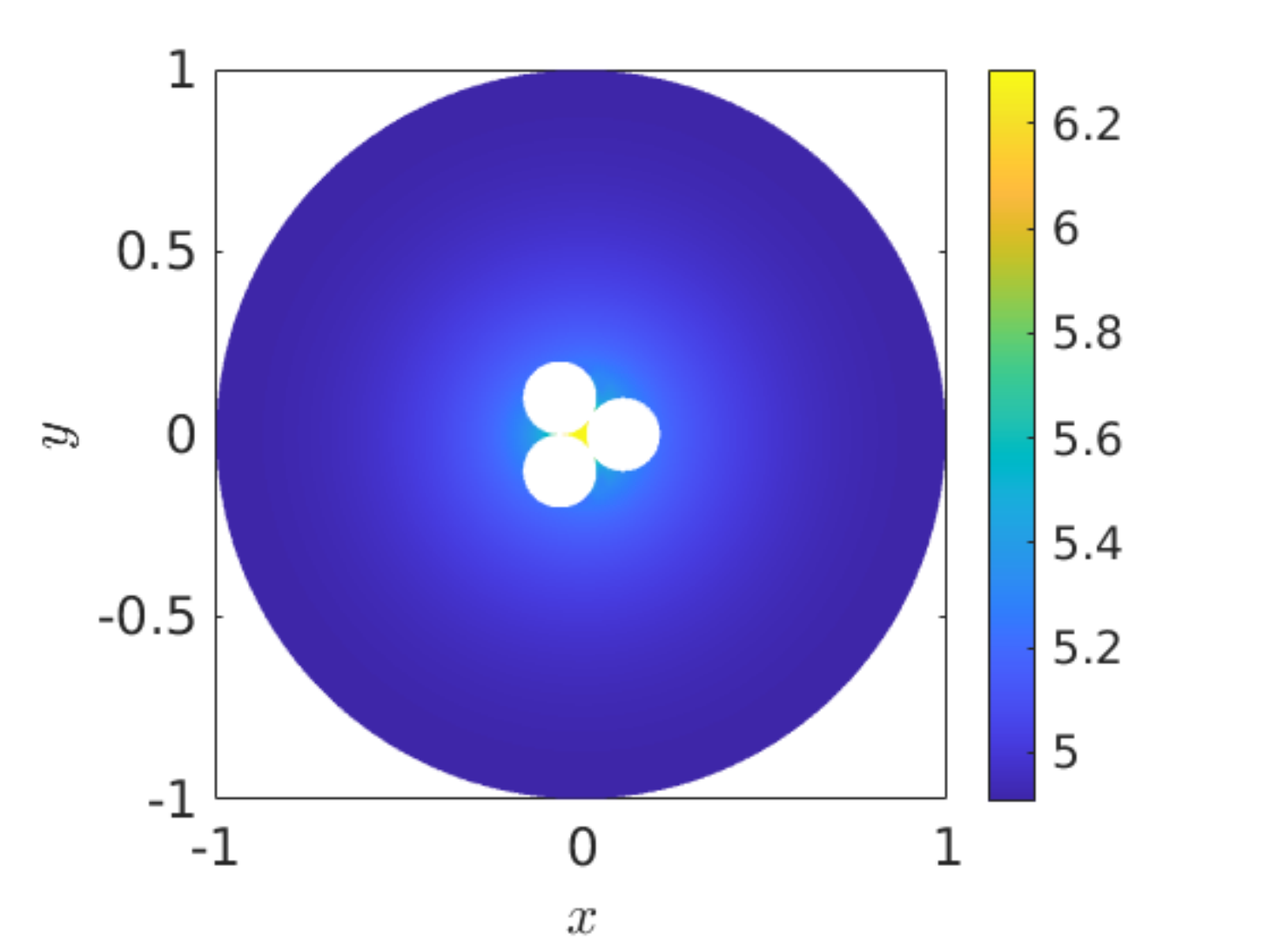  
		\end{subfigure}
	\end{subfigure}
	\begin{subfigure}[b]{.45\textwidth}
	    \begin{subfigure}[b]{1.\textwidth}
	        \centering
                \def\svgwidth{1\textwidth}
                                    \def\svgheight{4.2cm}
	        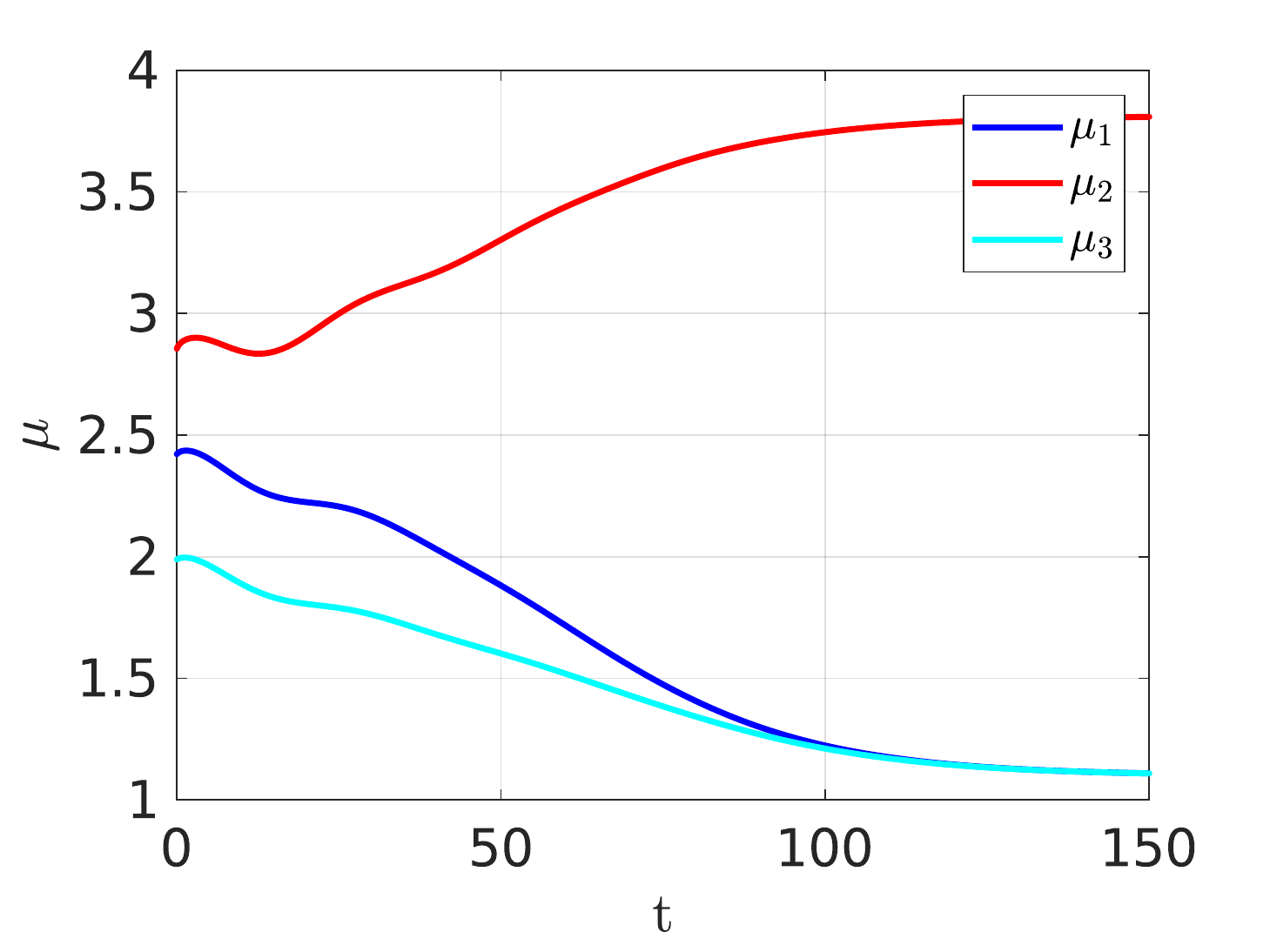
	    \end{subfigure}
	    \begin{subfigure}[b]{1.\textwidth}
	        \centering
                \def\svgwidth{1\textwidth}
                                    \def\svgheight{4.2cm}
	        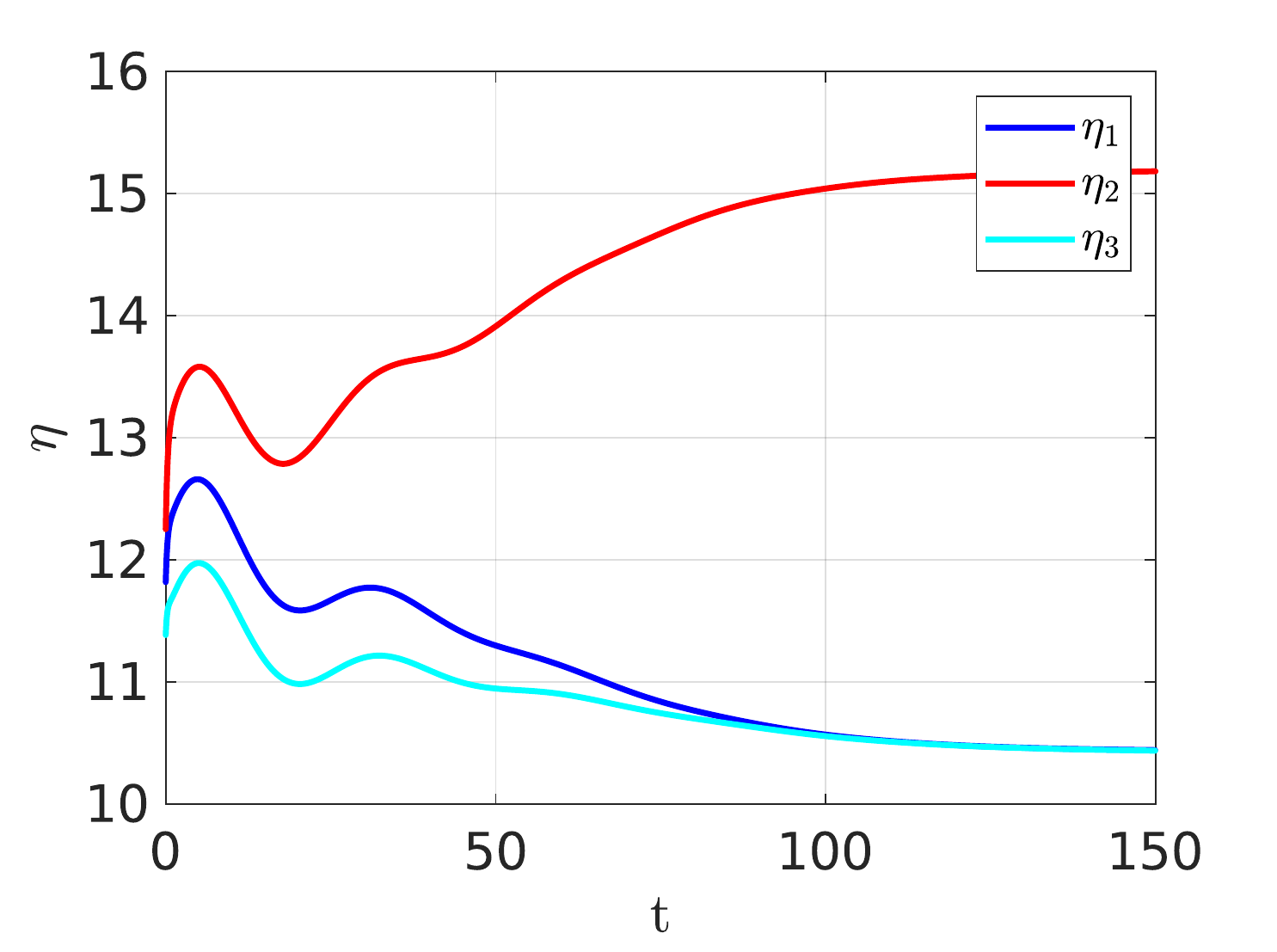
	    \end{subfigure}
	    \begin{subfigure}[b]{1.\textwidth}
    		\centering
                \def\svgwidth{1\textwidth}
                                    \def\svgheight{4.2cm}
			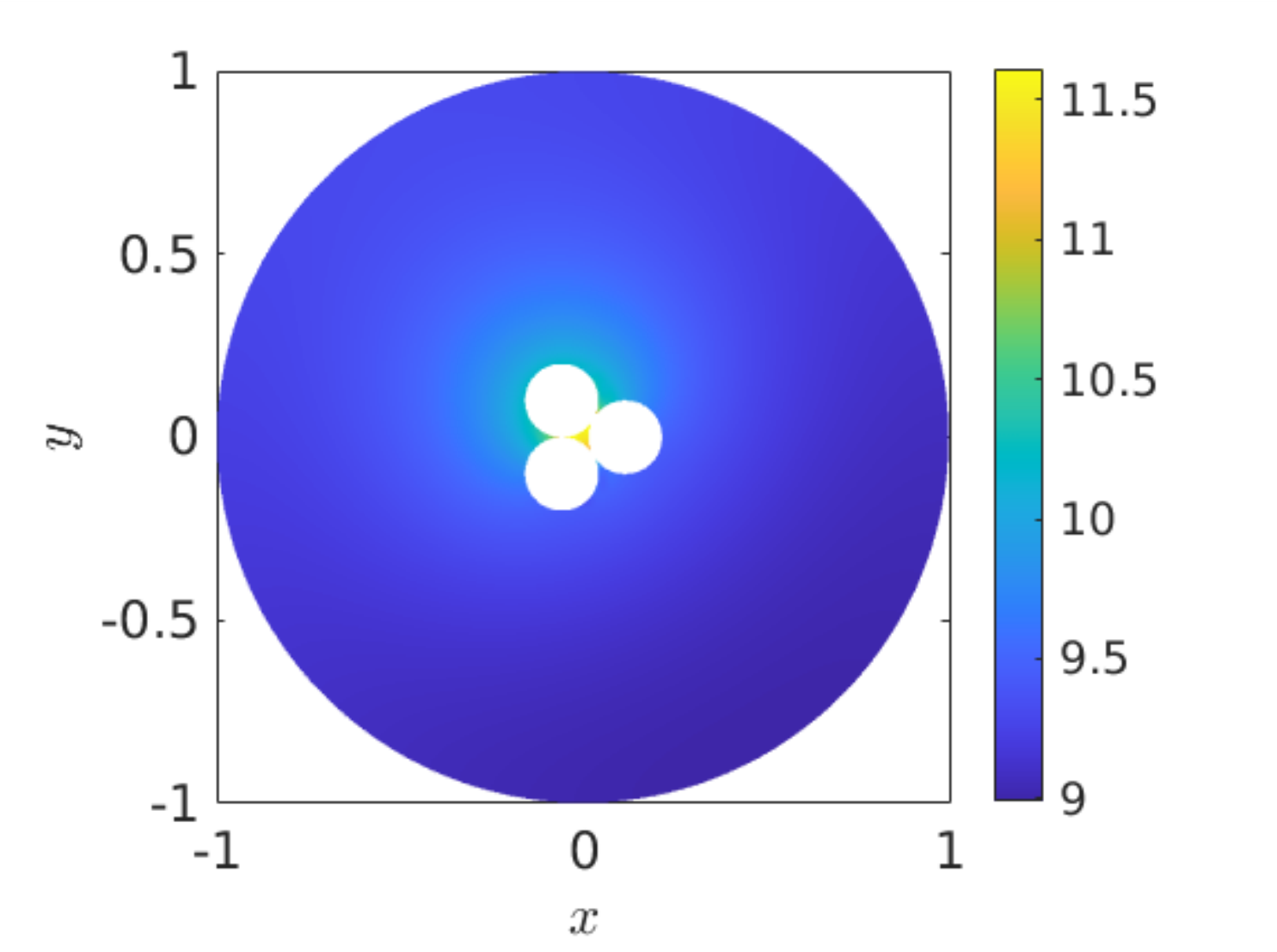  
		\end{subfigure}
	\end{subfigure}
	\caption{Full numerical simulation results of
          \eqref{eqsys:full} with FlexPDE \cite{flexpde} for GM
          kinetics (\ref{cell:GM}) with three closely spaced cells
          with cell centers located on the vertices of an equilateral
          triangle centered at the origin. The ring radius is
          $r=0.2{\sqrt{3}/6}$.   Left:
          convergence to a symmetric steady-state when $\rho=5$.
          Right: convergence to an asymmetric steady-state when
          $\rho = 10$.  Parameters:
          $D_u=D_v=5, \sigma_u =\sigma_v=0.6, d_u=0.05,
          \varepsilon=0.099$. The cells have a minimum separation of
          $0.002$. The bottom two panels show the
            concentration of $v$. By plotting $v$ rather than $u$, the
            bottom right panel clearly shows the asymmetry in the three
            cells.}
	\label{fig:GM3close}
\end{figure}

Finally, in Figure \ref{fig:GM3close} we show that the stable
asymmetric steady-state patterns can also occur for three
closely-spaced cells when $\rho$ exceeds some threshold.

\setcounter{equation}{0}
\setcounter{section}{4}
\section{Discussion}\label{sec:discussion}

We have analyzed symmetry-breaking behavior associated with the
PDE-ODE bulk-cell model (\ref{eqsys:full}) where identical
two-component intra-compartmental reactions occur only within a
disjoint collection of small circular compartments, or ``cells'', of a
common radius within a bounded 2-D domain. In the bulk, or
extra-cellular, medium two bulk species with comparable diffusivities
and bulk degradation rates diffuse and globally couple the spatially
segregated intracellular reactions. The bulk species are coupled to
the intracellular species through an exchange across the compartment
boundaries, as modeled by a Robin boundary condition that depends on
certain membrane reaction rates. In the limit of a small cell radius,
we have used a singular perturbation methodology to derive a nonlinear
algebraic system (\ref{non:full}) characterizing all the steady-states
for the bulk-cell model (\ref{eqsys:full}). Moreover, the linear
stability properties of the steady-state solutions of the bulk-cell
model were shown to be determined by the nonlinear matrix eigenvalue
problem (\ref{lin:mat_eig}) of size $2m\times 2m$, where $m$ is the
number of compartments. A root-finding condition on the determinant of
this matrix yields the discrete eigenvalues of the linearization
(\ref{eqsys:pertfull}) around an arbitrary steady-state solution, as
defined by the set (\ref{TransDent}).

We have shown that the steady-state and linear stability theory
simplifies considerably for a {\bf symmetric cell arrangement}, as
characterized by Definition \ref{def:symm}, and when one of the
intracellular species has a linear dependence of the form
(\ref{g:linear}). In this more restricted scenario, we have shown that
a symmetric steady-state solution, in which the steady-states of the
intracellular species are the same for each cell, will exist if the
scalar nonlinear algebraic equation (\ref{symm:scalar}) has a
solution. We emphasize that since our bulk-cell model does not admit
spatially homogeneous steady-state solutions that can be analyzed by a
simple Turing-type linear stability approach \cite{turing}, this
symmetric steady-state solution of the bulk-cell model
(\ref{eqsys:full}) represents the {\bf base state} in our
construction. Instabilities and bifurcations associated with this base
state are challenging to analyze owing to the fact that the base state
is not spatially uniform. Asymmetric steady-state solutions, as
determined from (\ref{g:mu_solve}), were shown to bifurcate from the
symmetric steady-state solution branch whenever the algebraic
criterion (\ref{break:red_simp}) is satisfied at some point on the
symmetric branch. For a symmetric cell arrangement, the linear
stability properties of the symmetric and asymmetric steady-state
solution branches are characterized by (\ref{new:TransDent}) and the
roots of the nonlinear matrix eigenvalue problem (\ref{lin:glin_non}),
respectively.

We have implemented our steady-state and linear stability theory for a
specific symmetric cell arrangement in which two cells are equally
spaced on a ring concentric within a unit disk, and where we have
specified either Gierer-Meinhardt, Rauch-Millonas, or FitzHugh-Nagumo
intracellular reactions, which all have the simplified form in
(\ref{g:linear}). By using parameter continuation numerical software
\cite{matcont} to implement the asymptotic theory, we have shown that
the symmetric steady-state solution branch can undergo {\bf
  symmetry-breaking pitchfork bifurcations}, leading to linearly
stable asymmetric patterns, even when the two bulk diffusing species
have identical diffusivities and degradation rates.  Overall, we have
shown that it is the magnitude of the ratio of the reaction rates for
the two bulk species to the cell membranes that determines whether
stable asymmetric patterns can occur. This membrane reaction rate
ratio threshold condition for the emergence of symmetry-breaking
bifurcations is in marked contrast to the well-known large diffusivity
ratio threshold condition for pattern formation from a spatially
uniform state that is typically derived by a Turing stability analysis
for two-component activator-inhibitor RD systems. For FitzHugh-Nagumo
and Rauch-Millonas kinetics we have also shown that stable asymmetric
patterns can also emerge from a symmetric steady-state pattern at a
fixed, but large, membrane reaction rate ratio when a control
parameter in the intracellular kinetics is varied. Our theoretical
predictions of symmetry-breaking behavior leading to linearly stable
asymmetric patterns for a symmetric two-cell arrangement were
confirmed through full time-dependent PDE computations of
(\ref{eqsys:full}).

We now briefly relate our theoretical results to some qualitative
behavior that has been suggested in chemical and biological
applications. Firstly, compartmental-reaction diffusion models of the
form (\ref{eqsys:full}) could potentially be useful for theoretically
modeling the collective behavior that occurs for a microemulsion
consisting of Belousov-Zhabotinsky (BZ) chemical reactants that are
confined within small aqueous droplets that is dispersed in oil
\cite{tompkins} (see also \cite{epstein_drop}, \cite{membrane}). In
this experimental set-up, polar BZ reactants and a catalyst are
confined within small immobile droplets, while two non-polar
intermediate species generated during the reaction can be transported
across the droplet boundaries. These intermediate species diffuse
across the domain, with comparable diffusivities, and provide the
mechanism for inter-drop coupling \cite{tompkins}. The recent
experimental study of \cite{membrane} has suggested that it is the
relative magnitude of the membrane reaction rates of these
intermediates on the droplet boundaries that plays a key role for
determining pattern-forming properties for BZ
microemulsions. Secondly, with regards to the transport of biological
morphogens, it has been suggested in \cite{morphogen} that a
differential reaction rate ratio on the cell boundaries for two
morphogen species with comparable diffusivities can yield the large
{\bf effective} diffusivity ratio that is needed for pattern formation
and symmetry-breaking in tissues. This membrane attachment mechanism,
which reduces the effective diffusivity of one of the morphogens and
is referred to in \cite{morphogen} as a {\bf binding-mediated
  hindrance} diffusion process, may be relevant in many biological
applications. Moreover, detailed intracellular mechanisms in
biological cells, such as signaling pathways and gene expression rate
constants, may also play a pivotal role in large-scale pattern-forming
properties of biological tissues \cite{morphogen}. By way of
qualitiative comparison, our theoretical analysis of the 2-D bulk-cell
model (\ref{eqsys:full}) for a very simple 2-cell pattern has revealed
that a large membrane reaction rate ratio, together possibly with changes in a
parameter in the intracellular kinetics, can trigger the emergence of
stable asymmetric steady-state patterns that bifurcate from a symmetry
steady-state. However, owing to the complexity of the analysis needed
for (\ref{eqsys:full}), where certain Green's matrices were found to
be central to the analysis, it appears rather intractable analytically
to isolate via a simple scaling analysis an {\bf effective
  diffusivity} for the bulk species that incorporates the membrane
reaction rates.

Although we have only applied our theoretical framework to a simple
two-cell arrangement, it is rather straightforward to numerically
implement the steady-state and linear stability theory for a symmetric
cell arrangement with a much larger number of cells. For this
scenario, the symmetric steady-state solutions are again determined by
the scalar nonlinear algebraic equation (\ref{symm:scalar}) and the
linear stability properties of this steady-state are readily studied
by computing the union of all the roots of the scalar problems
$\sigma_j(\lambda)=0$, for $j\in \lbrace{1,\ldots,m \rbrace}$, in
(\ref{lin:root}) that comprise the set (\ref{new:TransDent}) that
approximates the discrete eigenvalues of the linearization
(\ref{eqsys:pertfull}) of (\ref{eqsys:full}) around the
steady-state. However, for an arbitrary spatial arrangement of a large
number of cells, one key impediment for implementing the linear
stability theory for steady-state solutions is with regards to
numerically computing the eigenvalues $\lambda$ from a root-finding
condition on the determinant of the full $2m\times 2m$ GCEP matrix
${\mathcal M}(\lambda)$ in (\ref{lin:mat_eig}). This matrix is
non-Hermitian, is not sparse, and has an intricate dependence on
$\lambda$ through the Green's matrices. In contrast to the
availability of efficient numerical solution strategies for nonlinear
matrix eigenvalue problems with special structure, as was discussed in
\cite{guttel}, \cite{betcke} and \cite{betcke2}, it appears to a
significant open challenge to develop efficient numerical methods to
determine all such eigenvalues $\lambda$ for which
${\mathcal M}(\lambda)$ is a singular matrix when $m$ is large. Recall
that if there are any such eigenvalues in $\mbox{Re}(\lambda)>0$, the
steady-state for (\ref{eqsys:full}) is unstable.

A few other open problems related to our analysis are as
follows. Firstly, it would be interesting to analyze symmetry-breaking
bifurcation for (\ref{eqsys:full}) on $\R^2$ where identical cells of
small radii are centered at the lattice points of an arbitrary Bravais
lattice.  In this periodic setting, it should be possible to analyze
symmetry-breaking bifurcations of a periodic steady-state solution by
using Floquet-Bloch theory, combined with the explicit analytical
formulae for the reduced-wave Bloch Green's function as derived in
\cite{bloch}. Secondly, it would be interesting to develop an
extension of our asymptotic approach to treat closely-spaced cell
configurations that are more relevant to modeling pattern-forming
properties in biological tissues. Our numerical results shown in \S
\ref{sec:gm_close} have suggested that only a smaller membrane
reaction rate ratio is needed to initiate symmetry-breaking behavior
for closely-spaced cells than for arrangements with more spatially
segregated cells. To theoretically analyze pattern-forming properties
of the bulk-cell model with closely-spaced cells, an extension of the
approach developed in \cite{anziam} to analyze the mean first passage
time for a cluster of small traps may be fruitful. Thirdly, it would
be interesting to formulate and analyze a related bulk-cell model
where the chemical reactions occur on the boundaries of a collection
of small compartments, rather than in the interior of the
compartments. In this scenario, chemical species produced on the
membrane can then detach and diffuse in the bulk medium. Such an
extension is relevant for analyzing collective behavior that occurs
for dynamically reactive solid pellets that are chemically coated and
are coupled through a bulk diffusion field (cf.~\cite{showalter},
\cite{taylor1}, \cite{tinsley1}, \cite{tinsley2}). Finally, it
  would be worthwhile to extend our 2-D analysis to a 3-D setting. For
  a 3-D bounded domain that contains a collection of small spherical
  compartments, the analysis would be rather different than in 2-D
  since the free space Green's function has a rapid decay at infinity
  instead of a logarithmic growth. This would suggest that the
  inter-cell coupling effect would be, in general, much weaker in 3-D
  than in 2-D.

\begin{appendix}
\renewcommand{\theequation}{\Alph{section}.\arabic{equation}}
\setcounter{equation}{0}

\section{Non-dimensionalization}\label{app:nondim}

Here we non-dimensionalize \eqref{dim:pde} to obtain the PDE-ODE
system \eqref{eqsys:full}. Let $[z]$ denote the unit of some variable
$z$. In the SI unit system, we have
\begin{equation*}
        \begin{array}{rclrclrclrclrcl}
                [U] &=& \frac{\mol}{\m^2}\,, & [D_U] &=& \frac{\m^2}{\s}\,, &
[\kappa_U] &=& \frac{1}{\s}\,, & [T] &=& \s\,, & [X] &=& \m \,,\\
                \left[M_j\right] &=& \mol\,, & [\kappa_R] &=& \frac{1}{\s}\,, &
[\mu_c] &=& \mol\,, & [\beta_{U,1}] &=& \frac{\m}{\s}\,, & [\beta_{U,2}] &=&
\frac{1}{\m\,\s}\,.
        \end{array}
\end{equation*}
Letting $L$ denote the length-scale of the domain, we introduce the
dimensionless variables
$u\equiv {L^2 U/\mu_c}$, $v\equiv {L^2 V/\mu_c}$, $t\equiv\kappa_R T$,
${\bf x}\equiv {X/L}$, $\mu_j \equiv {M_j/\mu_c}$, and
$\eta_j \equiv {H_j/\mu_c}$. Then, we obtain that
\begin{equation*}
   \begin{array}{rcl}
        \kappa_R \,\partial_t U &=& \frac{D_U}{L^2}\, \Delta_{\bf x} U - \,
\kappa_U U\,, \\
        \kappa_R \,\partial_t V &=& \frac{D_V}{L^2} \,\Delta_{\bf x} V - \,
\kappa_V V\,,
   \end{array}\qquad 
   {\Large\Leftrightarrow}\qquad
   \begin{array}{rcll}
       \partial_t u &=& D_u \Delta_{\bf x} u - \sigma_u u, \quad &{\bf x} \in
\Omega\backslash\bigcup_{j=1}^m \Omega_j\,, \\
       \partial_t v &=& D_v \Delta_{\bf x} v - \sigma_v v, \quad & {\bf x} \in
\Omega\backslash\bigcup_{j=1}^m \Omega_j\,,
   \end{array}
\end{equation*}
where we have defined the dimensionless effective diffusivities
$D_u$ and $D_v$ and degradation rates $\sigma_u$ and $\sigma_v$ by
\begin{equation*}
        D_u \equiv \frac{D_U}{L^2\kappa_R}\,,\qquad  \sigma_u \equiv
\frac{\kappa_U}{\kappa_R}\,, \qquad
        D_v \equiv \frac{D_V}{L^2\kappa_R}\,, \qquad \sigma_v \equiv
\frac{\kappa_V}{\kappa_R}\,.
\end{equation*}
Since we assume that the common radius, denoted by $L_c$, of the cells
is much smaller than the domain length-scale $L$, we introduce
$\varepsilon\ll 1$ by $\varepsilon = {L_c/L} \ll 1$.

Next, by non-dimensionalizing the Robin boundary conditions in
(\ref{dim:pde}), we obtain
\begin{equation*}
        \begin{array}{rcl}
            \frac{D_U}{L}\frac{\mu_c}{L^2}\,\partial_{n_{\bf x}} u &=&
\beta_{U,1} \, \frac{\mu_c}{L^2}u - \beta_{U,2}\,\mu_c\mu_{j}\,, \\
            \frac{D_V}{L}\frac{\mu_c}{L^2}\,\partial_{n_{\bf x}} v &=&
\beta_{V,1}\, \frac{\mu_c}{L^2}v - \beta_{V,2} \,\mu_c\eta{j}\,,
        \end{array}
      \qquad {\Large\Leftrightarrow}\qquad
        \begin{array}{rcll}
            \varepsilon D_u \partial_{n_{\bf x}} u &=& d_{1}^u u - d_{2}^u
\mu_{j}, \quad  & {\bf x} \in \partial\Omega_j \,,\\
            \varepsilon D_v \partial_{n_{\bf x}} v &=& d_{1}^v v - d_{2}^v
\eta_{j},  & {\bf x} \in \partial\Omega_j\,,
        \end{array}
\end{equation*}
where we have defined $d_{1}^{u}$, $d_{2}^{u}$, $d_{1}^{v}$, and $d_{2}^{v}$ by
\begin{equation*}
  d_{1}^u \equiv\frac{\beta_{U,1}}{\kappa_R L}\varepsilon\,,
  \qquad d_{2}^u \equiv \frac{\beta_{U,2} L}{\kappa_R}\varepsilon\,, \qquad
  d_{2}^v\equiv \frac{\beta_{V,1}}{\kappa_R L}\varepsilon\,,
  \qquad d_{2}^v \equiv \frac{\beta_{V,2} L}{\kappa_R}\varepsilon\,.
\end{equation*}
Here, in order that there is an ${\mathcal O}(1)$ exchange across the
cell membranes we have assumed that $\frac{\beta_{U,1}}{\kappa_R L}$,
$\frac{\beta_{U,2} L}{\kappa_R}$, $\frac{\beta_{V,1}}{\kappa_R L}$ and
$\frac{\beta_{V,2} L}{\kappa_R}$ are all
$\O\left(\varepsilon^{-1}\right)$.

Lastly, we non-dimensionalize the intracellular reaction kinetics in
(\ref{dim:pde}) by
\begin{equation*}
        \begin{array}{lrrcl}
            &&\kappa_R\mu_c\, \frac{d}{dt}\mu_j &=& \kappa_R \mu_c \; f(\mu_j,
\eta_j) + \int_{\partial\Omega_j}\left(\beta_{U,1} \frac{\mu_c L}{L^2} \,u -
\beta_{U,2} \mu_c L \,\mu_j\right)\;dS_{\bf x} \,, \\
&&\kappa_R\mu_c\, \frac{d}{dt}\eta_j &=& \kappa_R \mu_c \; g(\mu_j,
\eta_j) + \int_{\partial\Omega_j}\left( \beta_{V,1} \frac{\mu_c L}{L^2} \,v -
\beta_{V,2} \mu_c L \,\eta_{j}\right)\;dS_{\bf x} \,,
        \end{array}
      \end{equation*}
which yields the dimensionless intracellular reactions
\begin{equation*}
            \frac{d\mu_j}{dt} = f(\mu_j, \eta_j) + \frac{1}{ \varepsilon}
\int_{\partial\Omega_j} (d_{1}^u\, u - d_{2}^u \,\mu_j)\;dS_{\bf x} \,, \qquad
\frac{d\eta_j}{dt} = g(\mu_j, \eta_j) + \frac{1}{ \varepsilon}
\int_{\partial\Omega_j} (d_{1}^v \,v - d_{2}^v \,\eta_{j})\;dS_{\bf x} \,,
\end{equation*}
for each $j\in\{1,...,m\}$. This completes the derivation of
(\ref{eqsys:full}).

\section{Reduced-wave Green's function for the unit disk}
\label{app:green}

When $\Omega$ is the unit disk, the reduced-wave Green's function
$G_{\omega}({\bf x}; {\bf \xi})$ and its regular part, satisfying
(\ref{R_Green}) can be determined analytically using separation of variables as
(see equations (6.10) and (6.11) of \cite{gou2d})
\begin{subequations} \label{cyclic:g}
  \begin{eqnarray} \label{cyclic:Glam} && G_\omega({\bf x};{\bf \xi})
    = \frac{\,1}{\,2 \pi} K_0\left(\omega|{\bf x}-{\bf \xi}|\right)
    -\frac{\,1}{\,2 \pi} \sum^{\infty}_{n=0} \beta_n
    \cos\left(n(\psi-\psi_0)\right) \frac{K^{\prime}_n(\omega)}
          {I^{\prime}_n(\omega)} I_n\left(\omega |{\bf x}|\right)
                     I_n\left(\omega |{\bf \xi}|\right)\,,\\
  && R_{\omega}({\bf \xi}) = \frac{1}{2\pi} \left(\log{2}
   - \gamma_e - \log{\omega} \right) -\frac{\,1}{\,2 \pi}
     \sum^{\infty}_{n=0} \beta_n  \frac{K^{\prime}_n(\omega)}
   {I^{\prime}_n(\omega)} \left[ I_n\left(\omega|{\bf \xi}|\right) \right]^2
                                          \,, \label{cyclic:Rlam}
\end{eqnarray}
\end{subequations}
where $\gamma_e\approx 0.5772$ is Euler's constant, and $I_n(z)$ and $K_n(z)$
are the modified Bessel functions of the first and second kind of
order $n$, respectively. In (\ref{cyclic:g}), $\beta_0\equiv 1$,
$\beta_n\equiv 2$ for $n\geq 1$, while
${\bf x}\equiv |{\bf x}|(\cos\psi,\sin\psi)^T$, and
${\bf \xi} \equiv |{\bf \xi}|(\cos\psi_0,\sin\psi_0)^T$.

For a ring pattern where the cell centers ${\bf x}_k$ for
$k\in \lbrace{1,\ldots,m\rbrace}$, are equidistantly spaced on a ring
of radius $r$ concentric within the unit disk, as in
(\ref{ex:ring_centers}), all the Green's matrices used in the
steady-state and linear stability analysis are circulant and symmetric
matrices. As a result, each such matrix spectrum is available
analytically.

Following Appendix A of \cite{ridgway}, for an $m\times m$ circulant
matrix $\mathcal{A}$, with possibly complex-valued matrix entries, its
complex-valued eigenvectors $\tilde{{\bf v}}_j$ and eigenvalues
$\alpha_j$ are
$\alpha_j = \sum\limits_{k=1}^m \mathcal{A}_{1k} \omega_j^{k-1}$ and
$\tilde{{\bf v}}_j = \left(1,Z_j,\ldots,Z_j^{m-1}\right)^T$, for
$j \in \lbrace{1,\ldots, m\rbrace}$. Here
$Z_j \equiv \exp{\left(\frac{2\pi i(j-1)}{m}\right)}$ and
${\mathcal A}_{1k}$, for $k\in\lbrace{1,\ldots,m\rbrace}$, are the
elements of the first row of ${\mathcal A}$. Since ${\mathcal A}$ is
also a symmetric matrix, we have
${\mathcal A}_{1,j}={\mathcal A}_{1,m+2-j}$, for
$j\in\lbrace{2,\ldots,\lceil m/2\rceil\rbrace}$, where the ceiling
function $\lceil x \rceil$ is defined as the smallest integer not less
than $x$.  Therefore, $\alpha_j=\alpha_{m+2-j}$, for
$j\in\lbrace{2,\ldots,\lceil m/2\rceil\rbrace}$, so that there are
$m-1$ eigenvalues with a multiplicity of two when $m$ is odd, and
$m-2$ such eigenvalues when $m$ is even. As a result, we conclude that
$\frac{1}{2}\left[\tilde{{\bf v}}_j+\tilde{{\bf v}}_{m+2-j}\right]$
and
$\frac{1}{2i}\left[\tilde{ {\bf v}}_j-\tilde{{\bf v}}_{m+2-j}\right]$
are two independent real-valued eigenvectors of ${\mathcal A}$,
corresponding to the eigenvalues of multiplicity two. In summary, the
matrix spectrum of a circulant and symmetric matrix ${\mathcal A}$, where
the eigenvectors have been normalized by ${\bf v}_j^T{\bf v}_j=1$, is
\begin{subequations}\label{app:cyclic_mat}
\begin{eqnarray}
 && \qquad \alpha_j = \sum\limits_{k=1}^m \mathcal{A}_{1k}
    \cos\left(\theta_j (k-1)\right)\,, \quad j\in \lbrace{1,\ldots,m\rbrace}
    \,; \qquad
    \theta_j\equiv  \frac{2\pi (j-1)}{m}\,; \qquad
    {\bf v}_1=\frac{1}{\sqrt{m}}{\bf e} \,, \\
&&  {\bf v}_j = \sqrt{\frac{2}{m}} \left(1, \cos \left(\theta_j
        \right), \ldots,\cos\left( \theta_j (m-1) \right)\right)^{T}
      \,, \quad
          {\bf v}_{m+2-j} = \sqrt{\frac{2}{m}}
      \left(0, \sin \left( \theta_j \right),
      \ldots, \sin \left( \theta_j (m-1) \right) \right)^{T} \,,
\end{eqnarray}
\end{subequations}
for $j \in \lbrace{2,\ldots, \lceil m/2\rceil\rbrace}$, where
$\theta_j\equiv {2\pi (j-1)/m}$.  When $m$ is even, there is an
additional normalized eigenvector of multiplicity one given by
${\bf v}_{{m/2}+1}=m^{-1/2}(1,-1,1,\ldots,-1)^T$.

\end{appendix}


\vskip2pc

\vspace*{-1.1cm}
\bibliographystyle{plain}
\bibliography{2dturing}

\end{document}